\documentclass[fleqn,usenatbib]{mnras}

\usepackage{mathptmx}
\usepackage{txfonts}

\usepackage[T1]{fontenc}

\DeclareRobustCommand{\VAN}[3]{#2}
\let\VANthebibliography\thebibliography
\def\thebibliography{\DeclareRobustCommand{\VAN}[3]{##3}\VANthebibliography}

\usepackage{graphicx}	% Including figure files
\usepackage{amssymb}	% Extra maths symbols
\usepackage{placeins}
\usepackage{tabularx}
\usepackage{enumitem}
\usepackage{tabulary}
\usepackage{booktabs}
\usepackage{multirow}
\usepackage{siunitx}
\usepackage{orcidlink}
\usepackage[normalem]{ulem}
\DeclareUnicodeCharacter{2248}{$\approx$}
\DeclareUnicodeCharacter{0306}{$\check{c}$}

\setlist[itemize]{leftmargin=*}
\title[Dust in Milky Way-like simulated galaxies]{High-resolution synthetic UV--submm images for simulated Milky Way-type galaxies from the Auriga project}
\author[A. U. Kapoor et al.]{Anand Utsav Kapoor\orcidlink{0000-0002-5187-1725},$^{1}$\thanks{E-mail: anandutsav.kapoor@ugent.be}
Peter Camps\orcidlink{0000-0002-4479-4119},$^{1}$
Maarten Baes\orcidlink{0000-0002-3930-2757},$^{1}$
Ana Tr$\mathrm{\check{c}}$ka\orcidlink{0000-0001-7827-1562},$^{1}$
Robert J. J. Grand,$^{2}$
\newauthor
Arjen van der Wel\orcidlink{0000-0002-5027-0135},$^{1}$
Luca Cortese\orcidlink{0000-0002-7422-9823},$^{3}$
Ilse De Looze,$^{1}$
Daniela Barrientos\orcidlink{0000-0002-1710-1460}$^{1}$
\\
$^{1}$Sterrenkundig Observatorium, Universiteit Gent, Krijgslaan 281 S9, B-9000 Gent, Belgium\\
$^{2}$Max-Planck-Institut f\"{u}r Astrophysik, Karl-Schwarzschild-Str. 1,
D-85748, Garching, Germany \\
$^{3}$International Centre for Radio Astronomy Research, The University of Western Australia, 35 Stirling Highway, Crawley
WA 6009, Australia
}

\date{Accepted XXX. Received YYY; in original form ZZZ}

\pubyear{2021}

\begin{document}
\label{firstpage}
\pagerange{\pageref{firstpage}--\pageref{lastpage}}
\maketitle

% Abstract of the paper
\begin{abstract}
We present redshift-zero synthetic observational data considering dust attenuation and dust emission for the thirty galaxies of the Auriga project, calculated with the \texttt{SKIRT} radiative transfer code. 
The post-processing procedure includes components for star-forming regions, stellar sources, and diffuse dust taking into account stochastic heating of dust grains. This allows us to obtain realistic high-resolution broadband images and fluxes from ultraviolet to sub-millimeter wavelengths. 
For the diffuse dust component, we consider two mechanisms for assigning dust to gas cells in the simulation. In one case, only the densest or the coldest gas cells are allowed to have dust, while in the other case this condition is relaxed to allow a larger number of dust-containing cells. The latter approach yields galaxies with a larger radial dust extent and an enhanced dust presence in the inter-spiral regions.
At a global scale, we compare Auriga galaxies with observations by deriving dust scaling relations using SED fitting. At a resolved scale, we make a multi-wavelength morphological comparison with nine well-resolved spiral galaxies from the DustPedia observational database.
We find that for both dust assignment methods, although the Auriga galaxies show a good overall agreement with observational dust properties, they exhibit a slightly higher specific dust mass.
The multi-wavelength morphological analysis reveals a good agreement between the Auriga and the observed galaxies in the optical wavelengths. In the mid and far-infrared wavelengths, Auriga galaxies appear smaller and more centrally concentrated in comparison to their observed counterparts.
We publicly release the multi-observer images and fluxes in 50 commonly used broadband filters.
\end{abstract}

% Select between one and six entries from the list of approved keywords.
% Don't make up new ones.

%%%%%%%%%%%%%%%%%%%%%%%%%%%%%%%%%%%%%%%%%%%%%%%%%%

%%%%%%%%%%%%%%%%% BODY OF PAPER %%%%%%%%%%%%%%%%%%

\section{Introduction}
Modern astronomical surveys provide enormous amounts of observational data. In order to interpret these data, a parallel theoretical framework is required. The currently accepted paradigm of the $\Lambda-$cold dark matter \citep[e.g.][]{2013ApJS..208...19H, 2016A&A...594A..13P, 2020A&A...641A...6P} provides us with a way to confront observations by the means of numerical simulations. 
Numerical simulations of galaxy formation have taken a giant leap in the last decade. Improvement in computer power, coupled with improved numerical treatment has made it possible to handle baryonic physics with increasing complexity. State-of-the-art galaxy formation simulations have started to include the effects of magnetic fields, radiation fields, relativistic particles, etc. For a review, see \citet{2015ARA&A..53...51S,  2020NatRP...2...42V}.
The inclusion of baryons, on one hand, is essential for the study of galaxies, but on the other hand involves the treatment of a wide variety of physical processes spanning tens of orders of dynamic range. This necessitates the use of simplified sub-grid recipes for unresolved processes like star formation and stellar feedback or disregarding certain physical processes.

Recent hydrodynamical simulations reproduce observational galaxy populations to a remarkable degree; examples include EAGLE \citep{2015MNRAS.446..521S}, SIMBA \citep{2019MNRAS.486.2827D}, and Illustris-TNG \citep{2019MNRAS.490.3196P}. However, detailed predictions of these simulations still depend on the underlying implementation of sub-grid physics.
The fine tuning of the sub-grid physics models introduces free parameters, which are generally calibrated based on some key observables, such as the cosmic star formation rate density and/or the stellar mass content of galaxies at redshift zero. These properties tend to be global in nature. Each galaxy is considered to be a single point, neglecting internal structure and leaving the sub-grid models' accuracy or the calibration in doubt. A comparison at resolved spatial scales could reveal interesting differences between simulated galaxy populations with otherwise similar global properties. For example,  \citet{2019MNRAS.488.4400I} show that adjusting the equation of state of the gas without changing feedback or initial conditions can control formation of giant star-forming clumps in massive disks at redshifts $\sim 1-2$ without affecting the global galaxy properties. Similarly, different stellar and active galactic nuclei (AGN) feedback schemes, which are essential in controlling gas cooling and star formation, could lead to a different gas composition and thermodynamical structure in and around galaxies \citep{doi:10.1146/annurev-astro-081913-040019}. Once again, a comparison at resolved scales could be used to differentiate and test galaxy formation models. 

The success of a galaxy formation model at resolved scales would eventually be determined by a comparison with observations, which makes it essential that we generate mock observational data from the simulations, taking in account the relevant physics involved in the transformation from intrinsic to observed quantities. This brings us to the need to account for the dust grains in the interstellar medium (ISM). Notwithstanding a mass contribution to the ISM of only about one percent, dust locks up a substantial fraction of all heavy elements and provides the primary source of opacity for non-ionizing photons, thus influencing a galaxy's spectral energy distribution (SED) at all wavelengths longer than 912~Å.  In a typical disk galaxy, dust grains reprocess nearly one-third of the stellar light  \citep{2018A&A...620A.112B, 2016A&A...586A..13V}. 
The reprocessed energy shows up in the mid-infrared (MIR) and far-infrared (FIR) wavelength ranges, providing yet another mode of comparison with observational data. 

Most cosmological hydrodynamical simulations do not include dust evolution in the galaxy formation physics, although important steps forward have recently been made in this direction  \citep{10.1093/mnras/stw253, 2017MNRAS.468.1505M, 10.1093/mnras/staa3249,2019MNRAS.487.4870V, 2020arXiv201005919G}. Simulated galaxies are, thus, usually post-processed by assuming a fixed percentage of the metal mass in the ISM as dust grains and solving the radiative transfer problem \citep[e.g.,][]{2011BASI...39..101W, 2013ARA&A..51...63S}. Recent forward modeling efforts at resolved scales \citep[e.g.,][]{10.1093/mnras/staa3765, 10.1093/mnras/stz3014, 2019MNRAS.483.4140R} target the impact of the presence of dust on the galaxy properties in the optical wavelengths. Such a strategy could be extended by including dust emission wavelengths, allowing a more comprehensive comparison, shedding light on the dust distribution in comparison with observations, and gaining important insights about the fidelity of the simulations in a wider spectral range.

The main purpose of the current work is to facilitate such comparisons by producing and publishing spatially resolved synthetic observables for a suite of high-resolution zoom-in hydrodynamical simulations, including the effects of both extinction and emission by dust. These data products enable a detailed, local scale, multi-wavelength comparison with observations across the full range from ultraviolet (UV) to sub-millimeter (submm) wavelengths. We use the Auriga suite of zoom-in hydrodynamical simulations \citep{2017MNRAS.467..179G}, which consist of a realistic galaxy population in a cosmological setting. The set of thirty galaxies reproduce a wide range of present-day observables, in particular, disk-dominated galaxies with appropriate stellar masses, sizes, rotation curves, star formation rates and metallicities.
We use the radiative transfer code \texttt{SKIRT} \citep{2015A&C.....9...20C, 2020A&C....3100381C} to capture the intricate interplay between the simulated galaxy’s constituents and generate high-resolution images in broadbands at UV to submm wavelength and at multiple observer positions.  

In Sect.~\ref{methodology.sec}, we provide some background on the Auriga simulations and the \texttt{SKIRT} radiative transfer code, and we describe how the Auriga results were exported to and post-processed by \texttt{SKIRT}, with appendices containing some related information. In Sect.~\ref{calibration.sec}, we explain our post-processing calibration strategy. Sect.~\ref{syntheticData.sec}, we describe the synthetic observables generated and their availability. Sect.~\ref{Results.sec} contains the results of the analyses carried out using the synthetic data including the derivation of global physical properties and a spatially resolved morphology study. In Sect.~\ref{discussion.sec}, we summarize and conclude and we suggest some possible uses of our data products.

\section{Methodology }
\label{methodology.sec}
We broadly follow the methodology pioneered by \citet{2016MNRAS.462.1057C, 2018ApJS..234...20C,2017MNRAS.470..771T} for the \texttt{SKIRT} post-processing of galaxies in the EAGLE simulations, and subsequently applied to other cosmological hydrodynamical simulations by for example \citet{10.1093/mnrasl/sly071, 10.1093/mnras/stz2134, 2019MNRAS.487.1844M, 2019MNRAS.483.4140R, 10.1093/mnras/stz1736, 2020MNRAS.492.5167V, 10.1093/mnras/staa1900,popping2021dustcontinuum} and Tr$\mathrm{\check{c}}$ka et al. (in prep.). We summarize the key features of our methodology here.

\subsection{The Auriga simulations}
%Why are the Auriga simulations ideal for our purposes?
In this paper we use the redshift-zero snapshots of the Auriga suite of cosmological simulations, which is a set of 30 zoom simulations aimed at the modeling of Milky Way-type galaxies in a full cosmological context \citep{2017MNRAS.467..179G}, carried out using the moving-mesh magnetohydrodynamics (MHD) code \texttt{AREPO} \citep{2010MNRAS.401..791S}.
The Auriga simulations follow a $\Lambda$ cold dark matter cosmology with parameters $\Omega_{\mathrm{m}}=\Omega_{\mathrm{dm}}+\Omega_{\mathrm{b}}=0.307, \Omega_{\mathrm{b}}=0.048, \Omega_{\mathrm{A}}=$
$0.693,$ and Hubble constant $H_{0}=100 h =67.77$
$\mathrm{km}\,\mathrm{s}^{-1} \mathrm{Mpc}^{-1},$ consistent with the \citet{2014} data release.

The host dark matter halos of the zoomed galaxy simulations were drawn from a dark matter only  simulation of comoving side length 100 cMpc.
Further selection criteria were imposed on the mass and the isolation of the host halos. The virial mass ranges between $1-2$ times $10^{12}~M_{\odot}$, consistent with recent determinations of the Milky Way mass \cite[see][and references therein]{2015MNRAS.453..377W}. The isolation criterion was imposed by ensuring that any halo more than 3$\%$ of the target halo mass is farther than 9 times its virial radius from the main halo.
A progressively coarser resolution at increasing distances from the target object is employed to increase the computation speed. This allows for a higher resolution simulation of the selected objects in a cosmological setting by maintaining the large-scale tidal field.
The selected halos were re-simulated by applying the so-called ‘zoom-in’ technique. The mass distribution in the Lagrangian region forming the main halo is identified in the initial conditions, sampled by a large number of resolution elements and then re-simulated.

%%Physics: Gas
The Auriga physics model uses primordial and metal-line cooling with self-shielding corrections. A spatially uniform UV background field \citep{2009ApJ...703.1416F} is employed. 
The ISM is modeled with a two-phase equation of state from \citet{2003MNRAS.339..289S}. Star formation proceeds stochastically in gas with densities higher than a threshold density ($n_{thr}=0.13~\text{cm}^{-3}$). The star formation probability in the candidate gas cells scales exponentially with time, with a characteristic time scale of $t_{\mathrm{SF}}=2.2~\text{Gyr}$.
%% Physics: Stars and Feedback
The single stellar population (SSP) of each star particle is represented by the \citet{2003PASP..115..763C} initial mass function (IMF).
Mass and metal returns from SNIa, AGB, and SNII stars are calculated at each time step and are distributed among nearby gas cells with a top-hat kernel. 
The number of SNII events equals the number of
of stars in an SSP that lie in the mass range 8-100 $\mathrm{M}_\odot$.
In order to mimic SNII events, a star forming gas cell could also be converted to a galactic wind particle instead of a star. The wind particle is launched in an isotropically random direction and interacts only gravitationally. It re-couples hydrodynamically as soon as it reaches a gas cell with a density of $5\%$ of the star formation density threshold, to mimic SN-driven winds emerging from star-forming regions. 
The model includes gas accretion by black holes, with AGN feedback in radio and quasar modes, both of which are always active and are thermal in nature.
%%Physics: MHD
Magnetic fields are treated with ideal MHD following \cite{2013MNRAS.432..176P}.
%% Suitability : resolution, properties
\begin{table}
\centering
\caption{Table of Auriga numerical resolution parameters at redshift zero. From left to right, the columns list resolution level, dark matter particle mass, typical baryonic particle mass, and softening length of collision-less particles.}
\label{AurigaResolution.table}
\begin{tabular}{lccc}
\hline Resolution level & ${m_{\mathrm{DM}}}~{\left[\mathrm{M}_{\odot}\right]}$ & ${m_{\mathrm{b}}}~{\left[\mathrm{M}_{\odot}\right]}$ & ${\epsilon}~{[\mathrm{pc}]}$ \\
\hline 4 & $3 \times 10^{5}$ & $5 \times 10^{4}$ & 369 \\
5 & $2 \times 10^{6}$ & $4 \times 10^{5}$ & 738 \\
3 & $4 \times 10^{4}$ & $6 \times 10^{3}$ & 184 \\
\hline
\end{tabular}
\end{table}

In this paper, we use the fiducial Auriga models at resolution level 4, the specifications of which are given in Table~\ref{AurigaResolution.table}. 

\subsection{\texttt{SKIRT} radiative transfer code}
\label{aboutSKIRT.ssec}
\texttt{SKIRT} is a public multi-purpose Monte Carlo radiative transfer code \citep{2015A&C.....9...20C, 2017A&C....20...16V, 2020A&C....3100381C} for simulating the effect of dust on radiation in astrophysical systems. It offers full treatment of absorption and multiple anisotropic scattering by the dust, computes the temperature distribution of the dust and the thermal dust re-emission self-consistently, and supports stochastic heating of dust grains. The code handles multiple dust mixtures and arbitrary three-dimensional (3D) geometries for radiation sources and dust populations, including grid- or particle-based representations generated by hydrodynamical simulations. The dust density distribution is discretized using one of the built-in dust grids, including octree, kd-tree \citep{2014A&A...561A..77S} and Voronoi \citep{2013A&A...560A..35C} grids.
In this work, we use \texttt{SKIRT}'s most recent version, \texttt{SKIRT}~9.
We next look at the steps used in the construction of the \texttt{SKIRT} input model.

\subsection{Data extraction from Auriga snapshots for \texttt{SKIRT}}
\label{dataExtraction.ssec}

\begin{figure*}
\centering
\includegraphics[width=0.85\textwidth]{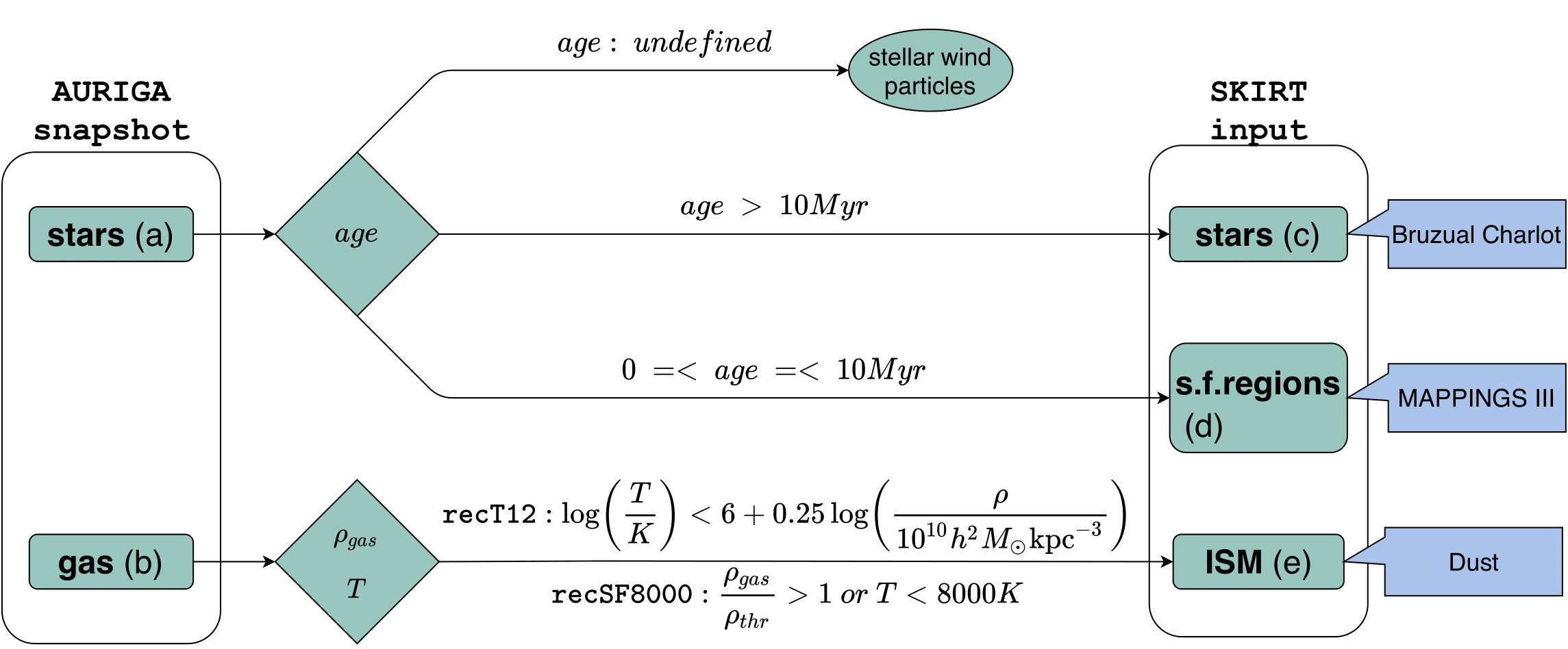}
\caption{Schematic overview of the \texttt{SKIRT} post-processing procedure used for the Auriga galaxies.} 
\label{SKIRTrecipe.fig}
\end{figure*}

For each galaxy, we extract the corresponding sets of star particles and gas cells, labeled as (a) and (b) in Fig.~\ref{SKIRTrecipe.fig}, from the Auriga simulation snapshots at redshift zero.
We use a cubical aperture centered at the galaxy center of mass to extract the data. The side of the cube is twice the radius at which the face-on stellar surface density within $\pm10~$\text{kpc} of the mid plane in the vertical direction falls to a value of $2\times10^5{\text{M}}_{\odot}/{\text{kpc}}^2$.
For three galaxies, AU-17, AU-18 and, AU-30, we use a higher value of cutoff surface density,  $6\times10^5{\text{M}}_{\odot}/{\text{kpc}}^2$, in order to avoid secondary structures in the data other than the main galaxy.
Any star particles or gas cells outside this region are ignored. A histogram of the aperture side lengths for the Auriga galaxies is shown in the upper-left panel of Fig.~\ref{OctreeGridProperties.fig}.

\subsubsection{Primary emission sources}
\label{DataExtraction_PrimarySources.sssec}
Star particles with an undefined age, which represent stellar wind particles, are ignored. In order to characterize of the primary sources of emission, each remaining star particle is assigned an SED depending on its intrinsic properties inherited from the simulation. The star particles with an age above 10 Myr are assigned an SED from the \citet{2003MNRAS.344.1000B} template library for the \citet{2003PASP..115..763C} IMF with the appropriate age, metallicity and initial stellar mass. Young star particles with an age below 10 Myr are assumed to be still enshrouded by dust, and receive a special treatment. 
%We assign a modified age to the SFR by adding a Gaussian noise to the original age values obtained for the star particles younger than 10 Myr.
A young star particle is assigned an appropriate SED from the \texttt{MAPPINGS III} SED family \citep{2008ApJS..176..438G}. These templates model both the H{\sc{ii}} region and the photo-dissociation region (PDR) surrounding the star-forming core, including the dust contained in those regions. The \texttt{MAPPINGS III} templates are parametrized by five parameters, and we determine these values for each young star particle in the following way. 
\begin{itemize}[leftmargin=*]
\item The metallicity, $Z$, is taken directly from the particle properties.
\item The star-formation rate (SFR) is determined from the mass assigned to the star-forming particle at birth assuming a constant SFR during the H{\sc{ii}} region’s lifetime of 10 Myr, following \citet{2008ApJS..176..438G}.
\item The H{\sc{ii}} region compactness, $\mathcal{C}$, is the prime parameter that sets the shape of the FIR continuum dust emission, and is hence a proxy for the dust temperature distribution \citep{2008ApJS..176..438G}. By fitting modified blackbody functions to the \texttt{MAPPINGS III} templates corresponding to different values of the compactness, we have determined the correspondence between dust temperature and $\mathcal{C}$. Subsequently, we use the observed dust temperature distribution in star forming regions in Local Group galaxies \citep{2019ApJ...874..141U}, and in hydrodynamical zoom simulations that take into account dust physics \citep{10.1093/mnras/staa3249}, to generate the distribution of compactness. This distribution can be approximated by a lognormal distribution with $\langle{\log\mathcal{C}}\rangle = 5$ and standard deviation 0.4. For each young stellar particle, we hence randomly sample a value for $\log{\mathcal{C}}$ from this distribution. 
\item The ambient ISM pressure only affects the strength of a number of emission lines, and does not affect the shape of the SED of the template. We calculate it for each star particle using equation (13) of \citet{2008ApJS..176..438G}.
\item The covering fraction of the PDR, $f_{\text{PDR}}$, is determined as $f_{\text{PDR}} = \exp(-t/\tau_{\text{clear}})$, where $t$ is the age of the stellar particle and $\tau_{\text{clear}}$ is the molecular cloud clearing or dissipation timescale. We treat $\tau_{\text{clear}}$ as a free parameter in our post-processing framework and its value is determined in Sect.~\ref{calibration.sec}.
\end{itemize}
The methodology presented here for assigning the H{\sc{ii}} region compactness and covering fraction, where we end up with a distribution for these properties rather than a single value, differ from those used in previously carried out post-processing of simulated galaxies.
%How these parameters are determined for each young star particle is discussed in Sect.~{\ref{calibration.sec}}.

\begin{figure}
\centering
\includegraphics[width=0.9\columnwidth]{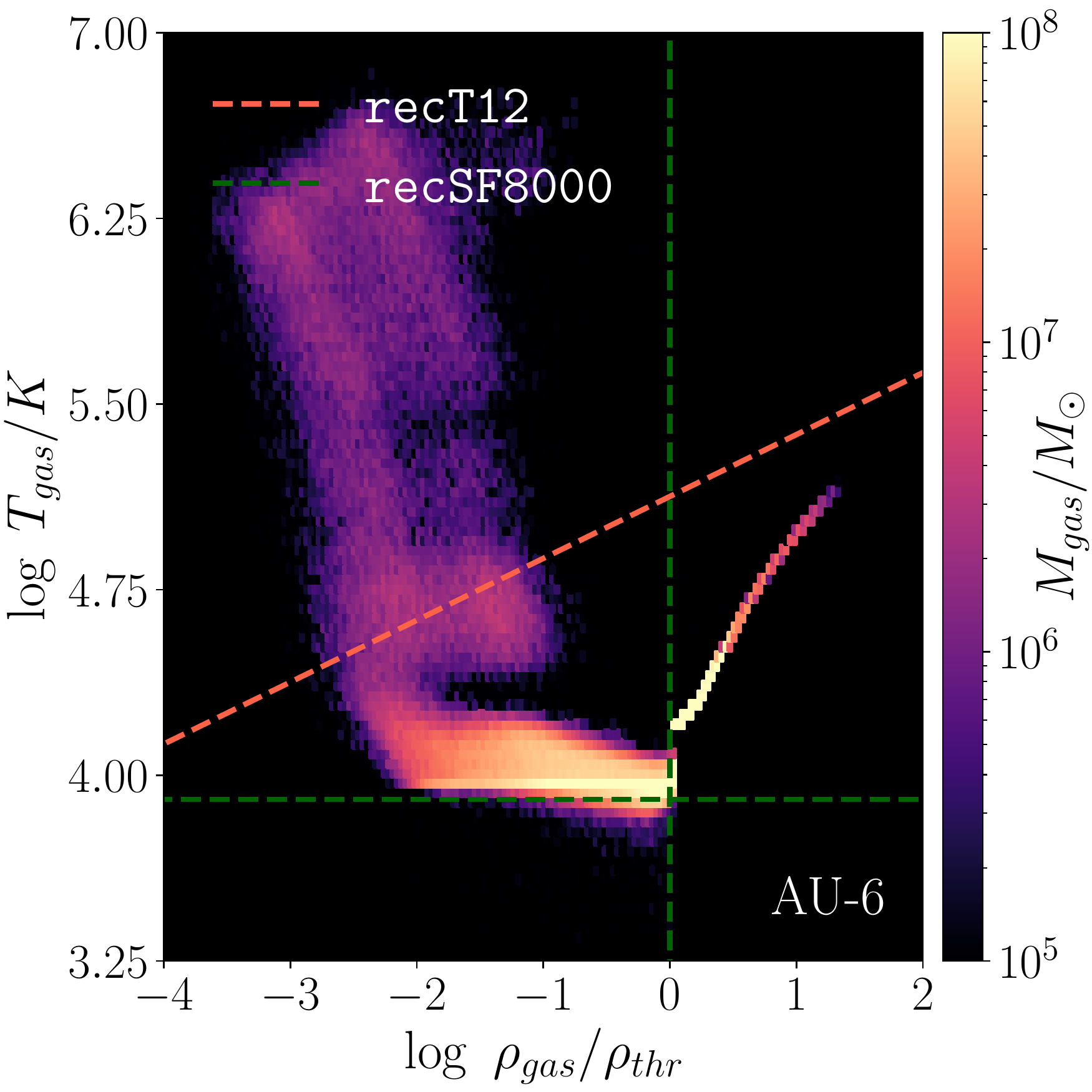}
\caption{Temperature-density phase diagram for the galaxy AU-6. The green lines represent \texttt{recSF8000} (Eq.~\ref{ISM_crit_SF8000.eqn}), where all gas to the right of the vertical line or below the horizontal line is used for dust allocation. The orange line represents \texttt{recT12} (Eq.~\ref{ISM_crit_torrey.eqn}), where all gas cells below the slanted line are eligible for dust allocation. The color indicates gas mass (in $\mathrm{M_{\odot}}$) binned in 150 bins for both $\log(T/K)$ and $\log(\rho_{gas}/\rho_{thr})$, where $\rho_{thr}$ is the star formation threshold density corresponding to $n_{thr}=0.13~\mathrm{cm^{-3}}$.} 
\label{ISMdemarcation_binned=gasMass.fig}
\end{figure}

\subsubsection{Dust Allocation}
\label{DustAllocation.sssec}

The next step in the post-processing framework is determining the distribution of interstellar dust in each simulated galaxy.
The Auriga simulations do not include dust physics, hence we need to use a recipe to infer the dust distribution based on the properties of the interstellar gas. We assume that a constant fraction of the metals in the eligible, dust-containing ISM (DISM) is locked up in dust grains. In other words, for each gas cell we set the dust density, $\rho_{dust}$, as follows:
\begin{equation}
\rho_{\text {dust }}=\left\{\begin{array}{ll}
f_{\text {dust }} Z \,\rho_{\text {gas }} & \text { if DISM } \\
0 & \text { otherwise, }
\end{array}\right.
\label{dust_allocation_crit.eqn}
\end{equation}
where $Z$ and $\rho_{gas}$ represent the metallicity and gas density given by the gas cell's properties in the Auriga snapshot. Furthermore, $f_\text{dust}$ refers to the dust-to-metal ratio, $M_{dust}/M_{Z}$, with $M_{dust}$ and $M_{Z}$ the dust mass and the metal mass in a given simulation cell. We treat $f_\text{dust}$ as a free parameter in our post-processing framework and its value is determined in Sect.~\ref{calibration.sec}.

We use two different recipes for selecting the DISM:
\begin{enumerate}[leftmargin=*]
\item Following \citet{2016MNRAS.462.1057C}, only gas cells with non-zero SFR or with a gas temperature, $T$, below 8000~K are treated as DISM. This can be written as
\begin{equation}
\mathrm{DISM} \iff \rho_{gas}>\rho_{thr}~\text{or}~T<8000~\text{K},
\label{ISM_crit_SF8000.eqn}
\end{equation}
where $\rho_{thr}$ is the Auriga star formation density threshold. In Fig.~\ref{ISMdemarcation_binned=gasMass.fig}, gas cells fulfilling these conditions are shown to the right of the vertical green line and below the horizontal green line.
The cut-off value of 8000 K is somewhat arbitrary, and in practice, the addition of the gas cells below 8000~K does not cause a considerable change in the dust geometry, so that the diffuse dust content of each galaxy is essentially traced by the star-forming gas alone, as in \citet{2019MNRAS.483.4140R}.
We refer to this recipe as \texttt{recSF8000} for the rest of this work.

\item We distinguish rotationally supported interstellar gas, settled in the disk, from the hot circumgalactic gas following \citet{2012MNRAS.427.2224T}:
  \begin{equation}
   \mathrm{DISM} \iff \log \left(\frac{T}{K}\right)<6+0.25 \log \left(\frac{\rho_{gas}}{10^{10} h^{2} M_{\odot} \mathrm{kpc}^{-3}}\right)
   \label{ISM_crit_torrey.eqn}
  \end{equation}
This demarcation is shown by the orange line in Fig.~\ref{ISMdemarcation_binned=gasMass.fig}.
This kind of dust allocation is also driven by the galaxy formation simulations following dust formation and destruction \citep[e.g., see][]{10.1093/mnras/stw253, 2017MNRAS.468.1505M}, which despite using simple dust-evolution models show the presence of dust in the lower density, non star-forming gas settled in the galactic disks aside from the higher density star-forming gas.
We call this recipe \texttt{recT12} from hereon. 
\end{enumerate}

Star-forming gas cells in the Auriga simulations are above the density threshold $\rho_{thr}$ and are assumed to be composed of hot and cold phases. The physics of the star-forming gas is governed by an effective equation of state \citep{2003MNRAS.339..289S}. The simulations report an effective temperature for the gas of such nature (as shown in Fig.~\ref{ISMdemarcation_binned=gasMass.fig}). This value is a mass weighted temperature obtained using the hot and cold phase temperatures, which are $\approx 10^6~\text{K}$ and $\approx 10^3~\text{K}$, respectively. This allows us to find a cold gas fraction in each cell. We associate dust only with the cold gas in such cells. In practice the cold gas fraction turns out to be nearly unity for most star forming cells.
For the lower density, non star forming gas, the gas fraction associated with dust is assumed to be unity.

\begin{figure}
\centering
\includegraphics[width=\columnwidth]{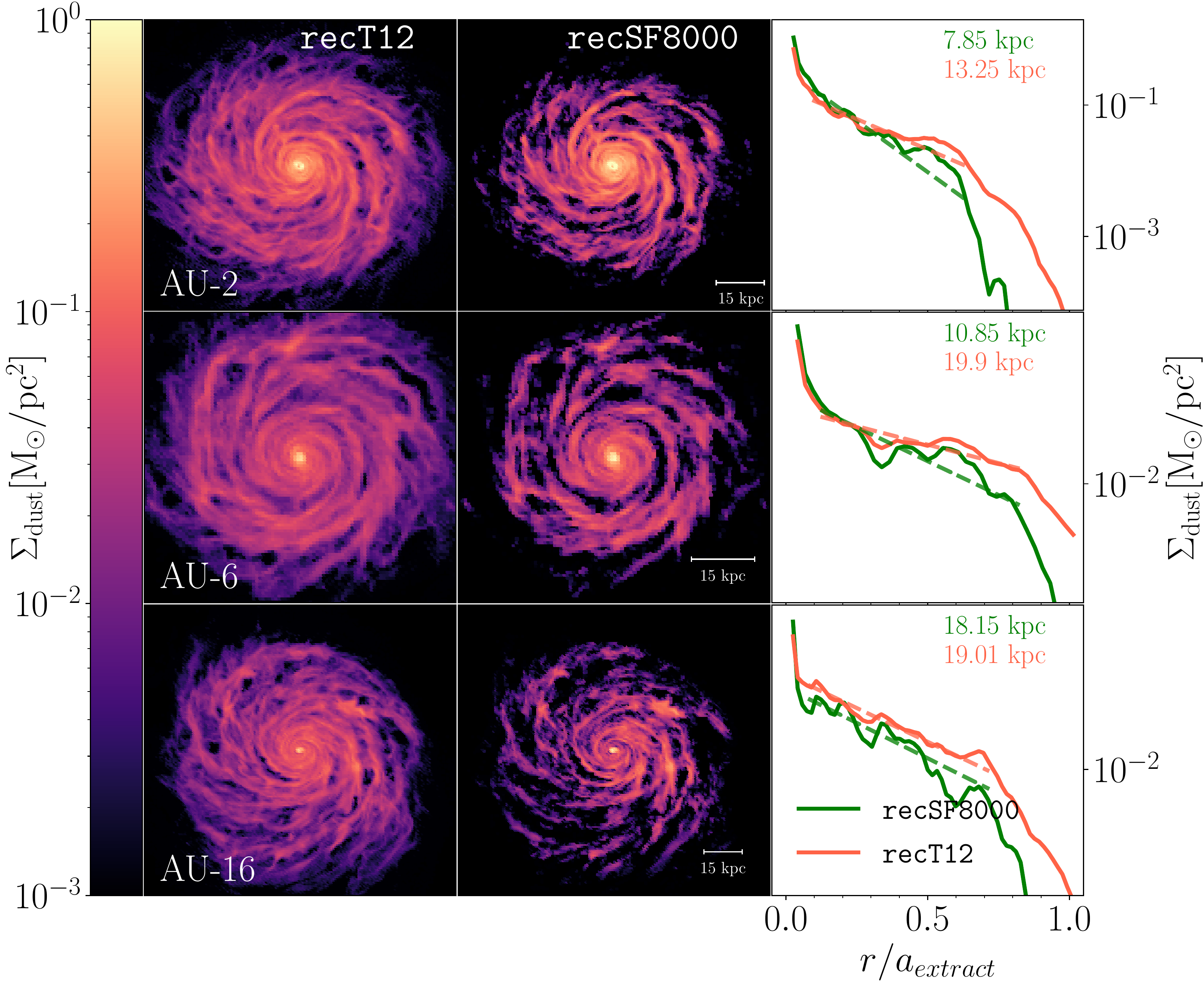} 
\caption{Dust surface density (in $\mathrm{M_{\star}/pc^2}$) computed directly from the simulation data for AU-2, 6, 16 based on the dust allocation schemes discussed in Sect.~\ref{DustAllocation.sssec} and the $f_{dust}$ values taken from Table~\ref{calibParameters.table}. All dust within $\mathrm{\pm 5~kpc}$ of the mid-plane is projected on a grid with a bin size of $0.75 \mathrm{kpc}$. The radial dust surface density distribution is also shown, along with the disk scale length $h$ obtained by fitting a function of the form $r_{0}e^{-r/h}$ ($r$ being the radius). The fits do not consider the central dust distribution, whose extent is determined by eye.
The impact of the change in dust allocation recipe can be clearly seen, with \texttt{recT12} showing a more diffuse and extended dust distribution in comparison to \texttt{recSF8000}. We remark that the lack of dust in the centers of these galaxies is likely because of the AGN quasar mode feedback, which causes low gas density holes to appear after a local dump of thermal energy.}
\label{recipeComaprison_dustSigma.fig}
\end{figure}

Fig.~\ref{recipeComaprison_dustSigma.fig} shows the impact of changing the dust allocation recipe on the dust surface density of three Auriga galaxies when viewed face-on. The dust morphology clearly differs: \texttt{recT12} shows a more diffuse dust distribution which is radially more extended and has a higher coverage of the inter-spiral region in comparison to \texttt{recSF8000}. 

The procedure described above completely sets the spatial distribution and properties of the stars and dust in each system, apart from two free parameters: the molecular cloud dissipation timescale $\tau_{\text{clear}}$ and the dust-to-metal ratio $f_{\text{dust}}$ in the interstellar medium. In Sect.~\ref{calibration.sec}, we calibrate these free parameters by comparing the global, spatially integrated fluxes of the entire sample of Auriga galaxies to a set of observed galaxies. 

\begin{table}
\caption{Input parameters of the SKIRT radiative transfer model for each of the Auriga components, in addition to the particle/cell positions. The procedure for deriving a dust distribution from the gas cells (item (e) in Fig.~\ref{SKIRTrecipe.fig}) is discussed in Sect.~\ref{DustAllocation.sssec}. The procedures for the particles representing stellar populations and star-forming regions (items (c) and (d) in Fig.~\ref{SKIRTrecipe.fig}) are discussed in Sect.~\ref{DataExtraction_PrimarySources.sssec}}
\label{inputTextFileParameters.table}
\begin{tabulary}{\columnwidth}{LLL}
\hline
 Param. \hspace{15mm} & Description & Origin \\
 & \vspace{.1cm} \centering \mbox{Dust} & \\
\hline
$\rho_\mathrm{gas}$ & Gas density & Simulation \\
$Z$ & Gas metallicity & Simulation \\
$T$ & Gas temperature & Simulation \\
SFR & Star formation rate of the gas & Simulation \\
$f_{\text{dust }}$ & Fraction of the metallic gas locked up in dust & Free parameter \\
 
\\
 & \centering \mbox{Stars} & \\
\hline
$h$ & Smoothing length & Assumed distribution \\
$M_{\text {init }}$ & Birth mass of the stellar population & Simulation \\
$Z$ & Metallicity of the stellar population & Simulation \\
$t$ & Age of the stellar population & Simulation \\

\\
 & \centering \mbox{SF regions} & \\
\hline
$h$ & Smoothing length & Calculated \\
$M$ & Mass of the H{\sc{ii}} region & Simulation \\
SFR & Star formation rate of the H{\sc{ii}} region & Calculated \\
$Z$ & Metallicity of the H{\sc{ii}} region & Simulation \\
$C$ & Compactness of the H{\sc{ii}} region & Assumed distribution \\
$P$ & Pressure of the ambient ISM & Calculated\\
%\tau_{\text {diss.}}$ & Molecular cloud dissipation time scale of the H{\sc{ii}} region & Free parameter\\
$f_{\text {PDR }}$ & Dust covering fraction of the PDR region & Free parameter\\
\hline
\end{tabulary}
\end{table}

\subsection{Radiative transfer on the Auriga galaxies: \texttt{SKIRT} configuration}
\label{SKIRTinput.sec}

\begin{figure*}
\centering
\includegraphics[width=0.9\textwidth]{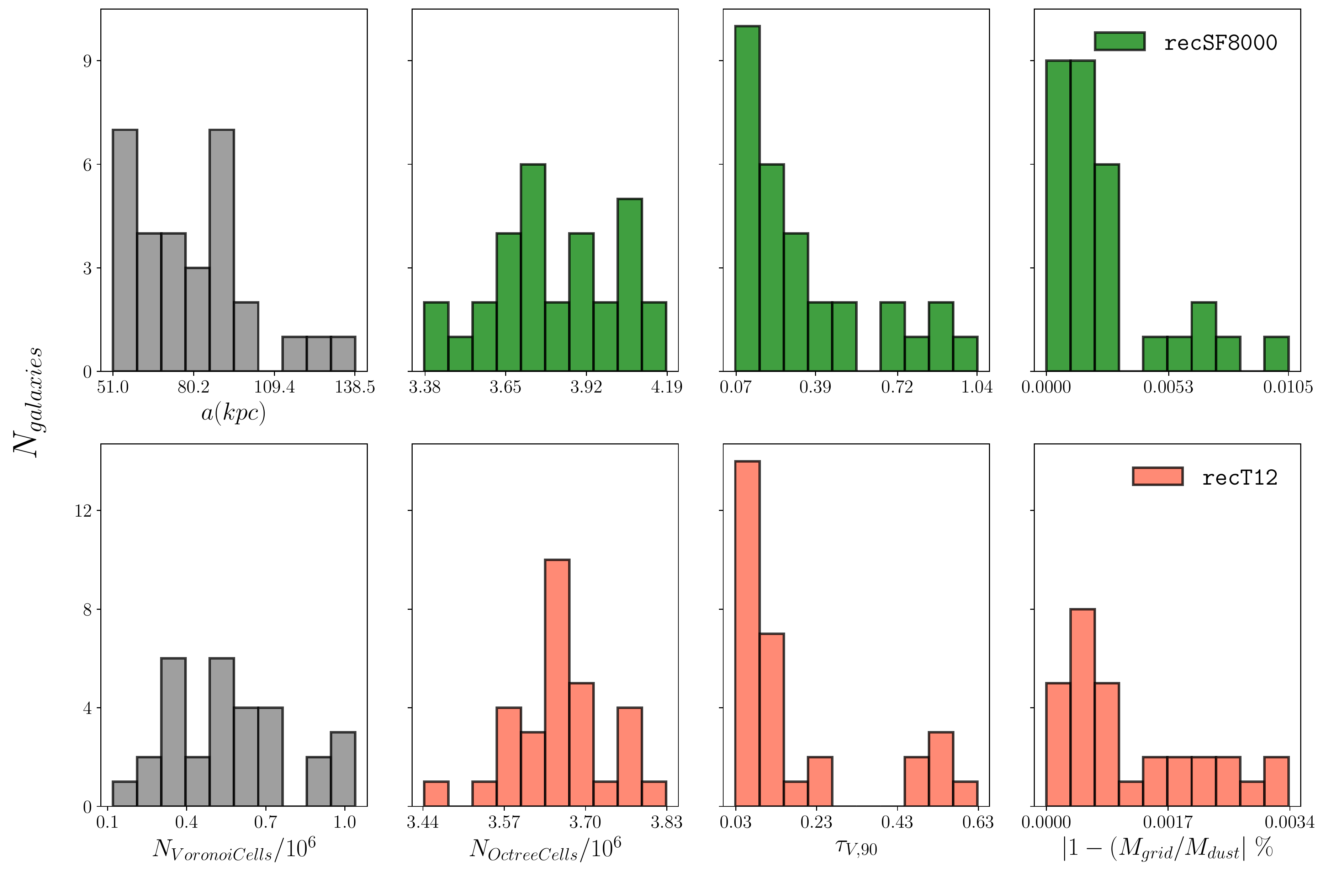}
\caption{The first column (in gray) shows the side length of the cubical aperture used to extract the simulation data and the number of simulation Voronoi cells within that aperture. 
The remaining panels show the distribution of the \texttt{SKIRT} dust discretization properties for each of the dust allocation recipes. From left to right: the number of cells in the octree dust grid constructed by \texttt{SKIRT}; the 90 per cent percentile V-band optical depth of the cells in that grid; and the discretization error on the total dust mass (i.e., the difference between the actual dust mass and the dust mass discretized on the grid).} 
\label{OctreeGridProperties.fig}
\end{figure*}

Apart from specifying the physical input model (Sects.~\ref{dataExtraction.ssec} and \ref{calibration.sec}) and the synthetic observables (Sect.~\ref{dataproducts.ssec}), we need to configure \texttt{SKIRT} to
perform the radiative transfer simulations on the Auriga galaxies, as described in this section. We use the same \texttt{SKIRT} configuration for both \texttt{recSF8000} and \texttt{recT12}, although the configuration parameters were determined using the \texttt{recSF8000} recipe.

\subsubsection{Dust spatial grid}
\label{SKIRTconfig_dustGrid.sssec}
%why do we choose an octree grid instead of the natural Voronoi grid?
The \texttt{SKIRT} radiative transfer procedure requires the dust density distribution of the system under study to be discretized on a spatial grid. Within each grid cell, the dust density and all other physical quantities, such as the radiation field, are assumed to be uniform. \texttt{SKIRT} supports many different dust grid types, including unstructured grids \citep{2013A&A...560A..35C} and adaptive hierarchical Cartesian grids such as octree or binary trees \citep{2013A&A...554A..10S, 2014A&A...561A..77S}. Given that the Auriga simulations are run with the \texttt{AREPO} hydrodynamics code on an unstructured voronoi mesh, it seems most logical to directly run the \texttt{SKIRT} post-processing using the same grid. Photon packet traversal through a Voronoi mesh is, however, inherently slower than through an octree grid. Moreover, it is not guaranteed that the Voronoi grid used for the hydrodynamics is optimal for radiative transfer post-processing. 
%When using the Cartesian grids, the dust density is interpolated using a scaled and truncated Gaussian kernel designed to approximate a finite-support cubic spline kernel from the voronoi input.
We ran tests using the native Voronoi grid and  octree grids with different subdivision characteristics, the results of which are given in Sect.~\ref{DustGrid.appendix}.
Our final simulations were run using an octree grid with a maximum of 12 allowed subdivisions and a maximum cell dust fraction value of $10^{-6}$.
Fig.~\ref{OctreeGridProperties.fig} provides some relevant statistics on the discretization of the diffuse dust density for the Auriga galaxies with these grid parameters.

\subsubsection{Dust Model}
\label{SKIRTconfig_dustModel.sssec}
%% From Peter's paper and SKIRT docs
Diffused dust in our simulations uses the \texttt{THEMIS} dust model described by \citet{2017A&A...602A..46J} and the references therein. In this model, there are two families of dust particles: amorphous silicates and amorphous hydrocarbons. 
For the silicates, it is assumed that half of the mass is amorphous enstatite, and that the remaining half is amorphous forsterite. The size distribution is considered as lognormal and the same ($a \simeq 10-3000~\mathrm{nm}$) for both populations of amorphous silicates with the distribution peak at $a_{\text {peak }} \simeq 140 ~\mathrm{nm}$.
For the amorphous hydrocarbon population, the size distribution is a combination of a power-law and a lognormal distribution.
The power law distribution is used for amorphous carbon particles with sizes $a \lesssim 20 ~\mathrm{nm}$, while the lognormal distribution is for larger grains, $a \simeq 10-3000~\mathrm{nm}$  with $a_{\text {peak }} \simeq 160~\mathrm{nm}$.

We use 15 grain size bins for each population of silicates and hydrocarbons to discretize the thermal emission calculations of the dust mix.

\subsubsection{Number of photon packets and wavelength grid}
\label{SKIRTconfig_numPacks_wavegrid.sssec}
Apart from a sufficiently high spatial resolution (Sect.\ref{dataproducts.ssec}), high-quality broadband images also require a sufficient signal-to-noise ratio (SNR). The SNR of the output images directly drives the number of photon packets required in the radiative transfer simulations, which, in turn, drives the simulation run time. We base the required number of photon packets in the simulations on the relative error statistic ($R$) described in \citet{2020A&C....3100381C}. 
The results of our tests are provided in Sect.~\ref{SEDphotonsConvergence.appendix}.
For the chosen instrument setup (Sect.~\ref{dataproducts.ssec}), it turns out that using $2\times10^{10}$ photon packets forms a good compromise between sufficient SNR and an acceptable simulation run time. 

For all simulations carried out for this work, we use a logarithmic wavelength grid with 40 points running from $.02~\mu m$ to  $10~\mu m$ for the storing the mean radiation field in each cell. For dust emission, a nested logarithmic grid is employed. The low resolution part of this nested grid has 100 points, running from $1~\mu m$ to  $2000~\mu m$, whereas the higher resolution part runs in the PAH emission range from from $2~\mu m$ to  $25~\mu m$ with 400 wavelength points, We refer to \cite{2020A&C....3100381C} for a discussion on the convergence of both these wavelength grids for a very similar simulation setup.

\section{Calibration of the model parameters}
\label{calibration.sec}

\begin{figure}
\centering
\includegraphics[width=.9\columnwidth]{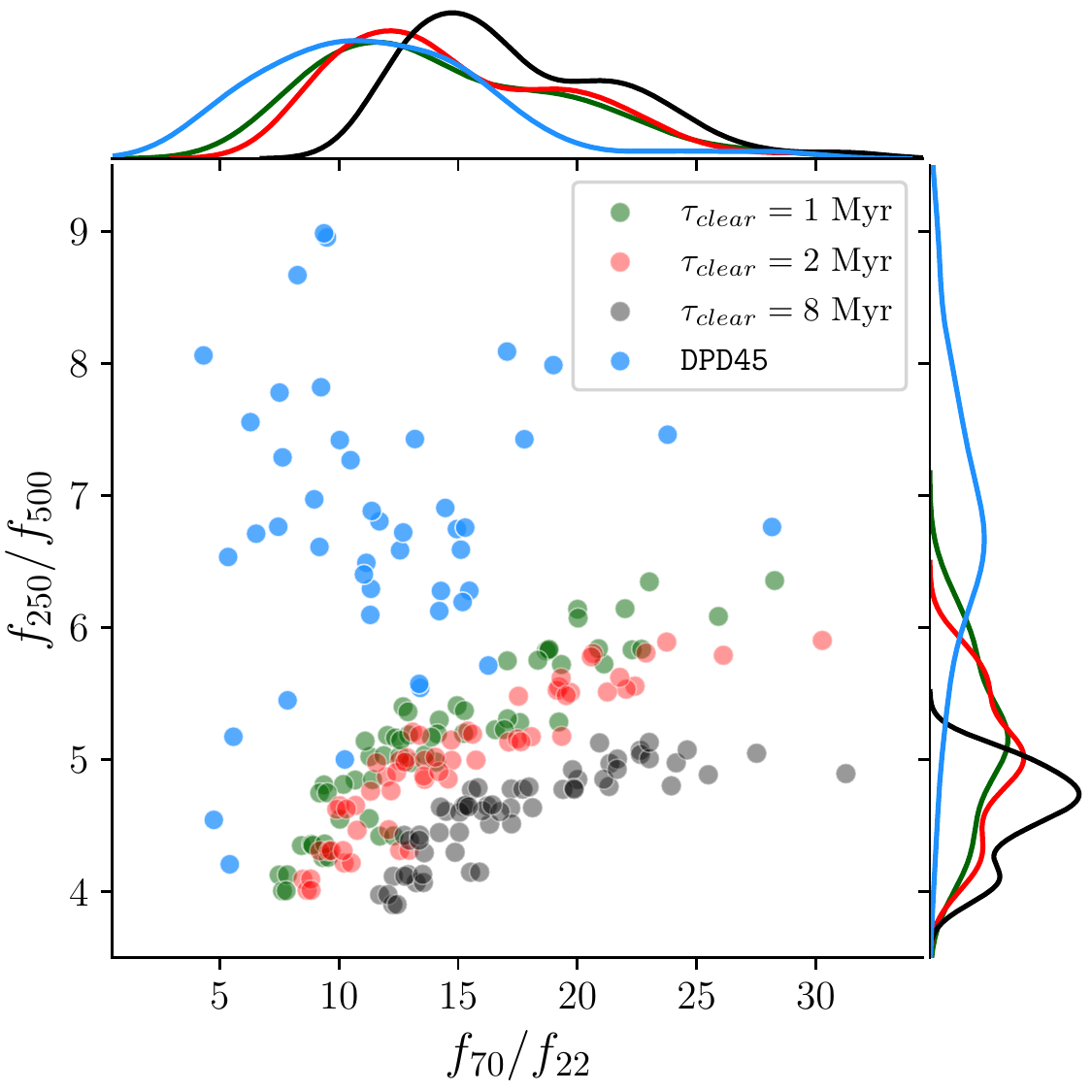}
\caption{Color-color relation between PACS-70/WISE-22 and SPIRE-250/SPIRE-500 bands. Larger values of $\tau_{clear}$ correspond to a higher contribution of cold dust both in the star forming particles, leading to higher $f_{70}$ fluxes relative to $f_{22}$, and in the diffuse ISM, leading to higher $f_{500}$ fluxes relative to $f_{250}$. This moves the Auriga data points away from those of \texttt{DPD45} with an increase in $\tau_{clear}$.
The Auriga points shown here use \texttt{recSF8000} with $f_{dust}=0.225$.} 
\label{tau&fdustParameterSpace_AU3.fig}
\end{figure}

\begin{figure*}
\centering
\includegraphics[width=0.9\textwidth]{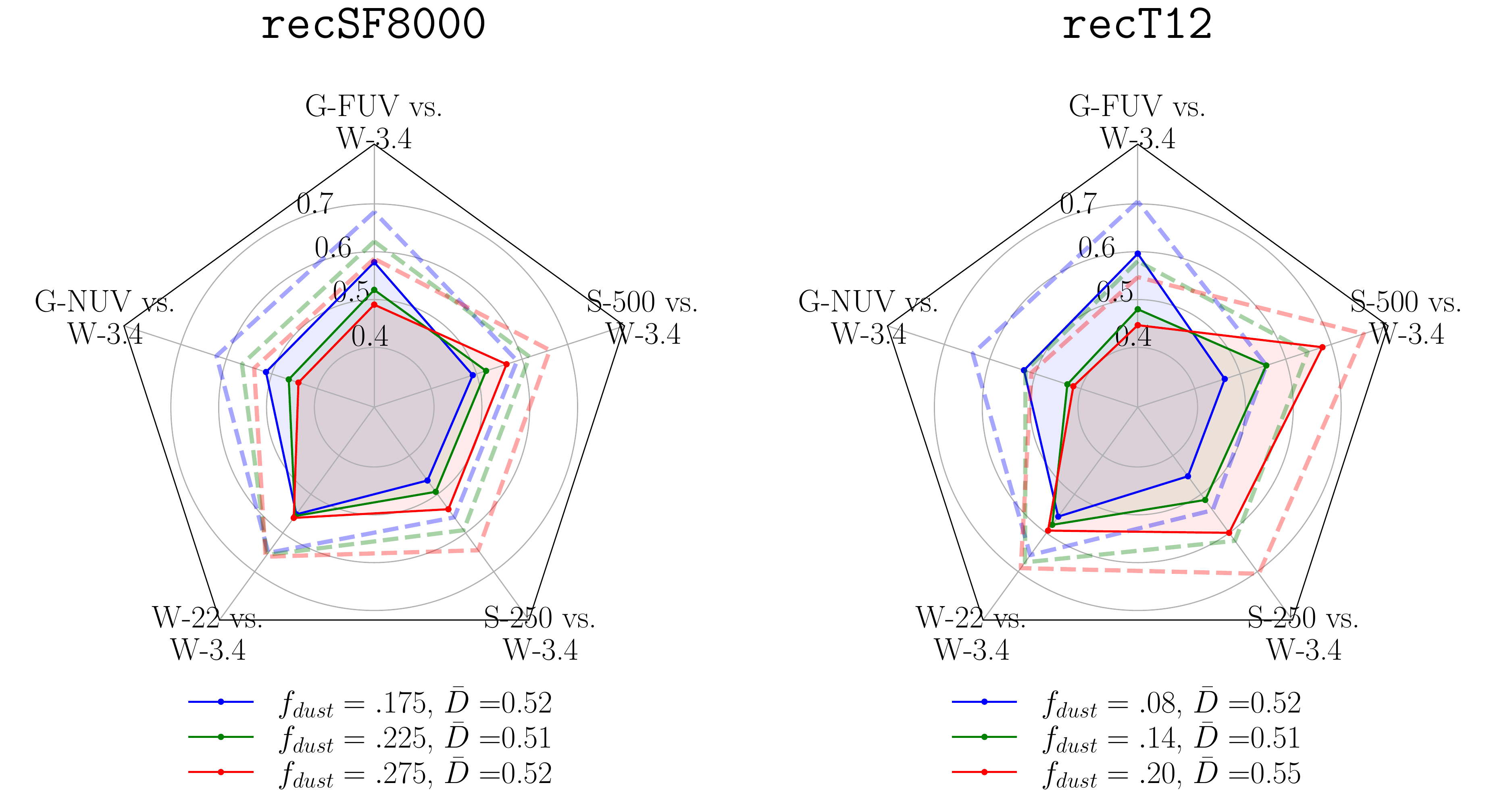}
\caption{Results of the 2D K-S test as described in Sect.~\ref{Calibration_fdust.ssec} for the two dust allocation recipes. The data points connected by the solid lines are the mean values, whereas the ones with dashed lines represent the one-sided  standard deviation. The legend lists the average $\bar{d}$ of the mean values for the five relations considered. We choose the $f_{dust}$ value where the mean values for all five relations are sandwiched between the extremes.} 
\label{KStest_spider.fig}
\end{figure*}

\begin{figure*}
\centering
\includegraphics[width=1\textwidth]{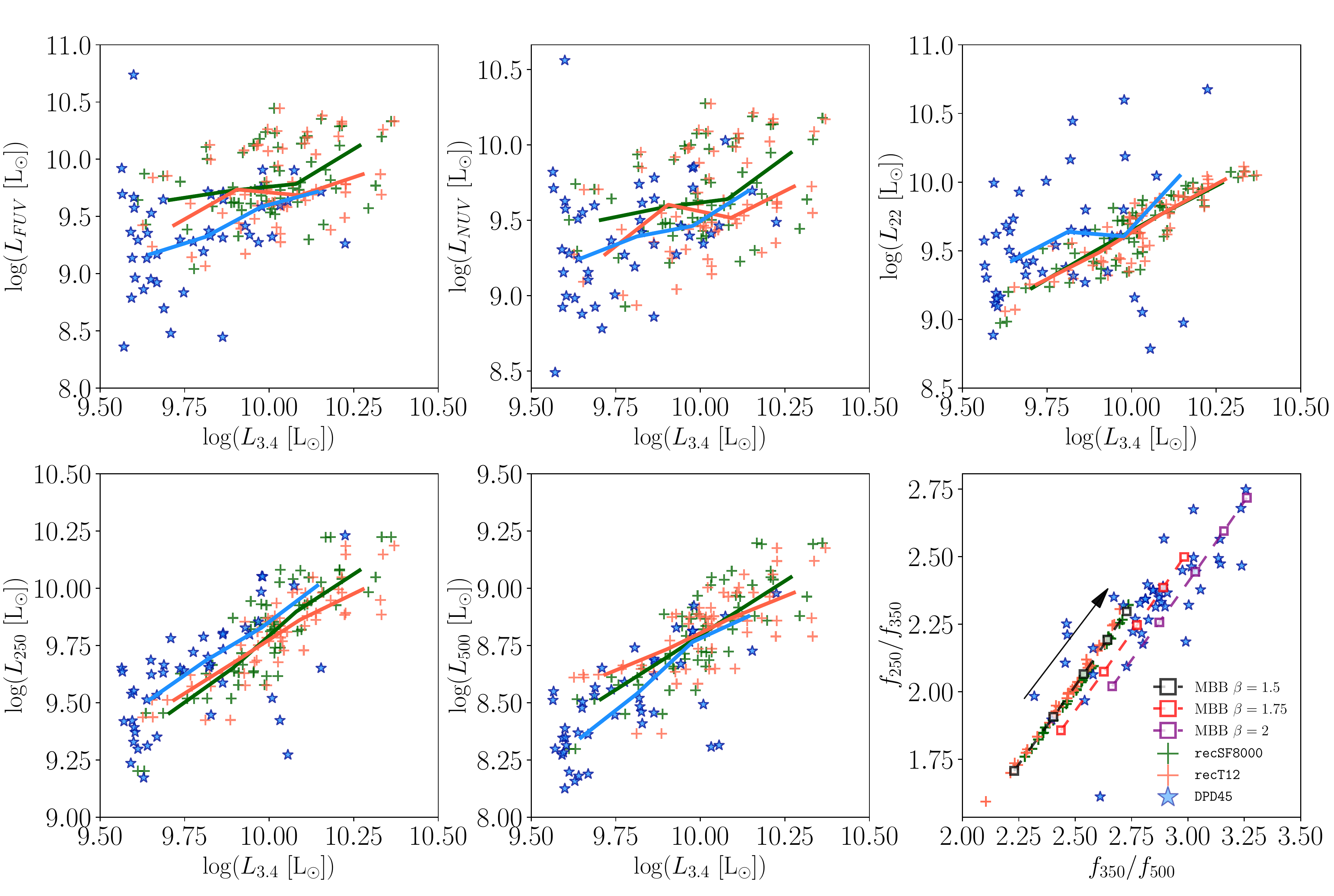}
\caption{Comparison of broadband luminosity scaling relations between the DustPedia sub-sample \texttt{DPD45} (see Sect.~\ref{calibration.sec}) and \texttt{SKIRT} post-processed Auriga galaxies using the calibration parameters listed in Tab.~\ref{calibParameters.table}. 
The Auriga sample for both \texttt{recSF8000} and \texttt{recT12} consists of sixty points, including an edge-on and face-on configuration for each of the thirty Auriga galaxies. The rolling median lines use three bins of equal width in $L_{3.4}$. The bottom right sub-panel also includes modified blackbody (MBB) relations for three $\beta$ values. In each case, the MBB temperature runs from $18-30~\mathrm{K}$ in the direction of the arrow, with a step-size of $3~\mathrm{K}$.} 
\label{ScalingRelations_BBand.fig}
\end{figure*}

As discussed in Sect.~\ref{methodology.sec}, our post-processing recipes contain two global free parameters that we calibrate by comparing spatially integrated luminosities  for the sample of Auriga galaxies to those of a set of nearby galaxies. 
For both dust allocation recipes, we generate synthetic broadband fluxes in the GALEX, SDSS, 2MASS, WISE and Herschel broadband filters for each of Auriga galaxies in face-on and edge-on configurations. We then convert these fluxes to luminosities.
We vary the molecular cloud dissipation timescale $\tau_{\text{clear}}$ and the dust-to-metal ratio $f_{\text{dust}}$ in the models to find the best match with the observational data available.   

For this comparison, we use the DustPedia galaxy sample \citep{2017PASP..129d4102D}. This sample contains 875 nearby galaxies with matched aperture photometry in more than 40 bands from UV to millimeter wavelengths \citep{2018A&A...609A..37C}. We apply the \texttt{CIGALE} SED fitting code \citep{2019A&A...622A.103B, 2011ascl.soft11004N} to each of the DustPedia galaxies with the same parameter settings as used by \citet{2018A&A...620A.112B}, \citet{2019A&A...624A..80N} and \citet{2020MNRAS.494.2823T} except that we use the \citet{2003PASP..115..763C} IMF for SSPs, consistent with the IMF used in the Auriga simulations.
In order to select DustPedia galaxies similar to the Auriga galaxies, we consider late type (Hubble Stage $>$~0), star-forming galaxies with ${\mathrm{SFR}}_{\mathrm{cig}}>~0.8~\mathrm{M_{\odot}yr^{-1}}$, where ${\mathrm{SFR}}_{\mathrm{cig}}$ is the SFR inferred using the \texttt{CIGALE} SED fitting code.
We impose an additional constraint based on WISE 3.4 $\mu$m band luminosities, where we select galaxies which fall in the WISE 3.4 $\mu$m luminosity range of the Auriga sample. This essentially imposes a stellar mass selection criterion because the 3.4 $\mu$m  band is a good proxy for stellar mass \citep{2013MNRAS.433.2946W}. This procedure leaves us with a Dustpedia sub-sample of 45 galaxies, which we call \texttt{DPD45}.
We mention that for any missing observational data, we substitute the best fit value obtained using \texttt{CIGALE}. For the seven bands used in calibration, i.e, GALEX-FUV, GALEX-NUV, WISE-1, WISE-4, PACS-70, SPIRE-250, SPIRE-500, the number of values replaced by SED fitting for \texttt{DPD45} are 1, 1, 0, 0, 36, 1, 1, respectively.

We calibrate the free parameters for our dust allocation recipes (see Sect.~\ref{DustAllocation.sssec}) by comparing broadband luminosity scaling relations between the post-processed Auriga galaxies and \texttt{DPD45}.

\subsection{Fixing the cloud dissipation time scale}
\label{Calibration_Tau.ssec}
As mentioned in Sect.~\ref{methodology.sec}, the cloud dissipation time scale is connected to the covering fraction of the PDR and thus determines the dust emission from the star-forming regions.
With increasing $\tau_{\text{clear}}$, the contribution of the dust emission from the star-forming regions modeled by the \texttt{MAPPINGS-III} templates in the galaxy SED is expected to rise \citep[see][Fig.\,6]{2008ApJS..176..438G}.
We expect this increased contribution to have influence the infrared (IR) colors of the galaxies.
To evaluate this effect, we study the color-color relation for the WISE-22, PACS-70, SPIRE-250 and SPIRE-500 bands as shown in Fig.~\ref{tau&fdustParameterSpace_AU3.fig}.
An increasing value of $\tau_{clear}$ leads to an increasing flux from the cold dust in the molecular cloud surrounding the H\textsc{ii} region. The contribution of this cool dust can be traced by the variation in the $f_{70}/f_{22}$ color, which is likely to increase with an increasing value of $\tau_{clear}$ \citep[again, see][Fig.\,6]{2008ApJS..176..438G}.
On the FIR side of the spectrum, we use $f_{250}/f_{500}$ color values to characterize the downwards slope of the dust continuum emission. This color shows sensitivity to the cold dust contents, with smaller color values indicating a flatter slope of the dust emission curve and thus a larger contribution from colder dust.
An increasing $\tau_{clear}$ value is likely to lead to a larger amount of cold dust in the ISM, manifesting itself as a reduction in the $f_{250}/f_{500}$ color value. 

Along with the \texttt{DPD45} data points, Fig.~\ref{tau&fdustParameterSpace_AU3.fig} shows the results for Auriga models with $\tau_{clear}=1,2~\textrm{Myr}$ and an extreme value of $8~\textrm{Myr}$. 
We find a limited impact of changing this parameter. None of the parameter values lead to a very good agreement with \texttt{DPD45}, although the results for lower values show a slightly better agreement. The large difference in the $f_{250}/f_{500}$ color between the Auriga galaxies and \texttt{DPD45} is likely to be a result of inadequate dust heating, which we discuss further along with the scaling relations of Fig.~\ref{ScalingRelations_BBand.fig}.

Observationally, $\tau_{\text{clear}}$ is not very well constrained but is expected to be of the order of a few Myr and is expected to be driven by early feedback mechanisms such as winds, radiation pressure, photoionization etc. The value quoted in \citet{2008ApJS..176..438G} is $1-2$ Myr, whereas in \citet{2020SSRv..216...50C, 2020MNRAS.493.2872C}, based on a sample of nine nearby disc galaxies, the duration of the feedback phase during which molecular clouds and H\textsc{ii} regions coexist is inferred to be 1–5 Myr.

We fix $\tau_{\text{clear}}=1$~Myr for both \texttt{recSF8000} and \texttt{recT12} based on the behavior of the \texttt{MAPPINGS-III} templates in our tests, while being in the range of values quoted in the literature. 
It is worth mentioning that although we did the parameter space exploration only for \texttt{recSF8000}, we do not expect a significant difference in results if we repeat this exercise for \texttt{recT12}.
We remark that while we have tested $\tau_{clear}<1~\textrm{Myr}$, we do not use those values in order to be consistent with the observationally inferred clearing time scales.
\subsection{Fixing the dust-to-metal ratio}
\label{Calibration_fdust.ssec}

Using $\tau_{\text{clear}}=1$~Myr, we now quantify the correspondence between \texttt{DPD45} and the Auriga mock observations as a function of $f_{\text{dust}}$.
We use a generalization to two-dimensional distributions of the well-known Kolmogorov–Smirnov test \citep{KOLMOGOROV1933SullaDistribuzione, Smirnov1948TableDistributions} as described in \citet{1992nrca.book.....P} following \citet{1987MNRAS.225..155F} and \citet{1983MNRAS.202..615P}. The K–S test computes a metric $D$ which can be interpreted as a measure of the ‘distance’ between two sets of two-dimensional data points, with smaller $D$ values indicating better correspondence.
We use the scaling relations between WISE-3.4, and each one of GALEX-FUV, GALEX-NUV, WISE-22, SPIRE-250 and SPIRE-500 to carry out the K-S test. We use the mean $D$ obtained from these scaling relations to quantify our choice of  $f_{\text{dust}}$.
%The actual comparison: we could show the normalised SEDs (figure 11) and some of the scaling relations from figure 14. Only for the final selection (run2)?
We note that the \texttt{DPD45} sample lacks galaxies at higher mass end of Auriga sample, i.e, with a WISE-3.4 luminosity above $10^{10}~L_{\odot}$ (median of the Auriga sample). Hence, we implement the K-S test as follows:
\begin{enumerate}[leftmargin=*]
    \item We generate a sub-sample out of the \texttt{DPD45} sample, where we keep all galaxies with WISE-3.4 luminosity above the median value of the Auriga sample, and we randomly select the same number of Dustpedia galaxies with WISE-3.4 luminosity below the median Auriga value.
    \item We randomly select the same number of Auriga galaxies as the Dustpedia sub-sample and apply the K–S test.
    \item We repeat this till the statistics converge, and use the mean and standard deviation of the test results to compare the different $f_{\text{dust}}$ values. 
\end{enumerate}

We select a final value for $f_{\text{dust}}$ based on the K–S test results shown in Fig.~\ref{KStest_spider.fig}. In the case of \texttt{recSF8000}, we choose $f_{\text{dust}}=~0.225$
considering its correspondence with the observational data in each of the chosen scaling relations, despite there being other values of $f_{\text{dust}}$  which show better agreement in some of the scaling relations. For example, a lower value of $f_{\text{dust}}$ shows a better agreement for the WISE-3.4 vs. SPIRE-500 scaling relation, or, a higher value of $f_{\text{dust}}$ is better for WISE-3.4 vs. GALEX-FUV.
A similar argument leads us to choosing $f_{\text{dust}}=~.14$ for \texttt{recT12}.
The final choice of the set of free parameters is shown in Table~\ref{calibParameters.table}.

It is worth noting that the $f_{dust}$ value is lower for the \texttt{recT12}
recipe than for \texttt{recSF8000}, which is to be expected because more gas cells are assumed to contain dust, while producing similar scaling relations.
In comparison, \citet{2019A&A...623A...5D} find a nearly constant $f_{dust}$ value for all  DustPedia galaxies with a gas fraction lower than $60\%$. The exact value is model dependent and falls in the range of $.092-0.214$.
Our calibrated $f_{dust}$ values lie in a similar range for Auriga galaxies, with gas fractions ranging from $10-40~\%$ and a median value of $22~\%$. 
We point out that there is some discrepancy between \texttt{DPD45} and the Auriga galaxies in the 
$\mathrm{sSFR-M_{\star}}$ plane, we discuss this in Sect.~\ref{results_CIGALE_DustScalingRelations.sssec}.
\begin{table}
\caption{Values of the calibration parameters used for recipes \texttt{recSF8000} and \texttt{recT12}}
\label{calibParameters.table}.
\centering
\begin{tabular}{lcl}
\hline 
Dust allocation recipe & $\tau_{\text{clear}}$(Myr) & $f_{\text{dust}}$ \\
\hline 
\texttt{recSF8000} & 1 & 0.225 \\
\texttt{recT12} & 1 & 0.14 \\
\hline
\end{tabular}
\end{table}

Fig.~\ref{ScalingRelations_BBand.fig} shows the broadband scaling relations of the post-processed Auriga galaxies using the final, calibrated free parameter values. 
In comparison with \texttt{DPD45}, there is good overall agreement.
We do note minor discrepancies in the FUV: Auriga galaxies have, on average, slightly higher UV luminosities than the observed galaxies. Furthermore, the color-color relation between $f_{350}/f_{500}$ and $f_{250}/f_{350}$ exhibits very little scatter and shows a lower global dust temperature in comparison to the galaxies in \texttt{DPD45}. 
The lack of scatter for the Auriga galaxies could be caused, at least in part, by our use of a single dust model with the same dust properties galaxy-wide, which is not very realistic. On the other hand, some of the scatter in the DPD45 galaxies might be the result of observational limitations, which we did not consider for the Auriga galaxies. We investigated this issue following the flux-limiting method used by \citet{2016MNRAS.462.1057C}, and found that, while the scatter in the Auriga data points does increase, the effect is minor.

The lower cold dust temperatures and flatter FIR slopes seen for the Auriga galaxies are most likely caused by inadequacies in the sub-grid treatment of the dust geometry on small scales. \citet{2016MNRAS.462.1057C, 2018ApJS..234...20C} have noted the limitations of the dust modeling which are likely responsible for similar discrepancies in synthetic observations of EAGLE galaxies.
Essentially, our dust modeling does not fully capture the clumpy nature of the dust density distribution. 
The \texttt{MAPPINGS-III} templates assume spherical symmetry for modeling the dusty region surrounding the H\textsc{ii} core, resulting in isotropic emission. Also, the diffuse dust structure in our models is tied to the gas structure in the simulations, which is governed by a multi-phase, `effective' equation of state. The pressure-supported gas is overly smooth on scales of a few hundred parsecs \citep{2019MNRAS.489.4233M, 2018MNRAS.473.1019B}. 
These approximations lead to an insufficient amount of dust being irradiated by the strong radiation fields present within and around star-forming regions, thus underestimating the temperatures obtained for the Auriga galaxies.
Future efforts may (and hopefully will) improve sub-grid models so that they more realistically reflect the turbulent structure of the dust density distribution in star forming regions and in the diffuse dust. However, that is beyond the scope of this work.

\section{Synthetic data products}
\label{syntheticData.sec}

\begin{figure}
\centering
\begin{tabular}{@{}c@{}}
\includegraphics[width=.4\columnwidth]{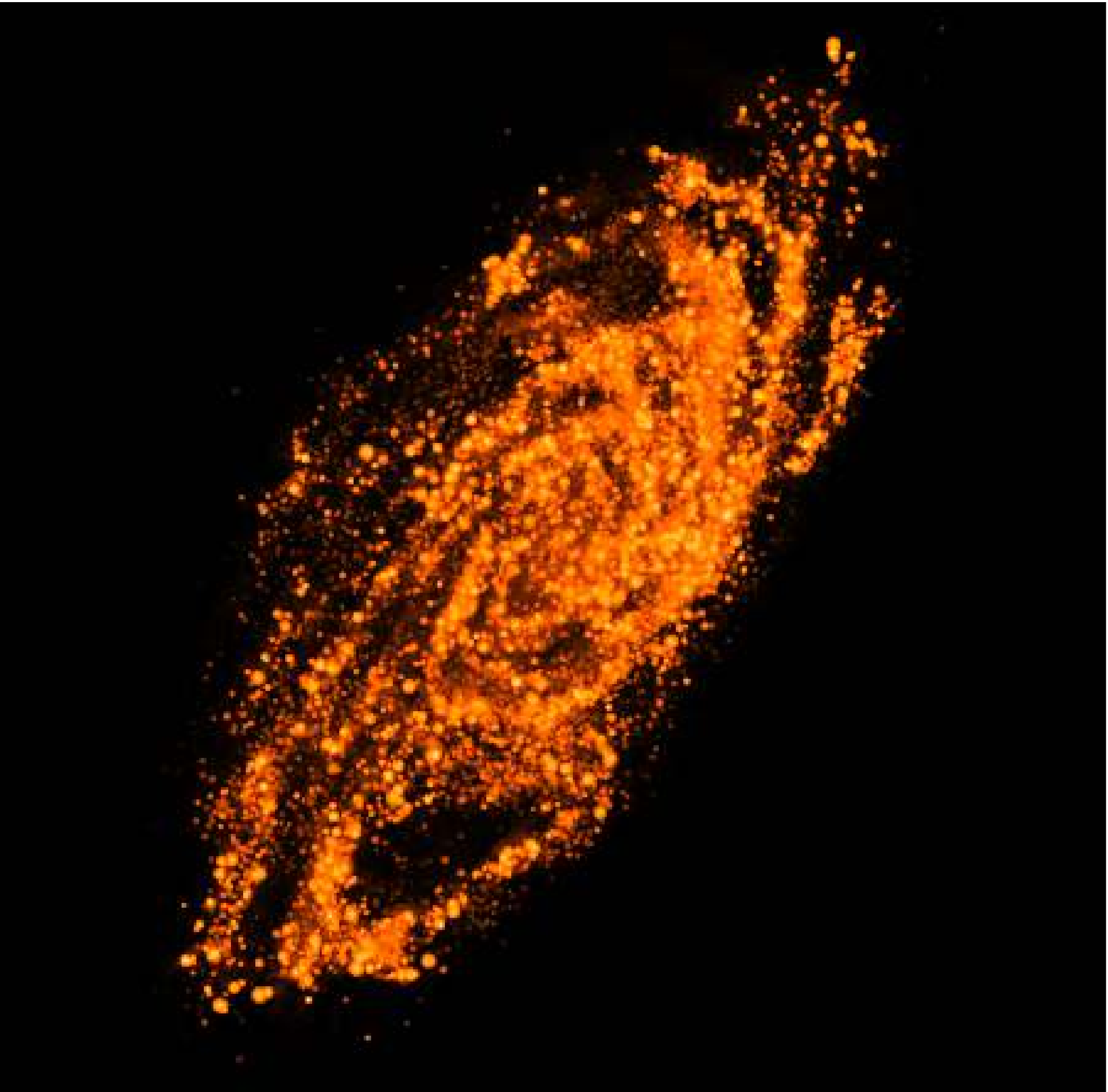} \\
\text{GALEX-NUV}
\end{tabular}
\begin{tabular}{@{}c@{}}
\includegraphics[width=.4\columnwidth]{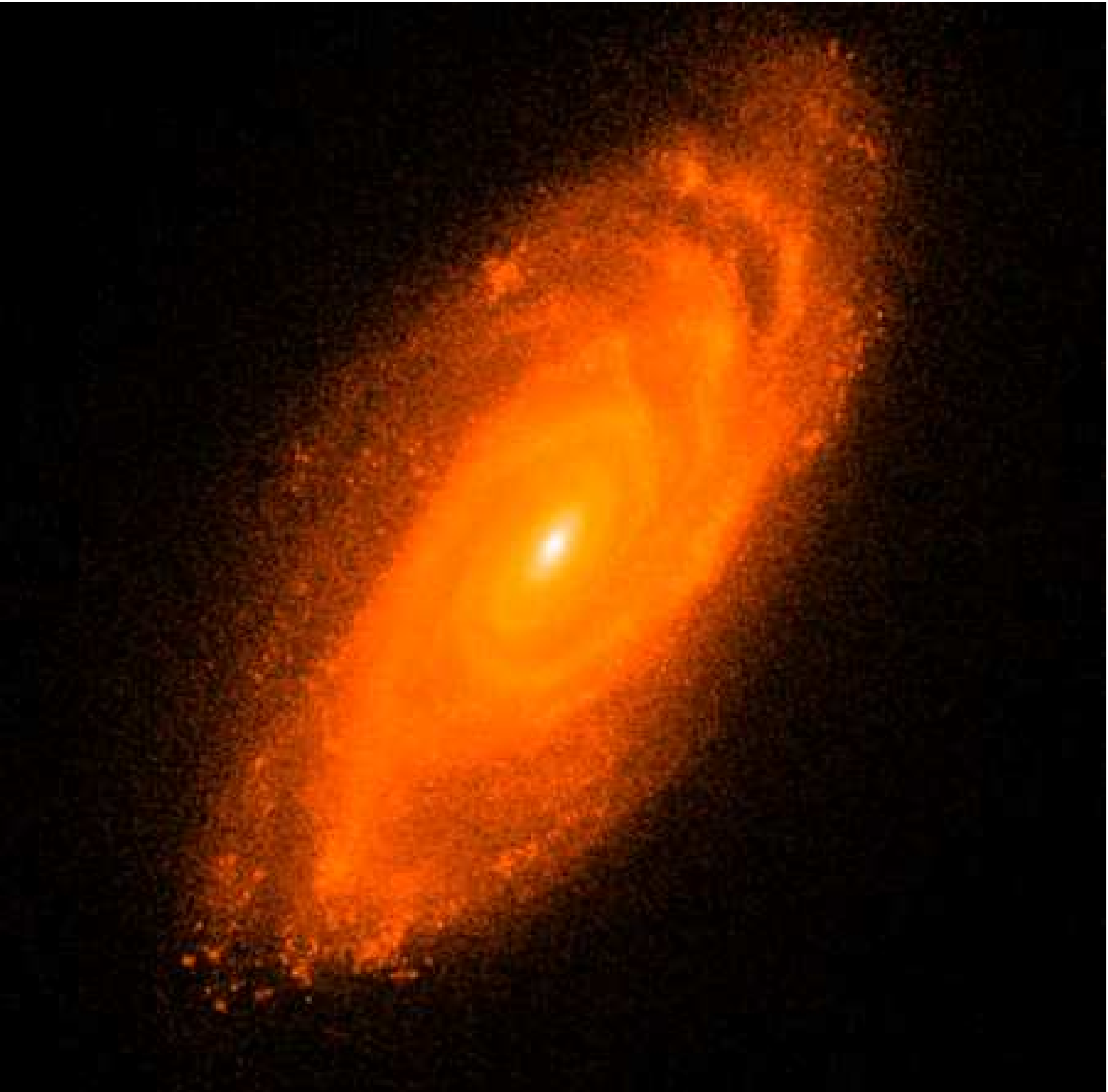} \\
\text{WISE-W1}
\end{tabular}
\begin{tabular}{@{}c@{}}
\includegraphics[width=.4\columnwidth]{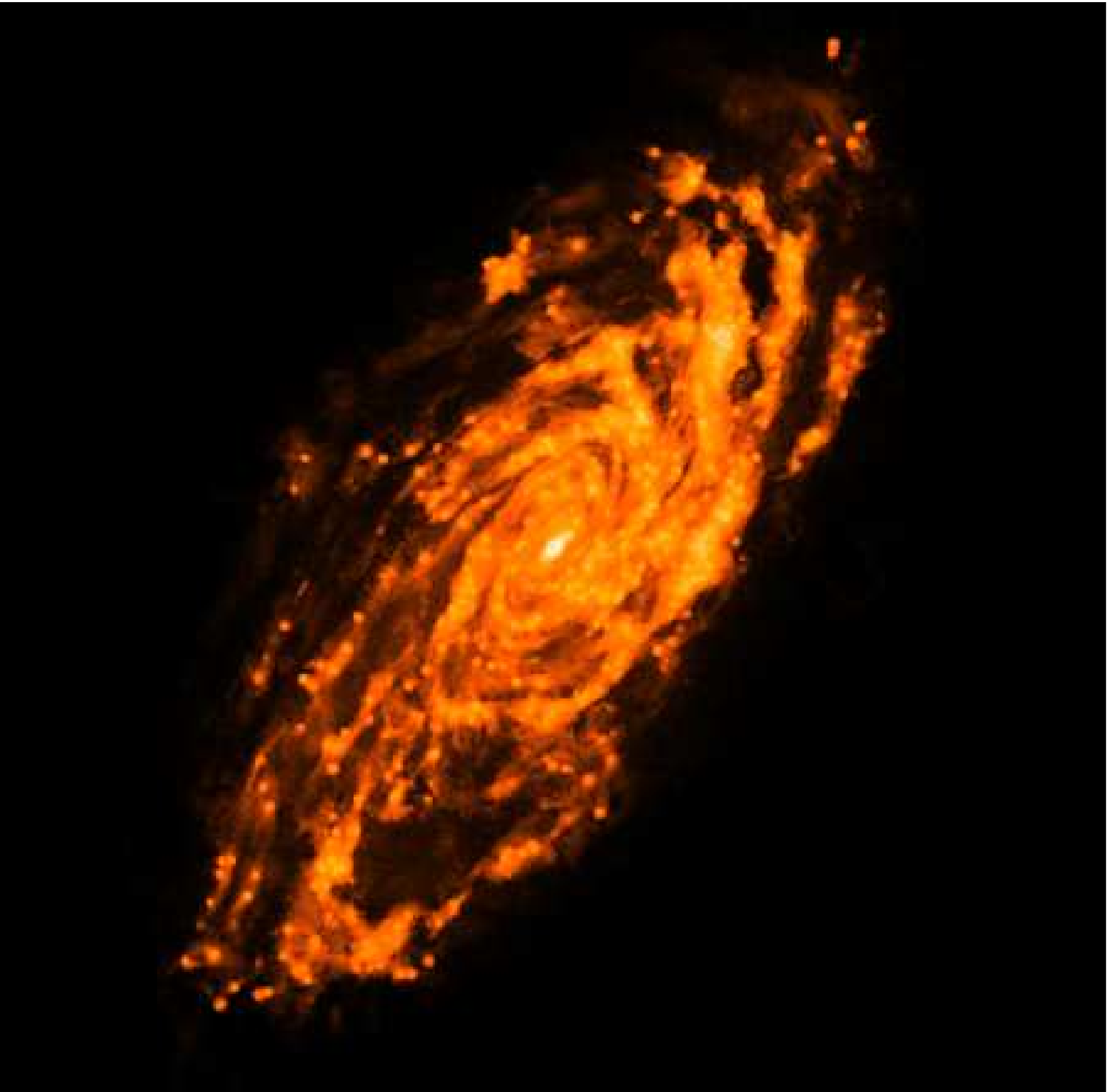} \\
\text{WISE-W4}
\end{tabular}
\begin{tabular}{@{}c@{}}
\includegraphics[width=.4\columnwidth]{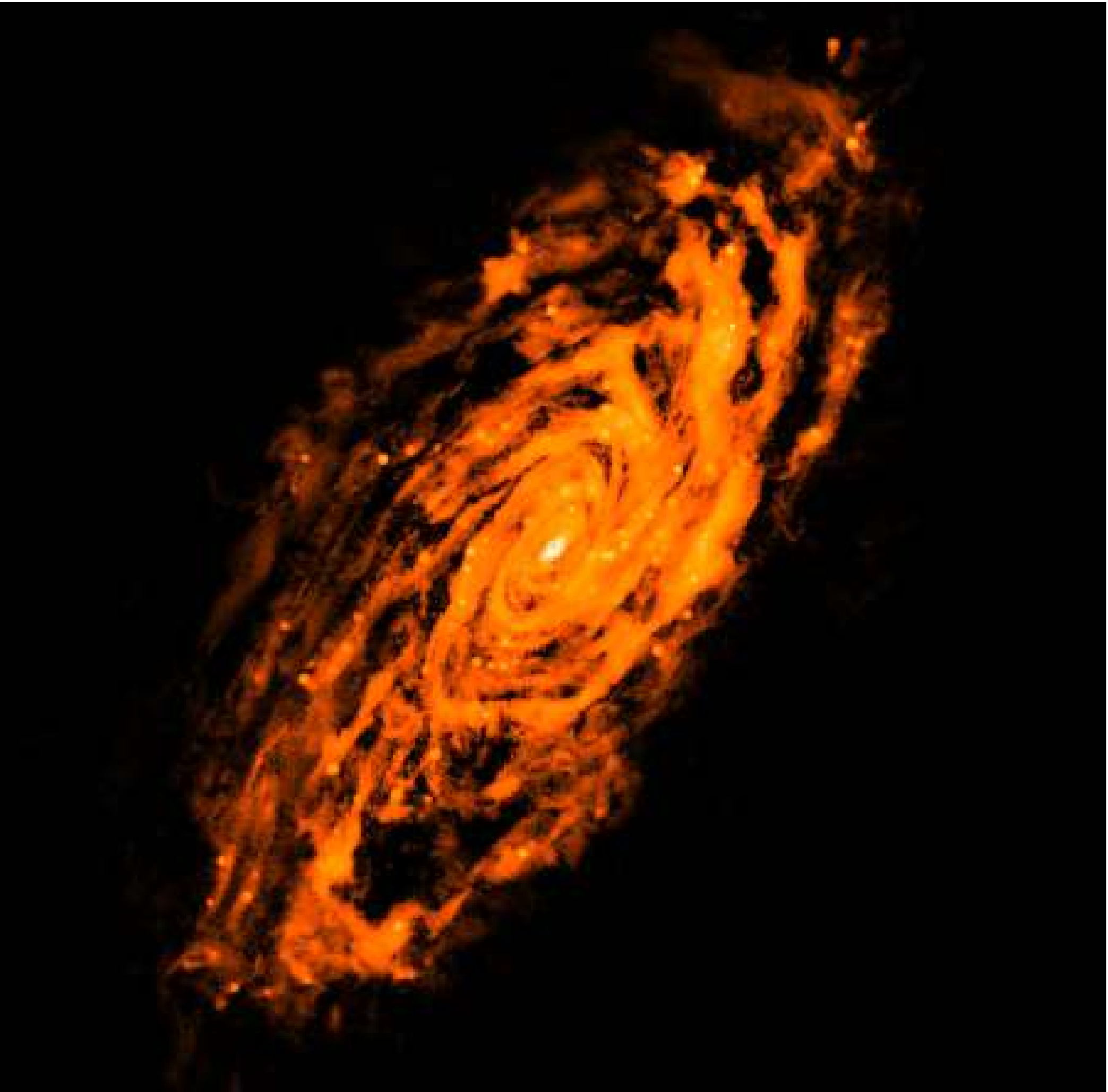} \\
\text{SPIRE-500}
\end{tabular}
\caption{An example of the broadband images of Auriga AU-16 in random orientation (see Sect.~\ref{dataproducts.ssec}) using the calibration defined in Sect.~\ref{calibration.sec} for \texttt{recSF8000}, using $2\times10^{10}$ photon packets. The images do not have the same color scale.} 
\label{BroadbandImages_Examples.fig}
\end{figure}

\begin{figure*}
\centering
\begin{tabular}{ccccc}
%%%%%%%%%%%%%% Row 1 and 2
\includegraphics[width=.1775\textwidth]{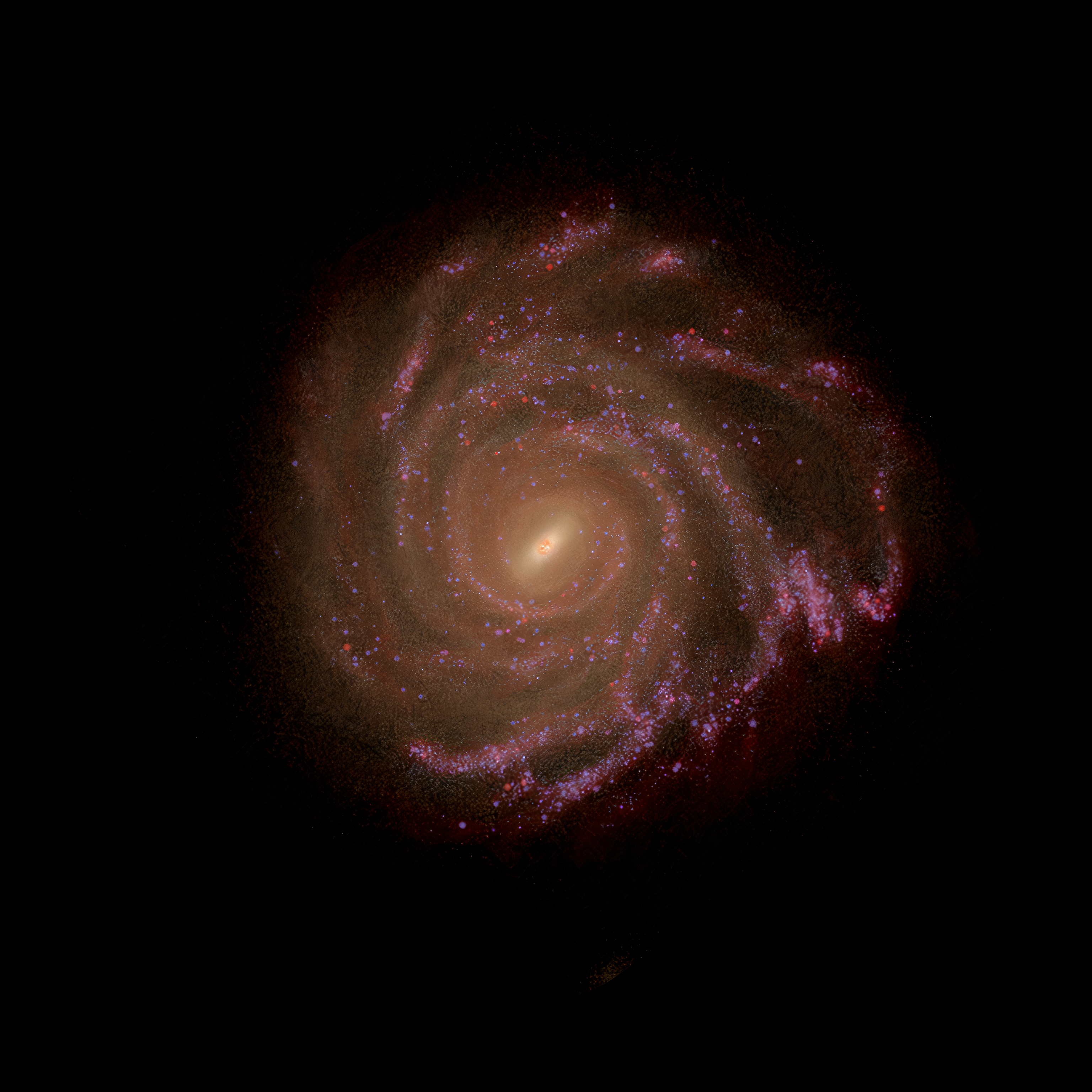} 
&
\includegraphics[width=.1775\textwidth]{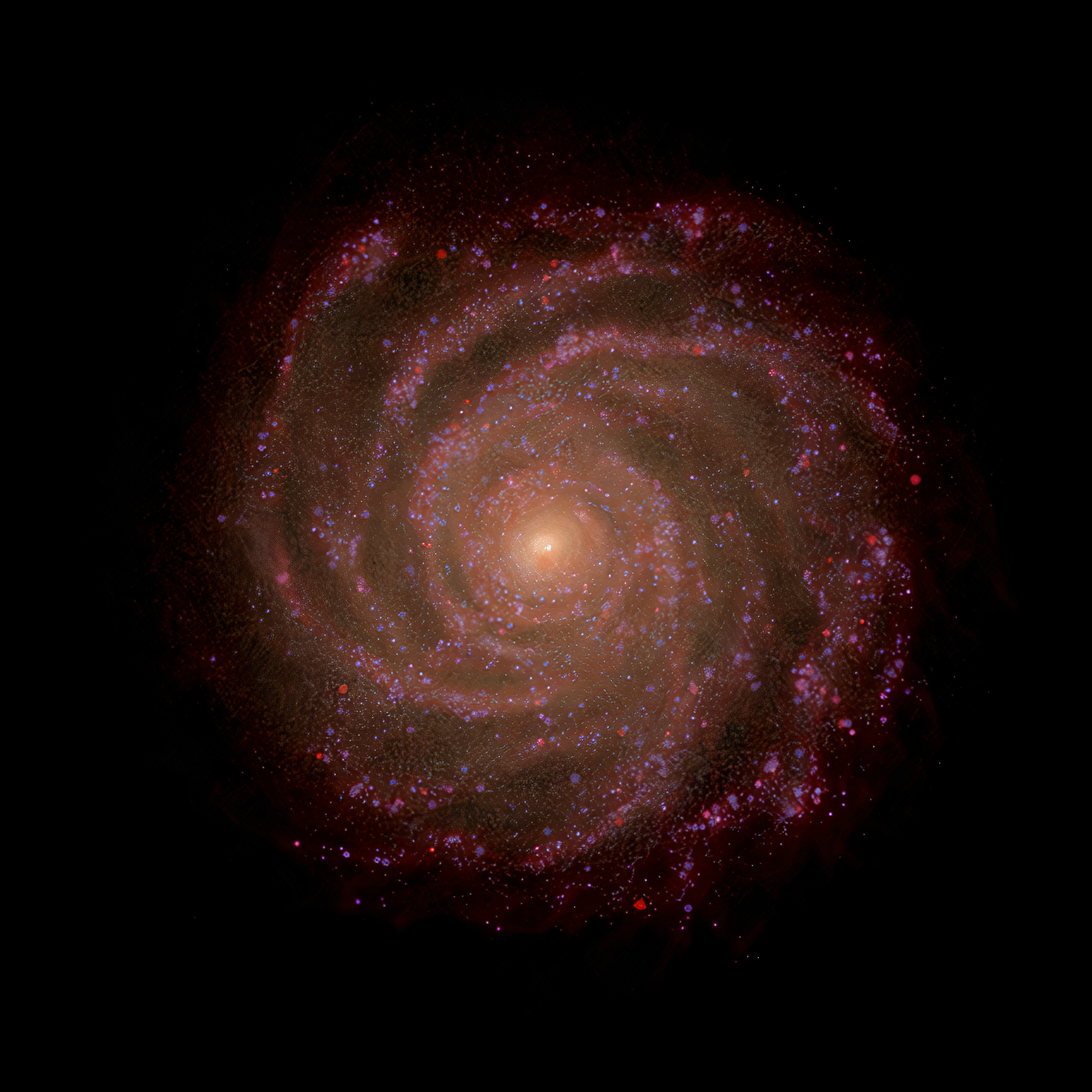}
&
\includegraphics[width=.1775\textwidth]{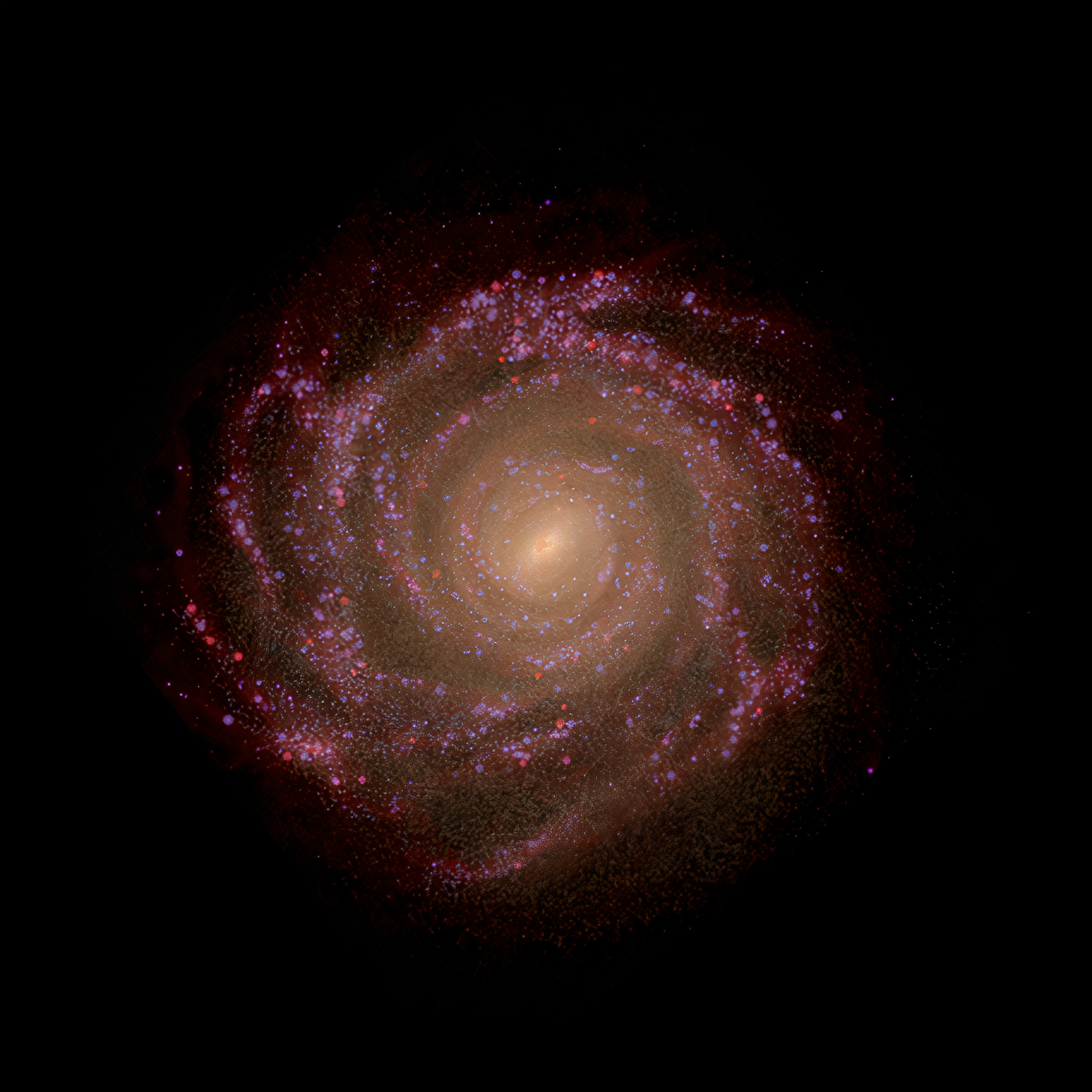}
&
\includegraphics[width=.1775\textwidth]{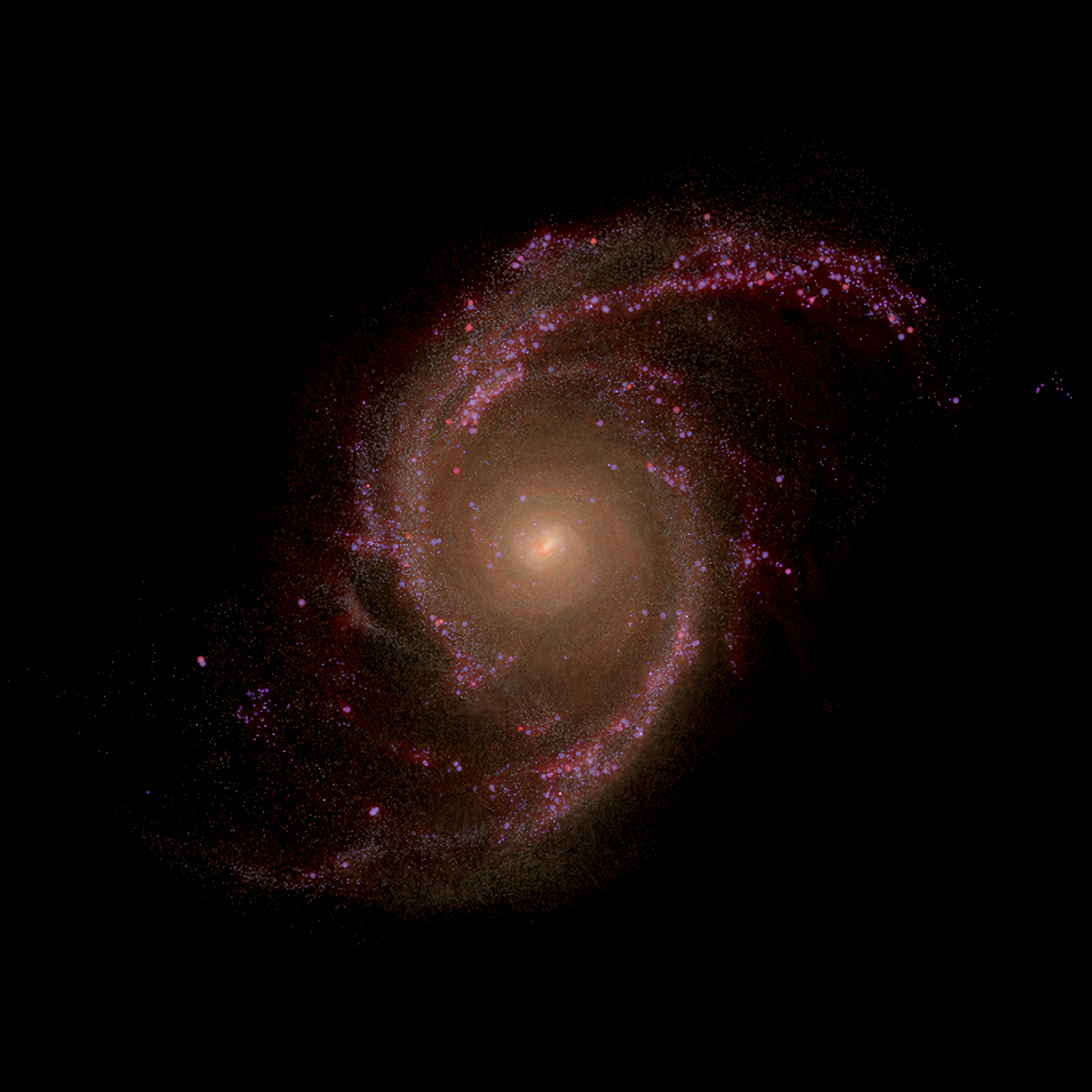}
&
\includegraphics[width=.1775\textwidth]{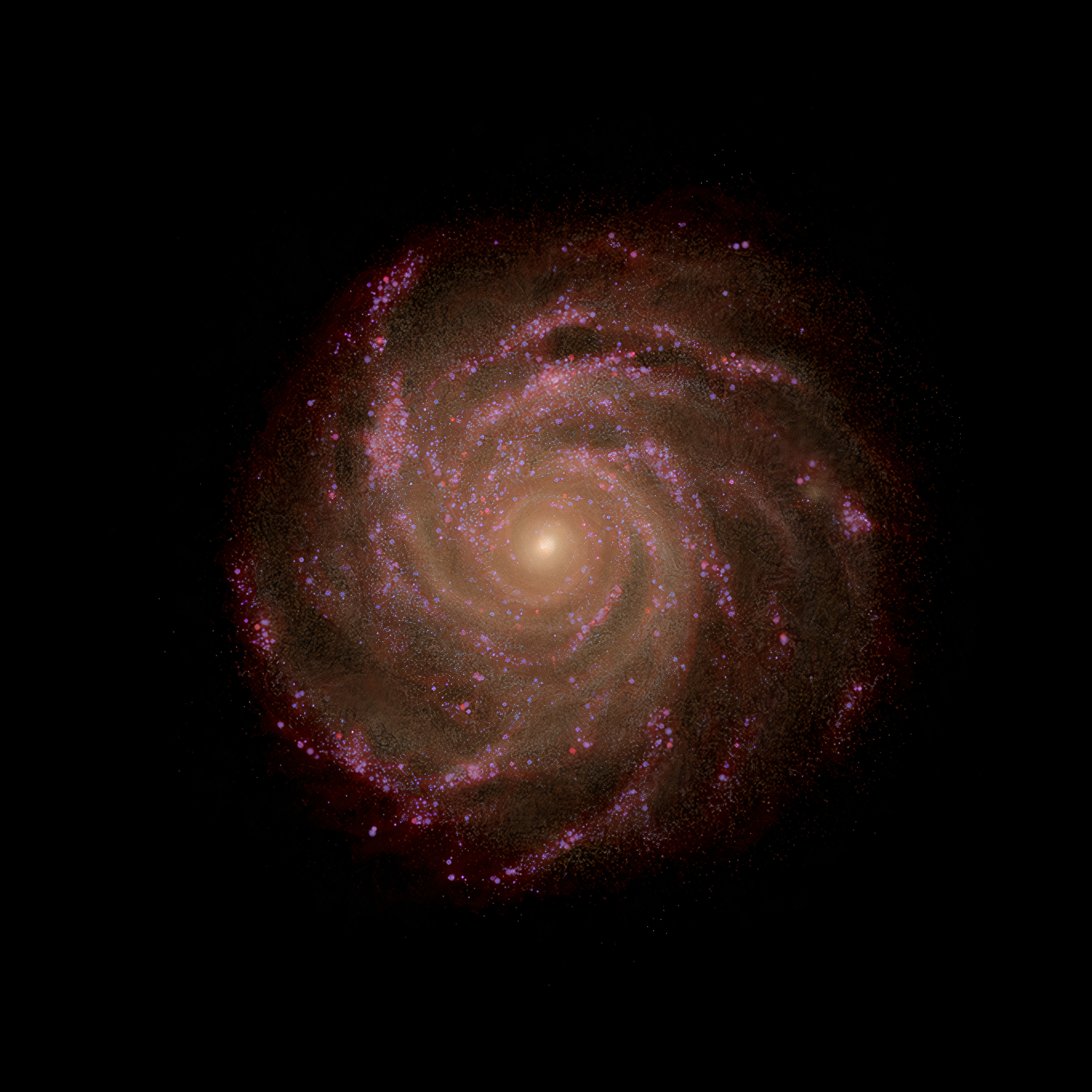} 

\\

\includegraphics[width=.1775\textwidth]{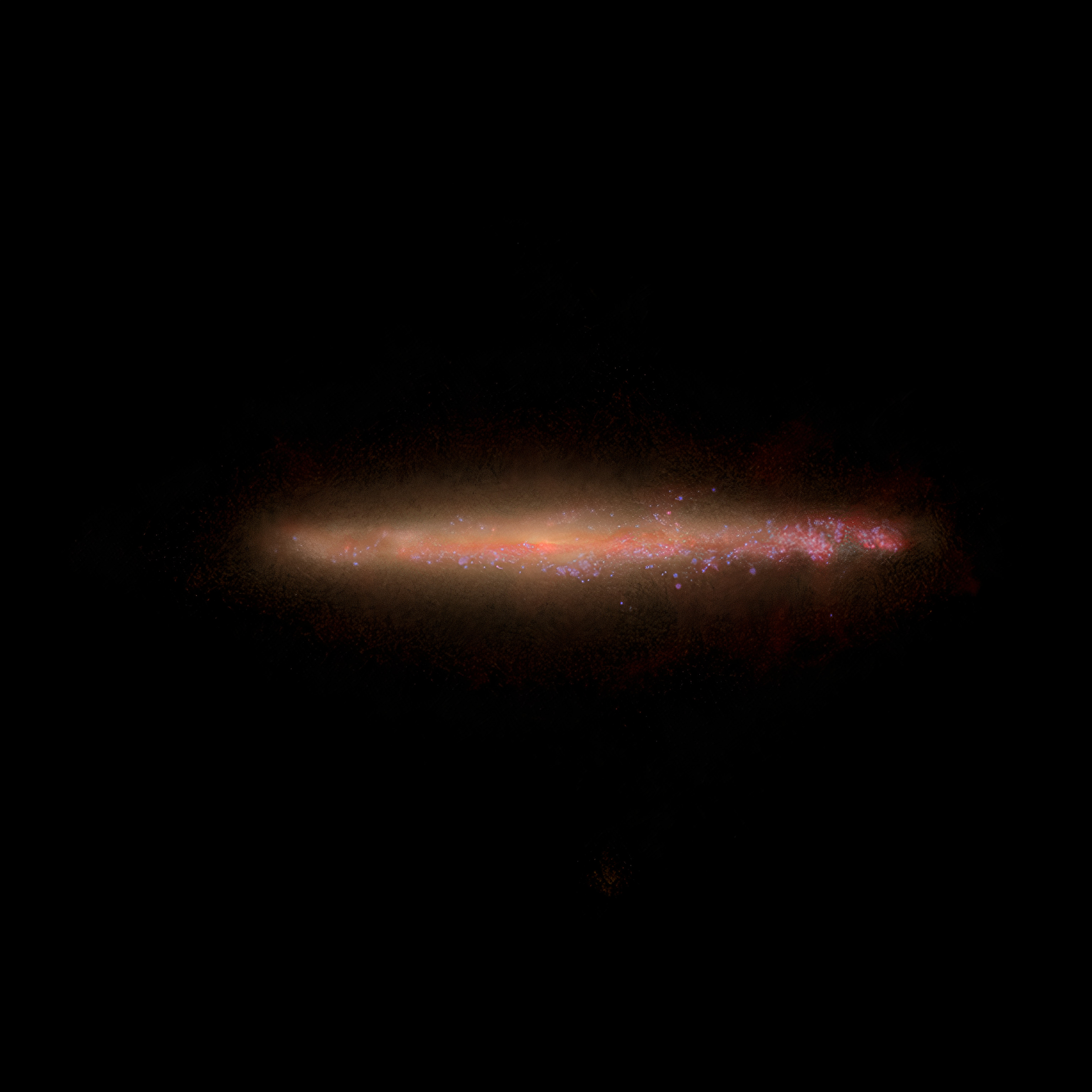} 
&
\includegraphics[width=.1775\textwidth]{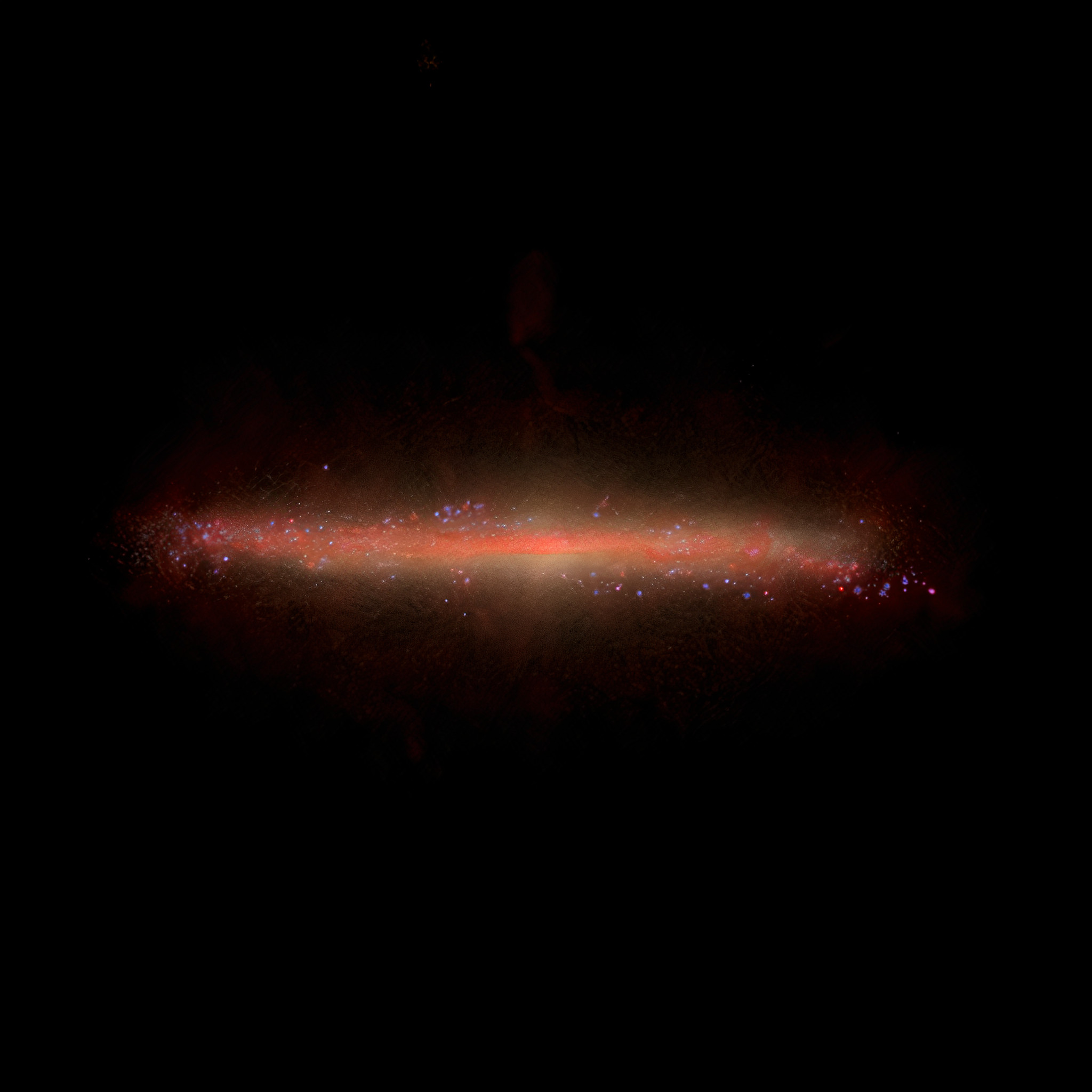}
&
\includegraphics[width=.1775\textwidth]{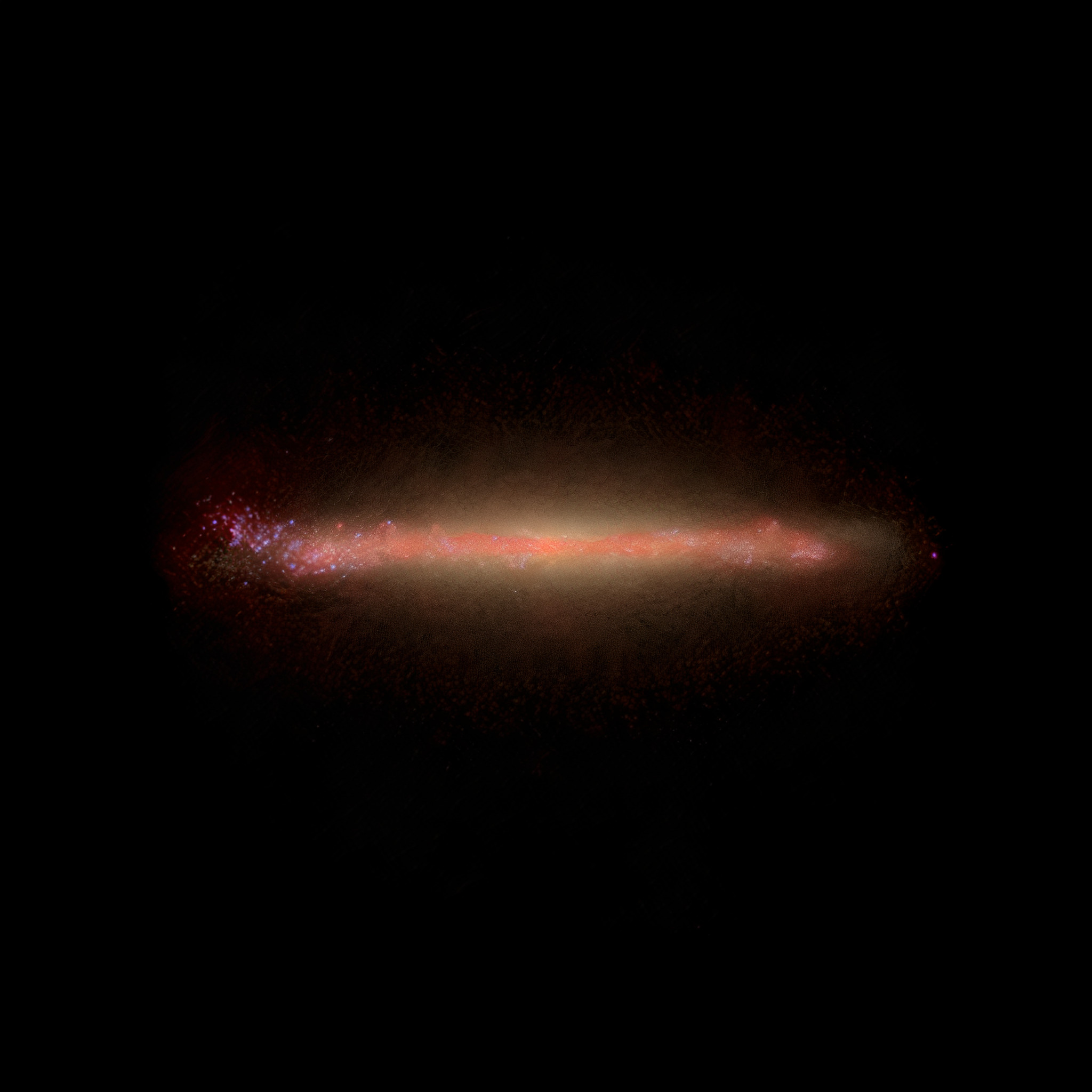}
&
\includegraphics[width=.1775\textwidth]{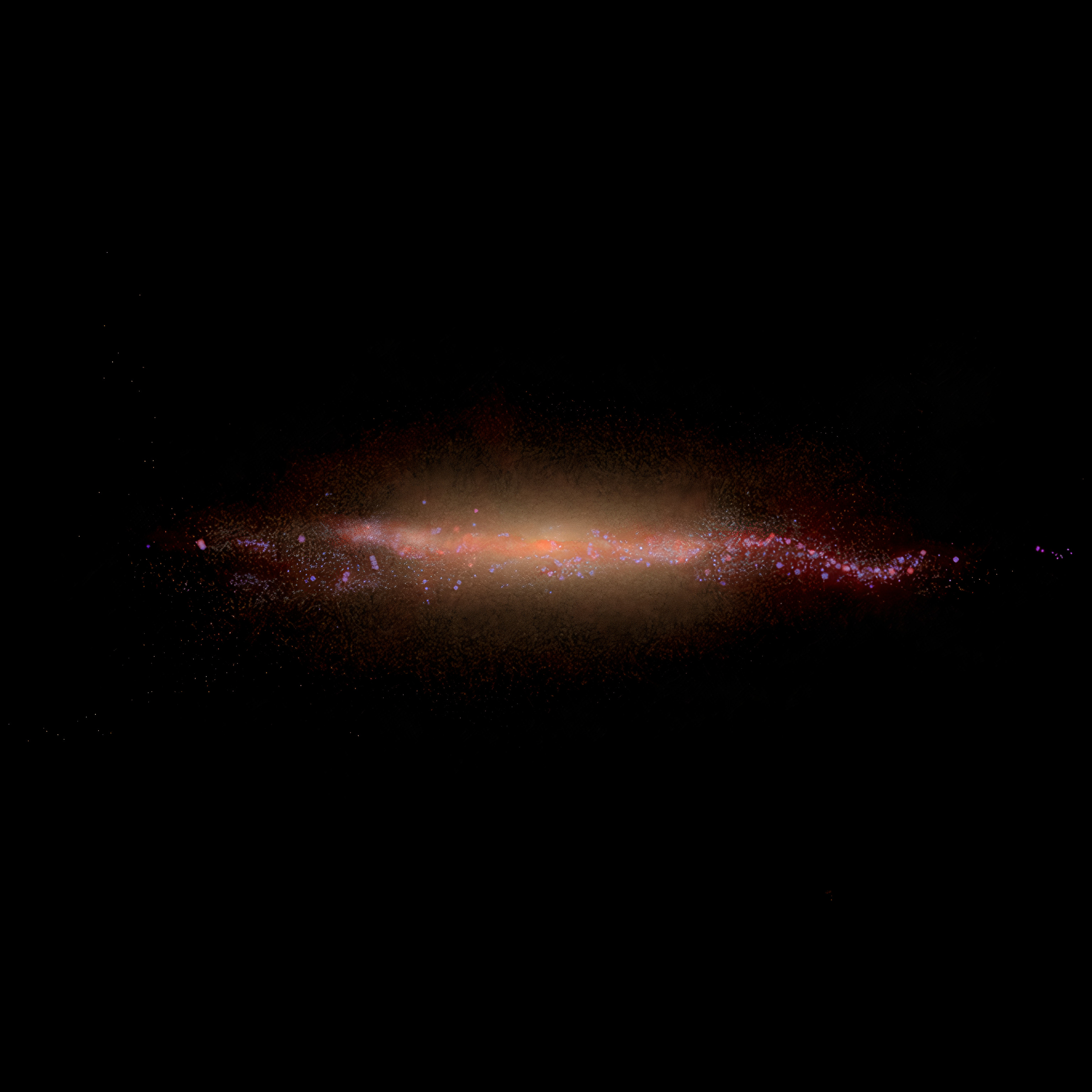}
&
\includegraphics[width=.1775\textwidth]{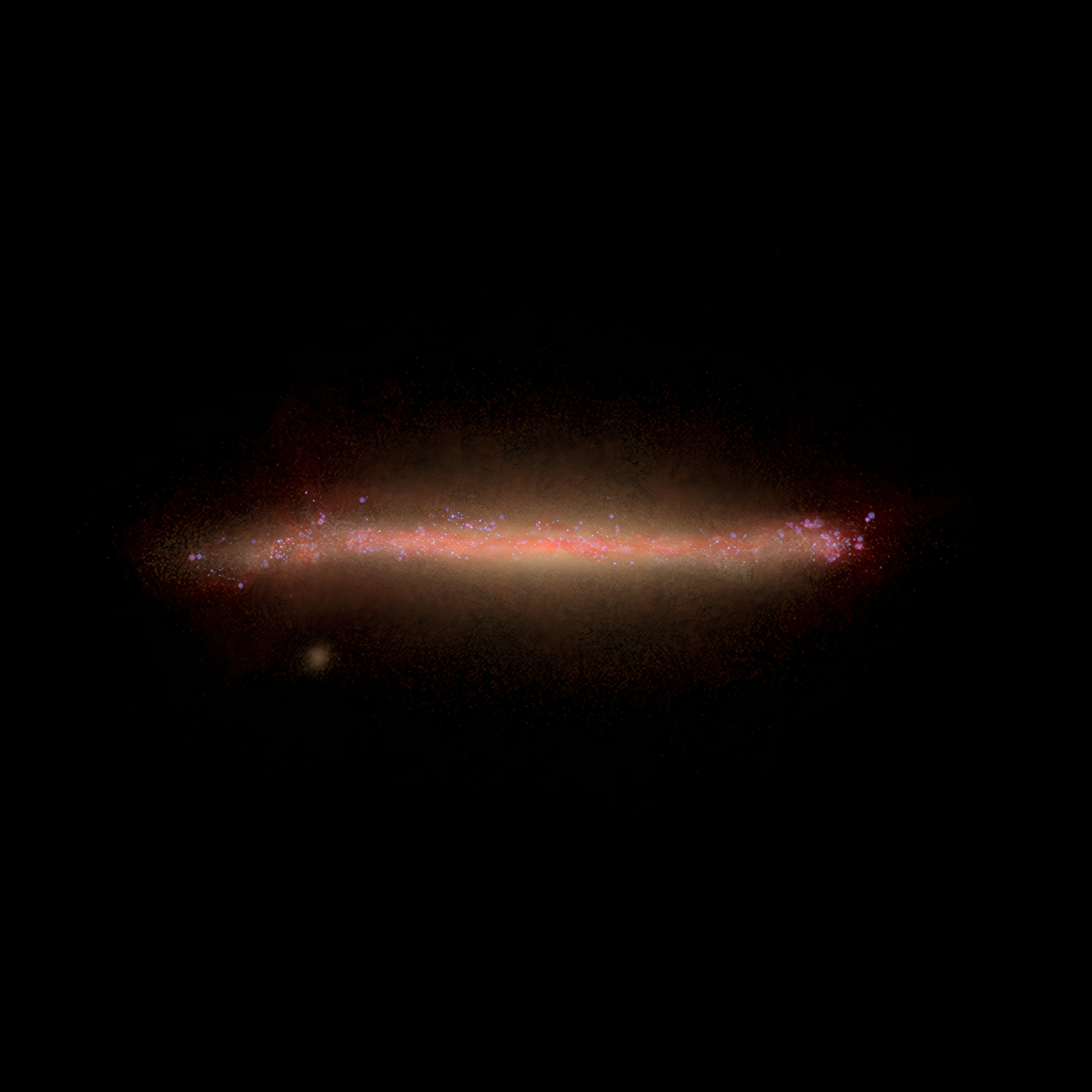} 

\end{tabular} 

\caption{Face-on and edge-on views of selected Auriga galaxies post-processed with the \texttt{recT12} recipe. Left to right: AU-2, AU-3, AU-6, AU-8, and AU-16. The images combine an optical view using SDSS$-u,~g,~r$ and
$z$ fluxes with additional blue for GALEX$-FUV$ flux and red for SPIRE$-250~\mu m$ flux. The resulting purple colors indicate star-forming regions, which strongly emit both in the FUV and FIR. The red colors indicate diffuse interstellar dust.} 
\label{compositeImages.fig}
\end{figure*}

%\subsection{Data products}
\label{dataproducts.ssec}

The main goal of our \texttt{SKIRT} post-processing of the Auriga galaxies is to produce and publish a set of synthetic observables. In this section we describe our choice for the number and positions of the synthetic observers, and the characteristics of the corresponding synthetic instruments. In principle, \texttt{SKIRT} can be run with an arbitrary number of observing positions, and with any number of synthetic instruments at each observing position. The only limitation is that each additional instrument implies an increase in both simulation run time and memory consumption.

We use 11 uniformly sampled points in $\cos i$, where $i$ is the inclination of the galaxy, i.e., the angle between the angular momentum vector and the direction towards the observer. This choice naturally leads to a finer grid close to edge-on positions \citep[see, e.g.,][]{2004A&A...419..821T}. For the three inclinations closest to edge-on, corresponding to $90^\circ$, $84.26^\circ$, and $78.46^\circ$, we place observers at three different azimuths. For barred galaxies which aren't significantly perturbed, these three positions allow views along, perpendicular to, and at an angle of $45^\circ$ with respect to the bar. For the remaining eight inclinations ($72.54^\circ,~ 66.42^\circ,~ 60^\circ,~ 53.13^\circ,~ 45.57^\circ,~ 36.87^\circ,~ 25.84^\circ,~ 0^\circ$) we use just a single observing position, at an azimuthal angle of $45^\circ$ with respect to the bar. We also consider an additional "random" viewpoint, corresponding to the native orientation of the galaxy in the simulation box. All together, this comes down to 18 observer positions per galaxy. 

%\subsubsection{Instrument characteristics}
%noindent

% As you describe in the document, for every observer, we consider spatially integrated SEDs with high spectral resolution ($\Delta\lambda/\lambda = ?$) and a set of broadband images. I would still go for the same broadband filters as in \citet{2018ApJS..234...20C}; see Table~\ref{bands.tab}. Concerning the FOV and spatial resolution of the images, I think we can be short and state that 60 kpc FOV and 2048 pixels is a choice that gives an ideal balance between high spatial resolution and still manageable in terms of memory and necessary run time. 
%Do we generate only the total images and SEDs, or do we also store the different contributions (direct emission, scattered emission, etc.)?

At every observer position, we place a UV--submm broadband instrument. This instrument provides us with both the spatially integrated flux and the high-resolution image for each band. We use the same set of 50 broadband filters as adopted by \citet{2018ApJS..234...20C}. This set includes most of the commonly used broadband filters for observations of nearby galaxies, together covering the entire UV--submm wavelength range.
Each individual broadband image has a resolution of 50~pc/pixel. Based on this pixel scale, the image field of view falls in one of two categories. If the extraction aperture side is smaller than or equal to $102.4~\text{kpc}$, we use a field of view of $102.4~\text{kpc}$, yielding an image with $2048\times2048$ pixels. If the extraction aperture is larger, we use a field of view of $153.6~\text{kpc}$, corresponding to an image with $3072\times3072$ pixels. The pixel scale of 50~pc/pixel is seven times smaller than the softening length of collision-less particles in the Auriga simulations (see Table~\ref{AurigaResolution.table}). This serves as a fair balance between high spatial resolution and the required memory and run time.

Fig.~\ref{BroadbandImages_Examples.fig} shows images in four selected broadbands for AU-16 in a random orientation.
Fig.~\ref{compositeImages.fig} shows composite images, combining fluxes from multiple bands, for selected Auriga galaxies.

%%%%%%%%%%%%%%%%%%%%%%%%%%%%%%%%%%%%%%%%%%%%%%%%%%%%%%%%%%%%%%%%%%%%%%%%%%%%%%%%%%%
%%%%%%%%%%%%%%%%%%%%%%%%%% CIGALE Analysis Results %%%%%%%%%%%%%%%%%%%%%%%%%%%%%
%%%%%%%%%%%%%%%%%%%%%%%%%%%%%%%%%%%%%%%%%%%%%%%%%%%%%%%%%%%%%%%%%%%%%%%%%%%%%%%%%%%

\section{Results and Analysis}
\label{Results.sec}

\begin{figure}
\centering
\includegraphics[width=\columnwidth]{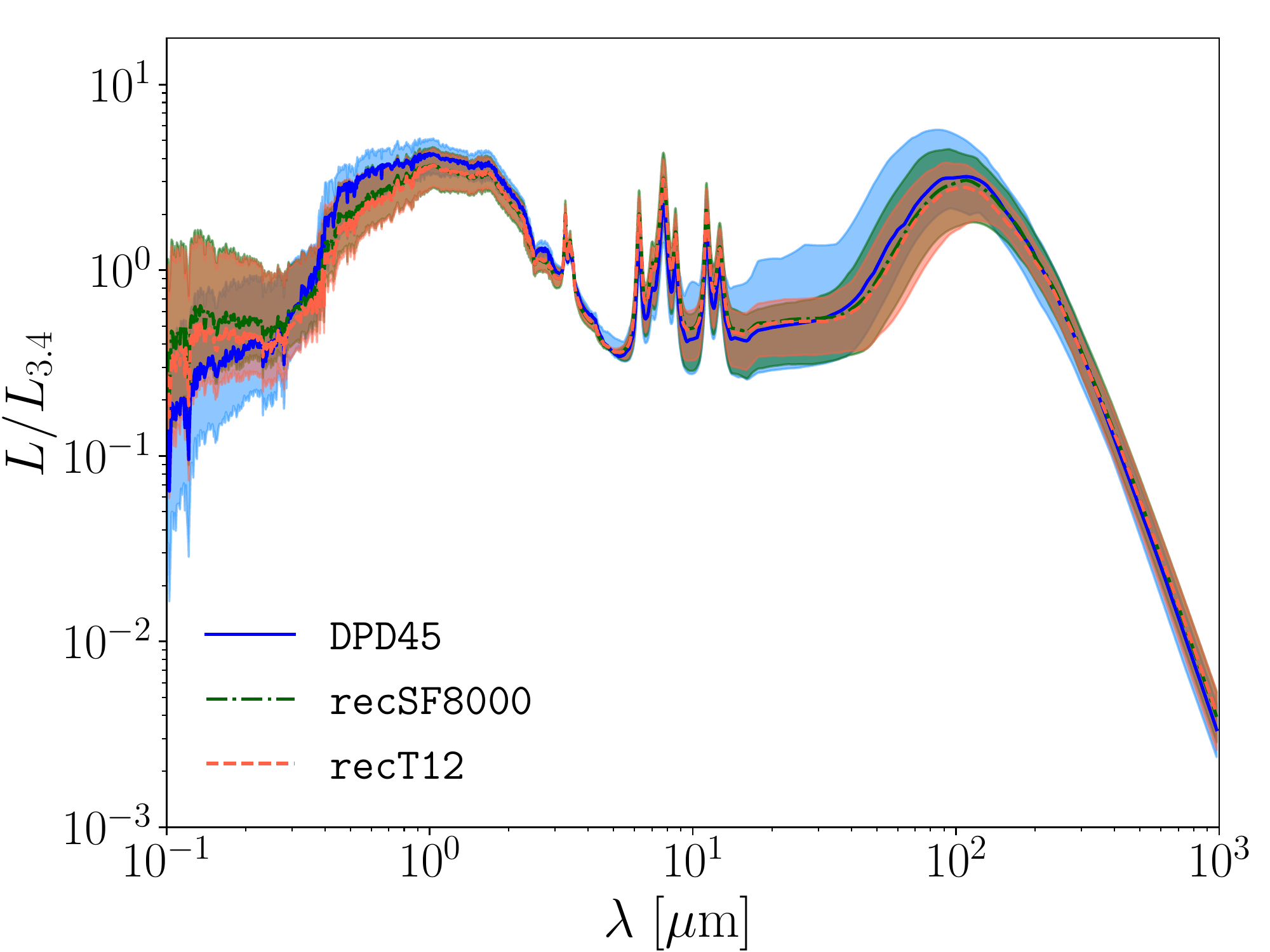}
\caption{Median Auriga SEDs obtained for each of the two dust allocation recipes using \texttt{CIGALE} fitting of the broadband data, along with those for \texttt{DPD45}. The shaded regions represent $16 \%-84 \%$ range for each of the datasets. The Auriga statistics have been obtained using both face-on and edge-on configurations for each galaxy.}
\label{SEDs_CIGALE.fig}
\end{figure}

\subsection{Global physical properties and dust scaling relations}
\label{results_globalProps_CIGALE.ssec}
In this section, we present the global physical parameters and dust scaling relations obtained using \texttt{CIGALE} fitting of the broadband fluxes for the two adopted dust recipes. As mentioned in Sect.~\ref{calibration.sec}, we use the same parameter settings as  \citet{2018A&A...620A.112B,2019A&A...624A..80N,2020MNRAS.494.2823T}, except that we use the \citet{2003PASP..115..763C} IMF.
We use these settings for both the Auriga and the DustPedia galaxies. The SED fitting is carried out at all inclinations for each of the Auriga galaxies (see Sect.~\ref{dataproducts.ssec}). We retain the best fitting values generated by \texttt{CIGALE} for all analyses in this work. A comparison between the Auriga input broadband fluxes and those derived by \texttt{CIGALE} is presented in Sect.~\ref{CIGALE.appendix}, showing a very good agreement between the two.
The average fitted SEDs are shown in Fig.\ref{SEDs_CIGALE.fig}.

\subsubsection{Comparison of intrinsic and inferred physical properties}
\label{results_CIGALE_inferred_vs_intrinisc.sssec}

\begin{figure}
\centering
\includegraphics[width=\columnwidth]{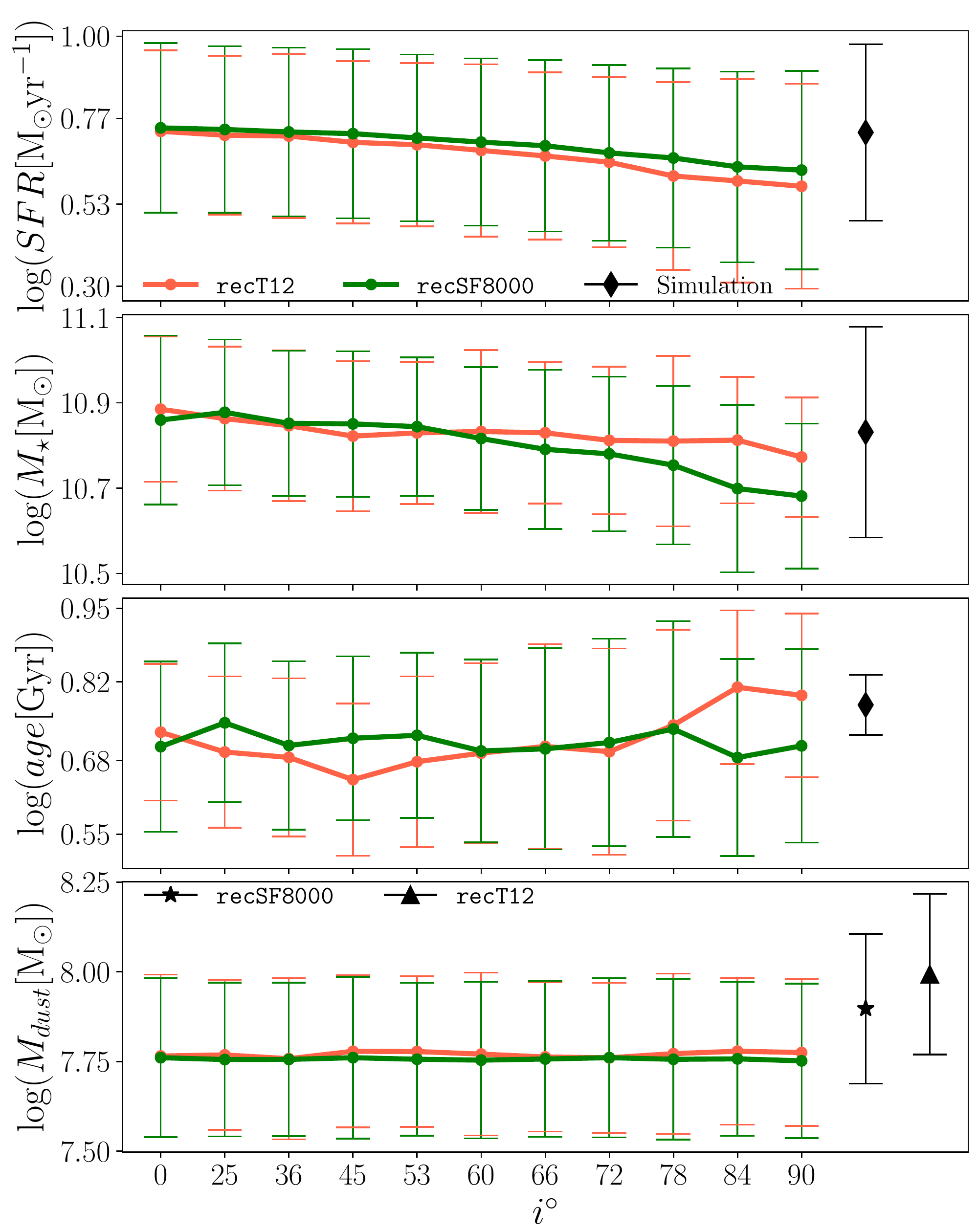}
\caption{Physical properties of the Auriga galaxies obtained using \texttt{CIGALE} broadband fitting as a function of inclination, for each of the two dust allocation recipes. The error bars represent the standard deviation in the values obtained for all 30 Auriga galaxies. The corresponding mass-weighted intrinsic values (calculated directly from the simulation data) are shown as error bars on the right-hand side of each panel. }
\label{inclinationEffects_CIGALE.fig}
\end{figure}

Fig.~\ref{inclinationEffects_CIGALE.fig} shows intrinsic values for a number of physical quantities derived directly from the simulation data and the corresponding values inferred using SED fitting as a function of inclination.
Both dust allocation recipes, on average, show a decline in inferred SFR from face-on to edge-on inclinations. The intrinsic SFR, i.e. the sum of the values assigned to the star-forming region particles during data extraction (see Sect.~\ref{DataExtraction_PrimarySources.sssec}), is in very good overall agreement with the derived values for the face-on configurations. When comparing the two dust allocation recipes, we find that the face-on SFR values are similar for both recipes, while the decline in average SFR as a function of inclination is sharper for \texttt{recT12} ($21\%$) than for \texttt{recSF8000} ($16\%$).
This decline is caused by the increased shrouding of star-forming regions by diffuse dust in the outer regions of the disk, which is present in larger amounts for \texttt{recT12}. This is also reflected by the radial extent of the galaxies in FIR, which is discussed in Sect.~\ref{results_Statmorph_comparison_with_DustPedia.sssec}. 
The inferred stellar mass also shows a decline with increasing inclination, now steeper for \texttt{recSF8000} ($36\%$) than for \texttt{recT12} ($23\%$). Similar to the SFR, the inferred face-on values show better agreement with the intrinsic values, calculated by summing the stellar particle masses during data extraction. 

The inferred luminosity-weighted stellar age is compared to the intrinsic mass-weighted age of the stellar particles in the third panel of Fig.~\ref{inclinationEffects_CIGALE.fig}. The inferred ages are slightly lower than the intrinsic values. The values for \texttt{recSF8000} are roughly constant with inclination. In contrast, \texttt{recT12} shows a decrease from $i=0^{\circ}$ to $i=45^{\circ}$, followed by an increase toward edge-on inclinations. 
The inferred dust mass does not vary significantly with inclination, which is expected because the emission at infrared wavelengths -- given the low dust opacity -- is essentially isotropic. The average inferred dust masses are very similar for both dust allocation recipes.

\subsubsection{Dust scaling relations}
\label{results_CIGALE_DustScalingRelations.sssec}

\begin{figure*}
\centering
\includegraphics[width=\textwidth]{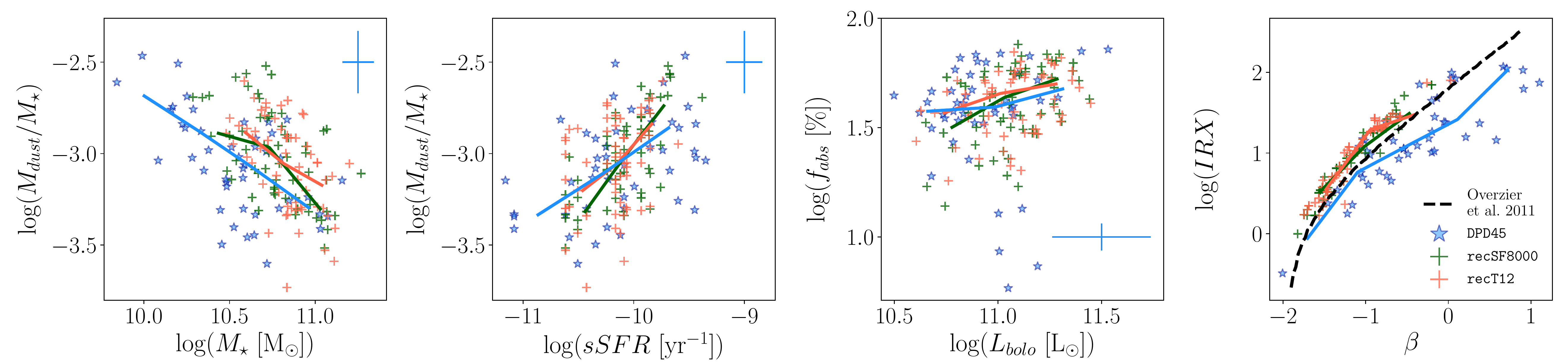}
\caption{Dust scaling relations obtained with \texttt{CIGALE} fits of the broadband data for the Auriga galaxies along with those for the DustPedia sub-sample \texttt{DPD45}. The rolling median for each sample is shown as a colored line, with the error bars corresponding to the medians of the errors for the DustPedia sample. For each Auriga galaxy, values for the face-on and edge-on configurations are shown here.}
\label{ScalingRelations_CIGALE.fig}
\end{figure*}

\begin{figure}
\centering
\includegraphics[width=.9\columnwidth]{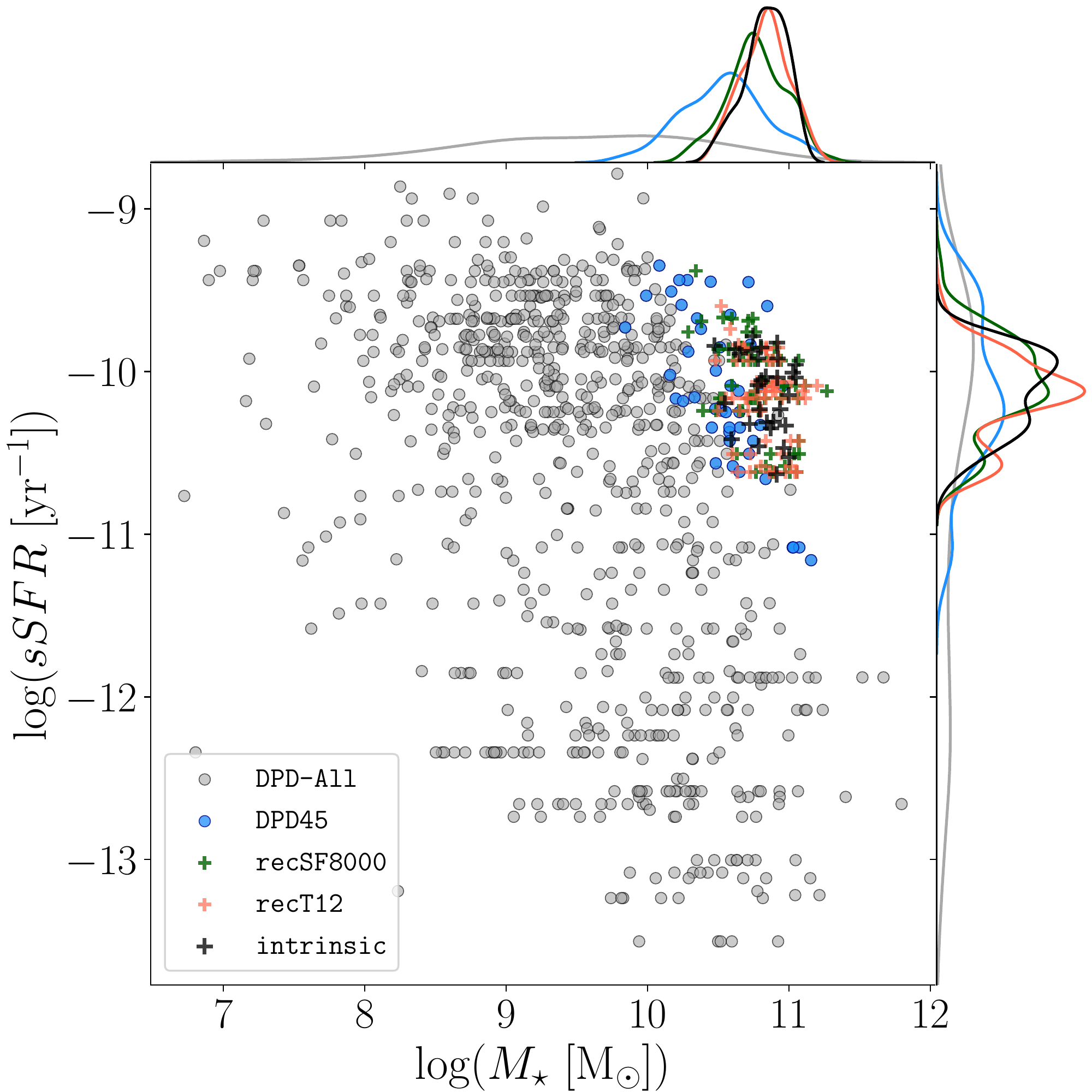}
\caption{Auriga galaxies occupy the top right corner of the $\mathrm{sSFR-M_{\star}}$ plane. The Auriga data is using face-on and edge-on configurations for each galaxy.  \texttt{DPD-All} represents the full DustPedia dataset (barring a few outliers). The data shown here has been obtained using \texttt{CIGALE} SED fitting for the observational (gray and blue) and the two post-processed datasets (green and orange), while the black markers represent the mass-weighted values taken directly from the simulation. Some effects of the discretization of the \texttt{CIGALE} model grid parameters are visible.}
\label{sSFR-MStar.fig}
\end{figure}

Fig.~\ref{ScalingRelations_CIGALE.fig} shows a number of dust scaling relations for the face-on and edge-on configurations for each of the Auriga galaxies along with those for the DustPedia calibration subset, \texttt{DPD45}.
The leftmost panel shows the relation between specific dust mass and stellar mass. The observed anti-correlation is reproduced for both dust allocation recipes. At the same time we find that on average, the Auriga galaxies are slightly more dusty in comparison to the DustPedia sub-sample considered. 
The disagreement between Auriga galaxies and \texttt{DPD45} can be understood by considering the specific star-formation rate-stellar mass ($\mathrm{sSFR-M_{\star}}$) relation shown in Fig.~\ref{sSFR-MStar.fig}. Auriga galaxies occupy the top right region in this plane, exhibiting a higher $\mathrm{sSFR}$ compared to DustPedia galaxies of similar stellar mass. 
The location of the Auriga galaxies in Fig.~\ref{sSFR-MStar.fig} is also reflected by the higher gas content of these galaxies compared to the observed galaxies of a similar stellar mass \citep[Fig.~12,][]{2017MNRAS.466.3859M}. 
Given that the specific star-formation rate and specific dust mass are tightly correlated \citep{2010MNRAS.403.1894D}, we deduce that the offset between \texttt{DPD45} and Auriga galaxies on the $\mathrm{M_{dust}/M_{\star}-M_{\star}}$ plane is partially a manifestation of the special position of the Auriga galaxies in Fig.~\ref{sSFR-MStar.fig}.
We note that the simulation star-formation rate and the stellar mass are well reproduced by the post-processing pipeline, as indicated by the comparison between the intrinsic values from the simulation and those obtained using our post-processing pipeline, shown in Fig.~\ref{sSFR-MStar.fig}. Therefore, it appears that the offset between \texttt{DPD45} and Auriga galaxies shown in Fig.~\ref{sSFR-MStar.fig} is intrinsic to the simulation and not introduced by our post-processing procedures.

The correlation between specific dust mass and specific star formation rate (second panel in the same figure) is very well reproduced by the Auriga galaxies. The \texttt{recT12} dust allocation recipe produces more dusty galaxies at the lower end of the specific star formation rate range, which is expected given that we allocate dust in some non-star-forming gas cells in that case.  

The third panel in Fig.~\ref{ScalingRelations_CIGALE.fig} shows the fraction of energy absorbed by dust, i.e. the ratio between the dust luminosity and the bolometric luminosity, $f_{\mathrm{abs}}=L_{\mathrm{dust}}/L_{\mathrm{bolo}}$. This ratio reflects the optical thickness of the galaxy, which depends on the amount, composition, and geometry of its dust contents. \cite{2018A&A...620A.112B} find a broad positive correlation between $f_{\mathrm{abs}}$ and the bolometric luminosity for late-type, disk-dominated, high $\mathrm{sSFR}$, gas-rich galaxies in the DustPedia sample. We find the same trend for the Auriga galaxies and our \texttt{DPD45} sample, with median values of $43.0\%$,  $41.7\%$ and $40.8\%$ for \texttt{recSF800}, \texttt{recT12} and \texttt{DPD45}, respectively.

\subsubsection{The IRX-beta relation and attenuation curves}
\label{results_IRXbeta_attenuationCurves.sssec}

We consider the infrared excess-UV slope, $\mathrm{IRX}-\beta$, relation \citep{1999ApJ...521...64M, 2011ApJ...726L...7O} in the rightmost panel of Fig.~\ref{ScalingRelations_CIGALE.fig}. Here $\mathrm{IRX}$ is the infrared excess defined as: \begin{equation}
\mathrm{IRX}=\log \frac{L_{\mathrm{dust}}}{L_{\mathrm{FUV}}}
\label{IRXdefinition.eqn}
\end{equation}
and $\beta$ is the UV slope given as:
\begin{equation}
   \beta=\frac{\log (f_{\mathrm{NUV}} / f_{\mathrm{FUV}})}{\log (\lambda_{\mathrm{NUV}} / \lambda_{\mathrm{FUV}})}-2
   \label{beta.eqn}
\end{equation}

We find that while the $\mathrm{IRX}$ values are in line with those of the DustPedia sub-sample, the UV slopes for the Auriga galaxies are slightly shallower in comparison to \texttt{DPD45}. The Auriga sample is on average $\approx60\%$ more massive than the \texttt{DPD45} sample, which could explain some differences in the $\mathrm{IRX}-\beta$ relation between the two samples, owing to changes in the dust attenuation curves as a function of stellar mass \citep{2019ApJ...872...23S} or due to the $\mathrm{M_{\star}-A_{V}}$ correlation \citep{2018ApJ...859...11S}. 

\begin{figure}
\centering
\includegraphics[width=\columnwidth]{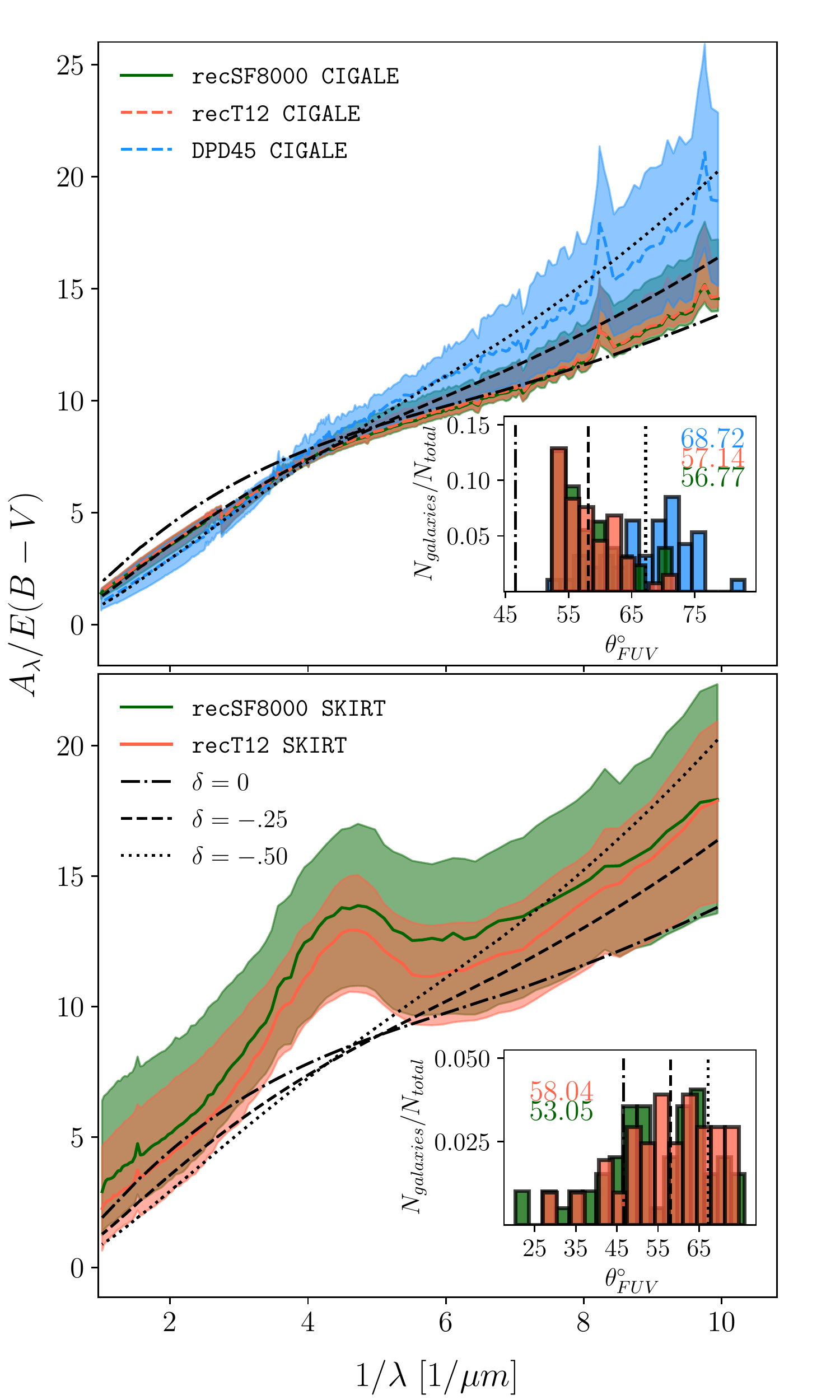}
\caption{Dust attenuation curves obtained for \texttt{CIGALE} fits (top) and directly obtained from the radiative transfer simulations (bottom). The lines represent the median values, while the corresponding shaded regions represent the $16 \%-84 \%$ range for each of the datasets. 
We also show the three power-law modified Calzetti attenuation curves ($\delta=0,-0.25, -0.5$) without a UV bump used for \text{CIGALE} fits \citep[for more discussion, refer to][]{2019A&A...624A..80N, 2019A&A...622A.103B}. The insets show the histograms of the UV slope of the attenuation curves between $\lambda=0.15~\mu m$ and $\lambda=0.1~ \mu m$. The  Auriga statistics have been obtained by using both the face-on and edge-on configurations for each galaxy.}
\label{attenuationCurves.fig}
\end{figure}

Comparing the two dust allocation recipes for the Auriga galaxies, we obtain very similar results for both \texttt{recSF8000} and \texttt{recT12}, with a low amount of scatter.
As described in \citet{2019ApJ...872...23S}, the diversity in galaxy attenuation curves drives the scatter in the $\mathrm{IRX}-\beta$ plane. These authors also show that the slope of the attenuation curve is tightly correlated with the position on the $\mathrm{IRX}-\beta$ plane, with attenuation curves with flatter slopes having a higher tilt in comparison to that found in \citet{2011ApJ...726L...7O}. 
We show the attenuation curves obtained for the Auriga galaxies in Fig.~\ref{attenuationCurves.fig}, along with modified Calzetti curves as defined by \citet{2019A&A...622A.103B}. It is clear that the Auriga galaxies, on average, have flatter attenuation curves in comparison with those of \texttt{DPD45}. Both dust allocation recipes show little diversity in the attenuation curves, which is reflected in the lack of scatter in the $\mathrm{IRX}-\beta$ plane. We stress that the UV bump is not modeled in the \texttt{CIGALE} SED fitting for the Auriga galaxies. This is consistent with the settings used in \citet{2019A&A...622A.103B}, who chose an attenuation law without a UV bump because the UV emission in their galaxy sample is covered by just the two GALEX bands, making it difficult to constrain the UV bump.
\citet{2018ApJ...869...70N} have shown the impact of dust geometry on galaxy attenuation curves. Because the two dust allocation recipes yield a different dust geometry, some variations in the dust attenuation curves are expected. Interestingly, the attenuation curves obtained with \texttt{CIGALE} fits show minimal differences between \texttt{recSF8000} and \texttt{recT12}.

Fig.~\ref{attenuationCurves.fig} also shows the attenuation curves calculated using high spectral resolution SEDs obtained directly from \texttt{SKIRT}. Here we see a more diverse range of attenuation curves, as well as a difference in the attenuation curve slope between the two dust allocation recipes. We find that Auriga galaxies post-processed with \texttt{recT12} exhibit slightly steeper attenuation curves, on average. 
It would be interesting to carry out a detailed study of the attenuation curves, both at global and local scales, and their variation with inclination and with physical properties.

%%%%%%%%%%%%%%%%%%%%%%%%%%%%%%%%%%%%%%%%%%%%%%%%%%%%%%%%%%%%%%%%%%%%%%%%%%%%%%%%%%%
%%%%%%%%%%%%%%%%%%%%%%%%%% StatMorph Analysis Results %%%%%%%%%%%%%%%%%%%%%%%%%%%%%
%%%%%%%%%%%%%%%%%%%%%%%%%%%%%%%%%%%%%%%%%%%%%%%%%%%%%%%%%%%%%%%%%%%%%%%%%%%%%%%%%%%

\defcitealias{Baes2020}{B20}

\begin{figure*}
\centering
\includegraphics[width=.9\textwidth]{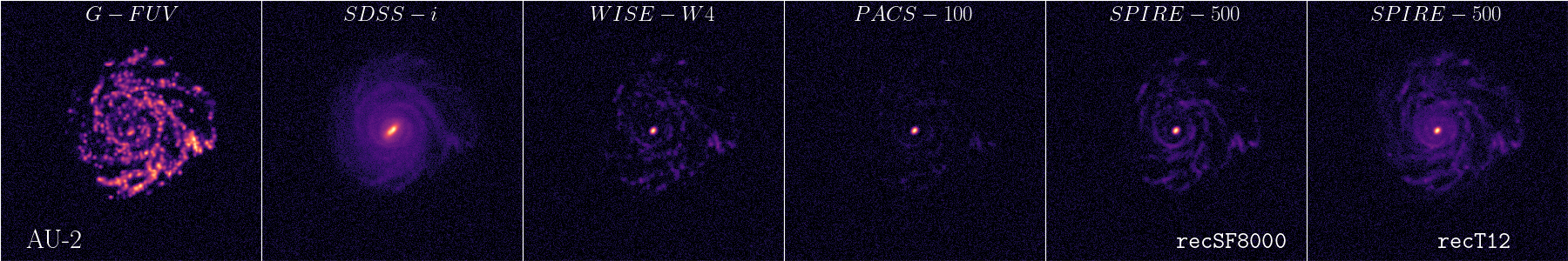}
\includegraphics[width=.9\textwidth]{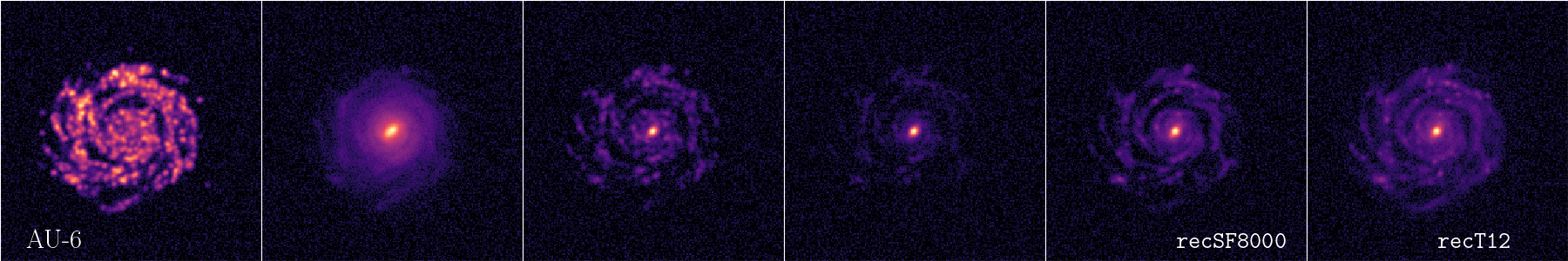}
\includegraphics[width=.9\textwidth]{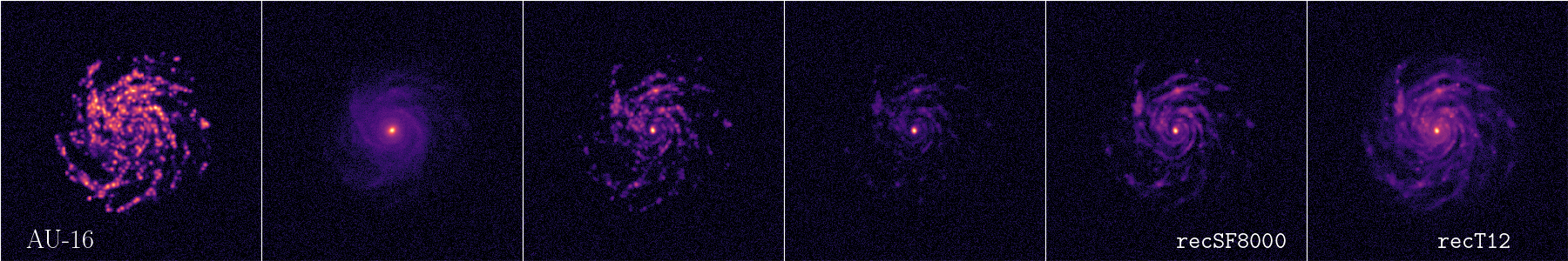}
\caption{Selected Auriga broadband images as they are used for determining the non-parametric morphology indicators with \texttt{StatMorph}. All images are convolved to the same resolution and are shown on a square root scale running from the image's minimum to maximum. The first four panels in each row are from \texttt{recSF8000}, only the SPIRE-500 images are shown for \texttt{recT12} as well. We note the prominence of the central component in the WISE-W4, PACS and the SPIRE bands, in contrast with the DustPedia galaxies shown in Figs.~1 and 2 of \citetalias{Baes2020}.} 
\label{StatMorph_images_simulations.fig}
\end{figure*}

\begin{figure*}
\centering
\includegraphics[width=\textwidth]{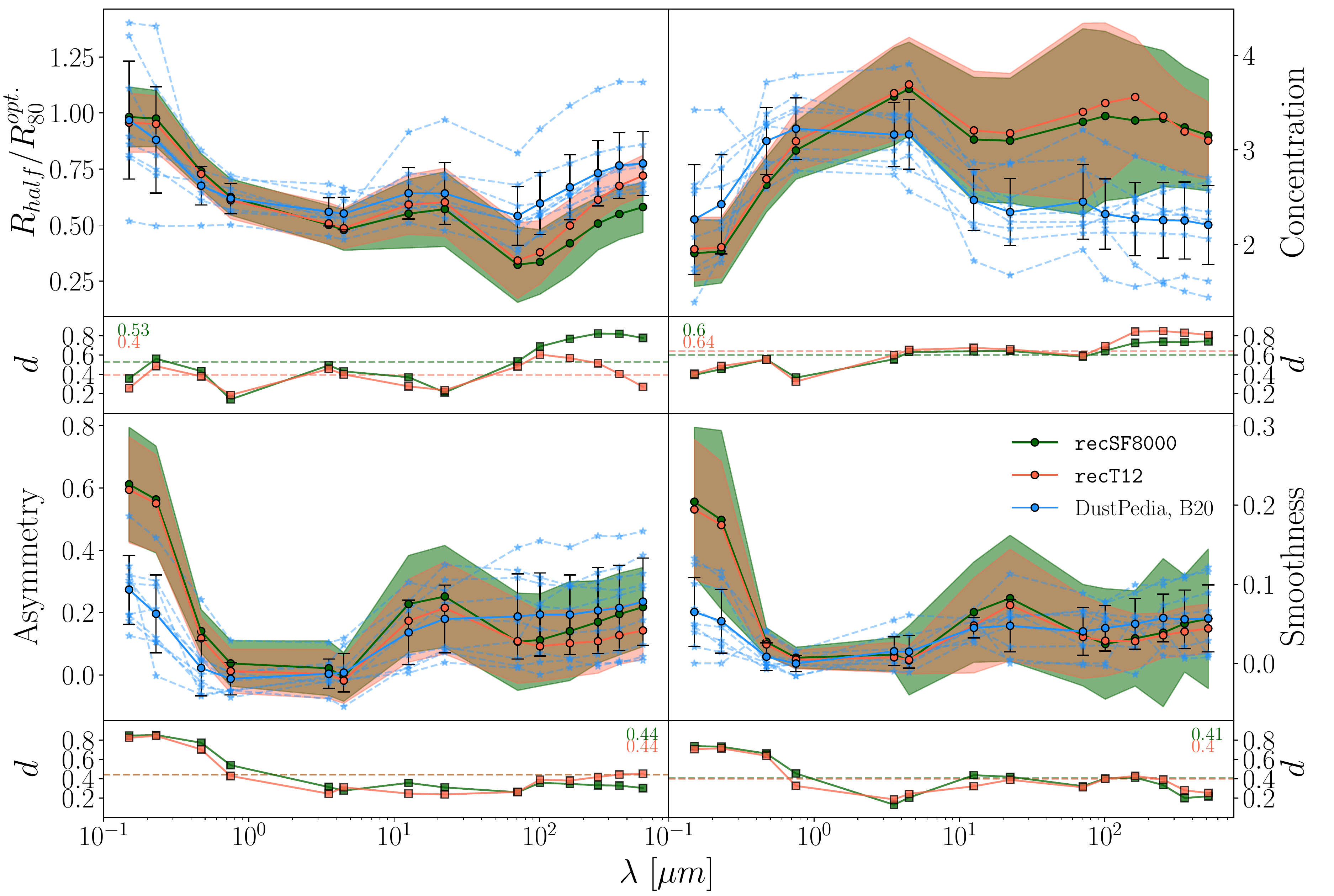}
\caption{Various non-parametric morphological parameters as a function of wavelength. The circular markers represent the mean values for the Auriga and DustPedia/\citetalias{Baes2020} samples. The shaded areas and the error bars represent the $\pm~1\sigma$ range for the Auriga data and the DustPedia/\citetalias{Baes2020} galaxies, respectively. The dashed blue lines represent individual DustPedia galaxies in \citetalias{Baes2020}. The smaller sub-panels under each panel show the K-S test distance $d$ as a function of wavelength, quantifying the similarity between synthetic and observed datasets. A smaller value of $d$ indicates better similarity.} 
\label{StatMorph_CASGMvsWavelength.fig}
\end{figure*}

\subsection{Multi-wavelength morphological analysis}
\label{statMorphResults.ssec}
In this section we compare the morphology of the Auriga galaxies to observations over the UV-submm range. This serves to evaluate the fidelity of the simulated galaxies in general and provides a way to discriminate between the two dust allocation recipes, given that these recipes will result in  different dust emission and attenuation values as a function of position in the galaxy. 
\subsubsection{Calculation of non-parametric morphological indices}
\label{StatMorph_setup.sssec}
The calibration process described in Sect.~\ref{calibration.sec} is solely based on comparing spatially integrated broadband fluxes with observations. As a means of comparing the resolved Auriga images with observations, we use the \texttt{StatMorph} package \citep{2019MNRAS.483.4140R}, a user-friendly Python implementation for the calculation of the most commonly used morphological statistics \citep{2004AJ....128..163L,  10.1111/j.1365-2966.2011.18938.x,  2011MNRAS.418..801H,2012MNRAS.423..197B}.

We obtain multi-wavelength sets of the concentration-asymmetry-smoothness (CAS) indices and of the elliptical half-light radius normalized by the optical $R_{80}$. We briefly discuss these four morphological indicators here.
\begin{itemize}[leftmargin=*]
    \item \textit{Concentration (C)}: The concentration parameter is a measure of how concentrated the central bulge component is with respect to the total flux of the galaxy. It is defined as $5 \times \log(R_{80}/R_{20})$, where $R_{20}$ and $R_{80}$ are the radii of circular apertures containing $20\%$ and $80\%$ of the galaxy’s light, respectively. Concentration has a strong correlation with color; early type galaxies tend to be redder and more centrally concentrated, whereas later type galaxies tend to be bluer and less centrally concentrated. It is also related to other intrinsic features such as velocity dispersion, Mg/Fe abundance ratio, and super-massive black hole mass.
    \item \textit{Asymmetry (A)}: The asymmetry index is obtained by subtracting the galaxy image rotated by $180^\circ$ from the original image. The asymmetry parameter could reveal merger events and interactions. Also, in normal star-forming galaxies, star formation in structures such as spiral arms lead to asymmetries. Therefore, the asymmetry parameter at optical wavelengths correlates with optical broadband color.
    \item \textit{Smoothness (S)}: The smoothness index is a  measure of high spatial frequency clumpiness and is computed by subtracting a lower resolution version of the galaxy image from the original galaxy image. A low value of the parameter indicates lower clumpiness or higher smoothness in the light distribution. Smoothness indices based on optical images tend to correlate weakly with optical color and SFR.
    \item \textit{Half light radius ($R_\mathrm{half}$)}: The half-light radius is calculated as the elliptical radius of the isophote that contains half of the light in the galaxy image. We normalize it with $R_{80}^\mathrm{opt}$, the radius of a circular aperture containing 80\% of the galaxy’s light in the Sloan Digital Sky Survey (SDSS) g band image.
\end{itemize}
For a consistent comparison with the observations, we use selected broadband images of a sub-sample of the Auriga galaxies.
We choose all galaxies with a disk-to-total mass ratio $D/T~>0.45$ \citep[see][Fig.7]{2017MNRAS.467..179G}, resulting in a sample of fourteen galaxies. 
The images are processed in a manner consistent with our observational sub-sample described in \cite{Baes2020} (B20 from hereon).

We cover the UV-submm wavelength range by utilizing the images in the broadbands listed here:
\begin{itemize}[leftmargin=*]
    \item UV: Galaxy Evolution Explorer (GALEX) far-UV (FUV) and near-UV (NUV) bands.
    \item Optical: Sloan Digital Sky Survey (SDSS) g band, 
    \item Near infrared (NIR): SDSS i band, Spitzer IRAC $3.6$ and $4.5~\mu \mathrm{m}$ bands.
    \item  Mid infrared (MIR): Wide-field Infrared Survey Explorer (WISE) $12$ and $22~\mu \mathrm{m}$ bands.
    \item Far infrared (FIR): three Herschel PACS bands and three Herschel SPIRE bands.
\end{itemize}

We process the images by re-gridding them on a 12" pixel scale, convolving to the SPIRE-500$\mu\mathrm{m}$ point spread function (PSF) using the convolution kernels of \citet{2011PASP..123.1218A} and adding a homogeneous Gaussian background noise, assuming a signal-to-noise ratio of 100. We do so for both the dust allocation recipes discussed in Sect.~\ref{DustAllocation.sssec}.

Finally, we generate a  $5 \times 5$ grid in inclinations and distances. The distance grid is uniform, with values lying between the extremes of those in \citetalias{Baes2020}.
The inclination grid is composed of five of the images available for each galaxy (see Sect.~\ref{dataproducts.ssec}), corresponding to inclinations of ($0^{\circ}, 25.84^{\circ}, 45.57^{\circ}, 60^{\circ}, 72.54^{\circ}$).
In all cases, the physical resolution is better than $0.78 ~\text{kpc/pixel}$, below the threshold value of $\sim 1~\text{kpc/pixel}$ where the accuracy of the morphological parameters begins to decline \citep{2000ApJ...529..886C}.
For each simulated galaxy, segmentation maps are based on the Galex-FUV, SDSS-$i$, WISE-$22~\mu \mathrm{m}$, PACS-$100~\mu \mathrm{m}$ and SPIRE-$350~\mu \mathrm{m}$ images.
In order to create the segmentation maps, we set the detection threshold at $1.2\sigma$ above the sky median, with the background level computed by using simple sigma-clipped statistics.
Finally, for all morphological measurements, we ensure that the \texttt{StatMorph}'s \texttt{flag} parameter, indicating a problem with the basic measurements, is zero.

\subsubsection{Comparison to DustPedia}
\label{results_Statmorph_comparison_with_DustPedia.sssec}

Fig.~\ref{StatMorph_images_simulations.fig} shows selected broadband images of some of the Auriga galaxies used for the morphological analysis.
We compare the Auriga morphological parameters to those determined for a set of nine well-resolved spiral galaxies from the DustPedia database as described in \citetalias{Baes2020}.
Based on the \texttt{StatMorph} setup described in Sect.~\ref{StatMorph_setup.sssec}, we show four non-parametric morphology indicators as a function of wavelength in Fig.~\ref{StatMorph_CASGMvsWavelength.fig}. We focus on the global trends observed for the Auriga sub-sample, represented by the circular markers in Fig.~\ref{StatMorph_CASGMvsWavelength.fig}.

The normalized half-light radius $R_\mathrm{half}/R_{80}^\mathrm{opt}$ has a characteristic behavior as a function of wavelength, with large values in the FUV and a gradual decrease over the optical regime to the NIR, followed by a subsequent increase in the MIR. This behavior is in excellent agreement with the observed sample.
As we move to the FIR from MIR, we find a sharp decrease in the size of the Auriga galaxies in the PACS-70 broadband, followed by an increase in the size with wavelength. The Auriga galaxies post-processed by \texttt{recT12}, on average, exhibit a larger radial extent in all FIR bands, which is in better agreement with the observed sample than the radii for the galaxies post-processed using \texttt{recSF8000}. 

The concentration index (top right panel of Fig.~\ref{StatMorph_CASGMvsWavelength.fig}), generally increases from UV to NIR wavelengths for the Auriga galaxies. This trend is also present for the observed galaxies, although the simulated galaxies are on average less centrally concentrated when observed in the UV, SDSS-g and i broadbands, while being more concentrated in the IRAC bands, representative of the old stellar population.
The difference in the concentration index between the Auriga and the observed galaxies becomes particularly pronounced as we move to the dust emission broadbands. The MIR emission is less concentrated than the NIR emission, similar to the observed galaxies. But at the same time, the central emission dominates much more strongly over the emission from the star-forming regions in the outskirts, unlike the observed galaxies.  The observed galaxies, on average, exhibit similar concentration values in the FUV and the MIR, the wavelength regimes dominated by emission from the star-forming regions. This, however, is not the case for the simulated galaxies.
The higher concentration index for the Auriga sample is also seen in the FIR broadbands, which does not show a significant change for the different dust allocation recipes.
It is clear that, given our fixed dust-to-metal ratio for assigning dust, the central regions of the Auriga galaxies are highly metal rich. 
We speculate that the high central metal content could partially be a result of stellar winds and AGN feedback not being able to drive out the central gas. The AGN feedback model used in the Auriga project, which consists of thermal feedback alone, could be responsible for high central gas-phase metal content. It would be useful, for example, to compare with the concentration indices of Illustris-TNG \citep{2019MNRAS.490.3196P} disky galaxies of comparable mass and SFR. The Illustris-TNG AGN feedback model \citep{2018MNRAS.473.4077P, 2018MNRAS.479.4056W} employs a kinetic feedback at low black hole accretion rates, which leads to a different central gas structure and composition \citep{2020MNRAS.493.1888T,2021arXiv210210752I}. The Illustris-TNG stellar wind feedback model \citep{2018MNRAS.473.4077P} is also significantly different in comparison to that used in Auriga. Therefore, a thorough study involving physics model variations is required to reveal the underlying causes of the high central metal content in Auriga galaxies.
Additionally, our dust allocation recipes do not model dust destruction due to various feedback mechanisms in the central regions of the simulated galaxies.

The asymmetry index is shown in the bottom left panel of Fig.~\ref{StatMorph_CASGMvsWavelength.fig}. The Auriga galaxies exhibit higher asymmetry in the UV wavelength bands in comparison to the observed set of galaxies. This could be caused by the rather patchy and discrete distribution of the star forming regions, as can be seen in Fig.~\ref{StatMorph_images_simulations.fig}. The asymmetry values in the UV are likely to go down if a star formation region re-sampling strategy would be used, such as the one discussed in \citet{2016MNRAS.462.1057C}.
The asymmetry in the NIR bands is lower than in the UV and optical bands and shows very good agreement with the observed galaxies.
The asymmetry rises as we move into MIR from NIR wavelengths; this trend and the range of values obtained are quite consistent with the observed data. 
Moving from MIR to FIR wavelengths, there is a dip in asymmetry, not seen in the observed data, consistent with a symmetric central component. We find a gradual rise in asymmetry with wavelength in the FIR bands, with the three PACS bands showing higher discrepancy than the SPIRE bands when compared with the observed galaxies. This is also reflective of the higher central concentration in the three PACS bands.
There is also an impact of the dust allocation on the asymmetry parameter, with the Auriga galaxies using \texttt{recT12} being more symmetric in comparison to those using \texttt{recSF8000}.

The bottom right panel of Fig.~\ref{StatMorph_CASGMvsWavelength.fig} shows the smoothness parameter, which traces the presence of small-scale clumpy structures outside of the central region. 
As expected the UV broadband images show the most clumpiness, in general, higher than the values found for the galaxies in \citetalias{Baes2020}. 
The smoothness values in most other broadbands are consistent with observations, with slightly underestimated clumpiness in the three PACS bands.
In the FIR, the consistency of the smoothness values with the observations is slightly unexpected given the overly smoothed nature of the pressure supported dense gas. The consistency of smoothness index of the Auriga galaxies with the observed galaxies is attributed to the dust in the star forming regions, which can be seen as brighter-than-average spots in Fig.~\ref{StatMorph_images_simulations.fig}.
The average clumpiness for the two dust allocation recipes is largely equivalent. In the FIR, this is contrary to expectation that galaxies with \texttt{recSF8000} would show higher clumpiness (see Fig.~\ref{recipeComaprison_dustSigma.fig}).

We note that the asymmetry and smoothness values are susceptible to the amount of Gaussian noise added to the image. In Sect.~\ref{StatMorph.appendix}, we normalize the noise to levels outside the central region and find a higher difference between the two dust recipes, \texttt{recSF8000} being more asymmetric in almost all broadbands, and more clumpy in the FIR bands.

The K-S test distance values averaged over all four relations and over specific wavelength regimes are given in Table~\ref{StatMorph_KStest.table}.
Based on these results, we conclude that galaxies using \texttt{recT12} show slightly better overall agreement with the observational data.

\begin{table}
\caption{Average K-S test values for \texttt{recSF8000} and \texttt{recT12} as a function of the wavelength regimes. The values are averaged over the four non-parametric indices considered in this work.}
\label{StatMorph_KStest.table}
\centering
\begin{tabular}{lccc}
\hline 
Dust allocation recipe & UV & Optical and NIR & MIR and FIR \\
\hline 
\texttt{recSF8000} & 0.62 & 0.44 & 0.49 \\
\texttt{recT12} & 0.59 & 0.42  & 0.47 \\
\hline
\end{tabular}
\end{table}

\section{Conclusions and outlook}
\label{discussion.sec}
We calculated synthetic observations in the wavelength range from UV to submm for simulated galaxies extracted from the Auriga suite of Milky Way-like galaxies at redshift zero using the radiative transfer code \texttt{SKIRT}. These include global fluxes and resolved broadband images at multiple observer positions and using two different dust allocation recipes. 
We compared our post-processing results to observations, both at a global scale, by means of the SED fitting code \texttt{CIGALE}, as well as at resolved scales through non-parametric morphological indicators. 
Based on these two comparisons, we conclude the following:
\begin{itemize}[leftmargin=*]
\item Auriga galaxies show an overall good agreement in terms of dust properties with the DustPedia sample used for comparison, although they have a slightly higher specific dust mass. 
\item The attenuation curves of the Auriga galaxies exhibit relatively shallow UV slopes in comparison to the observational sample.
\item Global properties such as galaxy stellar mass, SFR and mean stellar mass  derived using SED fitting for the two dust allocation recipes are quite similar but show a slightly different variation with inclination.
\item The optical and NIR morphologies of the Auriga galaxies are in very good agreement with the observational sample used, with images post-processed with \texttt{recT12} showing slightly better agreement.
\item The morphological indicators at MIR and FIR  show a lower level of agreement with the observational dataset, in particular, a higher average concentration and a smaller average size. Images post-processed with \texttt{recT12} show better agreement with observations when considering the FIR galaxy size, while \texttt{recSF8000} exhibits asymmetry values which are more in line with observations in the same wavelength regime. Both recipes show similar concentration and smoothness values.
\item We speculate that the high concentration observed at MIR and FIR wavelengths for the Auriga galaxies may be caused by the way AGN feedback is implemented. A comparison of Milky Way-like galaxies obtained using different AGN feedback implementations would serve as an interesting test.
\end{itemize}

We publish the synthetic data for any interested third party to study the dust-related properties of simulated Milky Way-like galaxies at redshift zero.
We list some of the possible applications here:

\begin{itemize}
\item {\em{Spatially resolved dust scaling relations}}: Using integrated UV--submm fluxes of observed galaxies, several studies have revealed strong correlations between the global properties of the dust and stellar components in galaxies \citep[e.g.,][]{2012A&A...540A..52C, 2020A&A...633A.100C}. These dust scaling relations provide interesting tests for  cosmological hydrodynamics simulations \citep{2020MNRAS.494.2823T}. \citet{2014A&A...567A..71V} presented dust scaling relations in M31 on sub-kpc scales, showing interesting relation between, for example, dust-to-stellar mass ratio and the NUV--{\em{r}} color, or dust-to-stellar mass ratio and stellar mass surface density. The different regions and pixels in Andromeda seem to obey the same scaling relations as global galaxy scaling relations in the HRS survey. A high-resolution panchromatic imaging data set such as the one presented here allows investigating to which physical scale these spatially resolved dust scaling relations hold.
\newline
\item {\em{Spatially resolved SED fitting}}: Panchromatic SED fitting is a powerful method to interpret the global energy output of galaxies. Modern panchromatic SED fitting codes \citep[e.g.,][]{2008MNRAS.388.1595D, 2016MNRAS.462.1415C, 2017ApJ...837..170L, 2019A&A...622A.103B} are based on an energy balance assumption, i.e., the energy absorbed by dust at short wavelengths is balanced by the energy emitted thermally in the infrared regime. Panchromatic SED fitting is also being applied to local scales \citep{2014A&A...567A..71V, 2018MNRAS.479..297W}. The reliability of SED fitting on spatially resolved scales has been investigated by \citet{2018MNRAS.476.1705S}. The data set presented here allows to extend this study and investigate to which spatial scale the dust energy balance is preserved. 
\newline
\item {\em{Dust heating and inverse radiative transfer}}: The various stellar populations in galaxies contribute to the heating of the dust grains in a different way \citep{2012MNRAS.419.1833B, 2015MNRAS.448..135B, 2019A&A...624A..80N}. Our team has developed an advanced radiative transfer modeling technique to investigate the importance of young and old stellar populations to dust heating, and has applied it to several nearby galaxies \citep{2014A&A...571A..69D, 2017A&A...599A..64V, 2020arXiv200501720V, 2019MNRAS.487.2753W, 2020A&A...637A..24V, 2020A&A...637A..25N}. In order to thoroughly test the underlying assumptions and assess the reliability and limitations of this approach, it would be interesting to apply the same technique to a synthetic data set. 
\newline
\item {\em{Advanced dust mass maps}}: The \texttt{PPMAP} code \citep{2017MNRAS.471.2730M} has been developed to infer the dust distribution in star-forming regions \citep{2019MNRAS.483...70C, 2019MNRAS.483..352M, 2019MNRAS.489..962H} and galaxies \citep{2019MNRAS.489.5436W} from a set of FIR/submm images at different resolutions. A synthetic data set of highly resolved images allows to test the reliability of this method for application on galaxy-wide scale, and to evaluate the sensitivity of the results on the availability of imaging data in different bands.

\end{itemize}

\section*{Data Availability}
\label{dataproductsAvailability.sec}
The broadband images as well as spatially integrated SEDs for \texttt{recT12} for all 18 observers (see Sect.~\ref{syntheticData.sec}) is publicly available at \url{www.auriga.ugent.be}. 
The data for \texttt{recSF8000} is available upon request.

\section*{Acknowledgments}
This project received funding from the European Research Council (ERC) under the European Union's Horizon 2020 research and innovation programme, grant agreement No. 683184 (Consolidator Grant LEGA-C).

AUK, MB, DBA and AT acknowledge the financial support of the Flemish Fund for Scientific Research (FWO-Vlaanderen), research projects G039216N and G030319N.

IDL acknowledges support from ERC starting grant 851622 DustOrigin.

LC is the recipient of an Australian Research Council Future Fellowship (FT180100066) funded by the Australian Government.

The radiative transfer simulations carried out for this work used the Tier-2 facilities of the Flemish Supercomputer Center (\url{https://www.vscentrum.be/}) located at the Ghent University.

%%%%%%%%%%%%%%%%%%%% REFERENCES %%%%%%%%%%%%%%%%%%

% The best way to enter references is to use BibTeX:
\bibliographystyle{mnras}
\bibliography{main} % if your bibtex file is called example.bib

%%%%%%%%%%%%%%%%%%%%%%%%%%%%%%%%%%%%%%%%%%%%%%%%%%

%%%%%%%%%%%%%%%%% APPENDICES %%%%%%%%%%%%%%%%%%%%%

\appendix
\section{Convergence tests}

\begin{figure}
\centering
\begin{tabular}{@{}c@{}}
\includegraphics[width=.495\columnwidth]{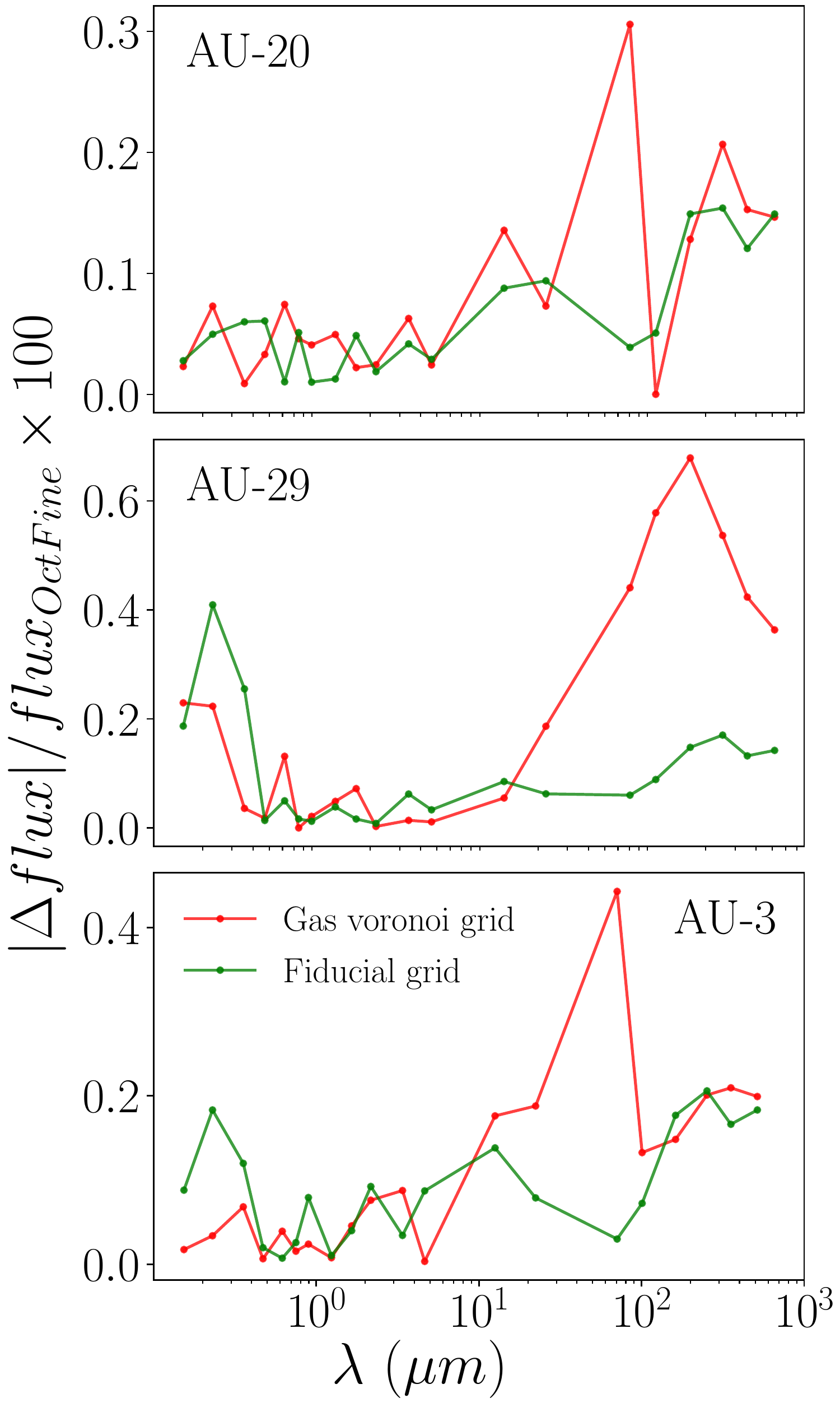} \\
\text{}
\end{tabular}
\begin{tabular}{@{}c@{}}
\includegraphics[width=.495\columnwidth]{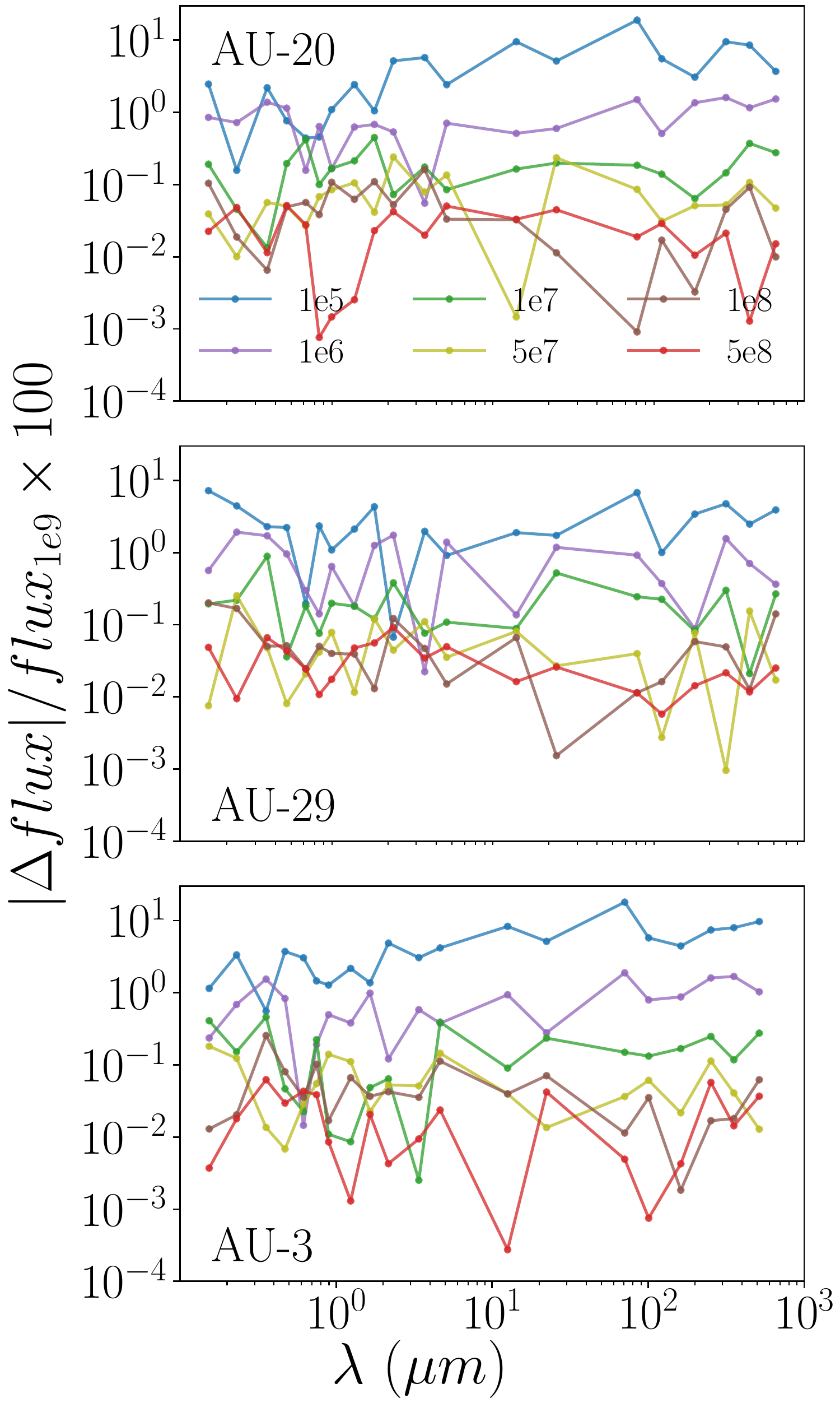} \\
\text{}
\end{tabular}
\caption{Left column: Comparison for simulations with  different dust discretization grids as described in Sect~\ref{DustGrid.appendix}. The flux error percentage is defined in comparison to a much finer grid with about ten times more cells.
Right column: Comparison between the broadband fluxes produced by \texttt{SKIRT} simulations for three Auriga galaxies with a varying number of photon packets as described in Sect.~\ref{SEDphotonsConvergence.appendix}. The flux error percentage is determined in comparison to a simulation with $10^{9}$ packets. The legend shows the number of photon packets used.} 
\label{Photon_and_grid_convergence.fig}
\end{figure}

\begin{figure}
\centering
\includegraphics[width=\columnwidth]{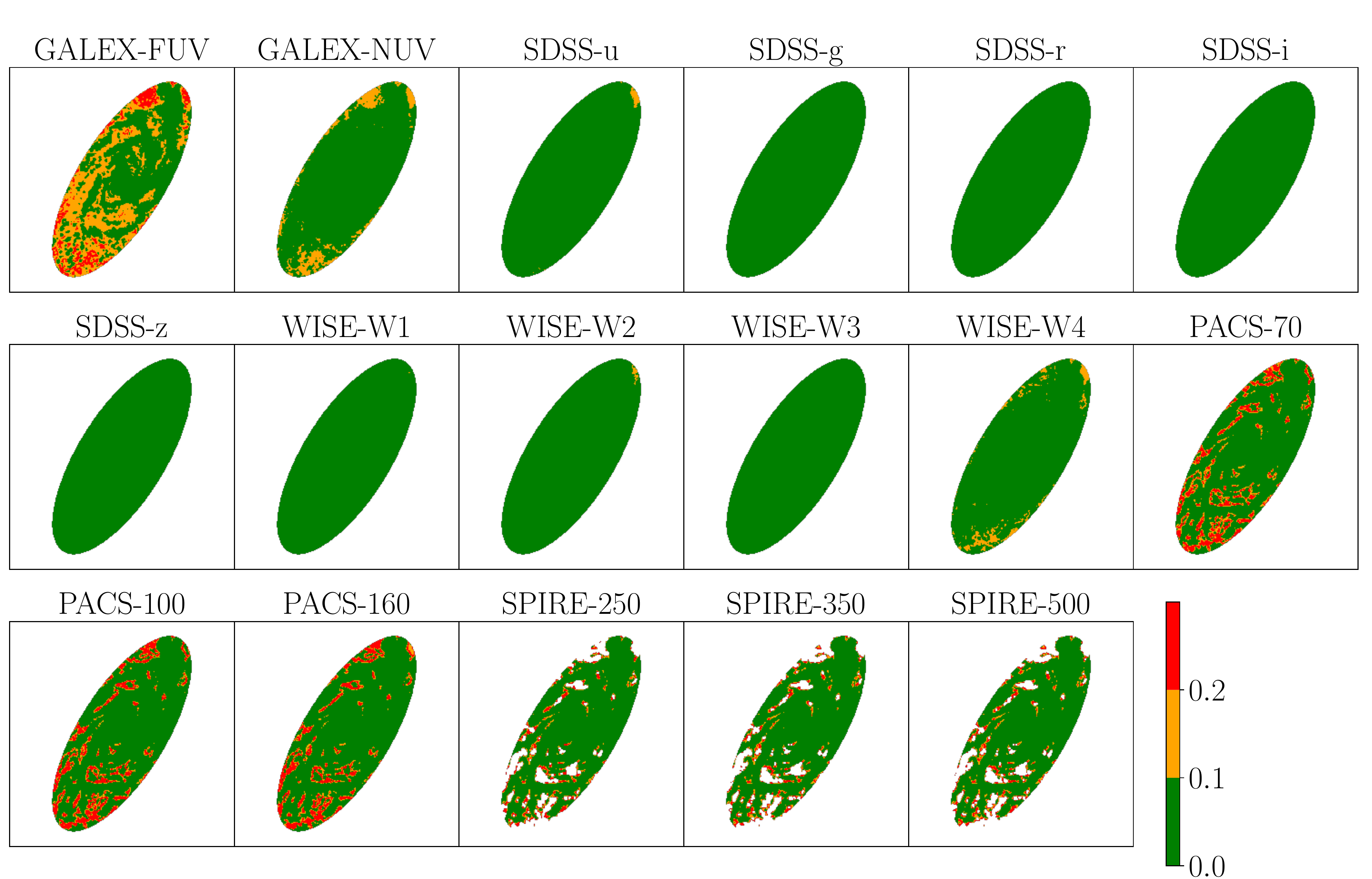}
\caption{Relative error ($R$) maps for Auriga galaxy AU-16 in different broadband filters for a nearly face-on view. This simulation uses our fiducial setup with $2\times10^{10}$ photon packets and including instruments with $2048\times2048$ pixels. The color scale indicates reliable data (green, $R<0.1$), questionable data (orange, $0.1<R<0.2$) and unreliable data (red, $R>0.2$).} 
\label{RStatistic.fig}
\end{figure}

\subsection{Dust grid selection}
\label{DustGrid.appendix}

In order to select the most efficient dust grid for the simulations, we run tests using the native Voronoi grid imported from the Auriga simulation and re-sampled octree grids with different subdivision characteristics. 
We compare the results to an extremely high-resolution octree grid. 
This grid has an imposed maximum dust threshold fraction of $10^{-7}$ in any given cell and is allowed to have cells with a refinement up to level 20 of the octree.
We use $5 \times 10^{8}$ photon packets for this test, which is significantly more than our fiducial setup (see Sect.~\ref{SEDphotonsConvergence.appendix}).
These tests show that an octree grid with a maximum of 12 allowed subdivisions and a maximum cell dust fraction value of $10^{-6}$ offers an optimal balance between accuracy and speed. 
Fig.~\ref{Photon_and_grid_convergence.fig} shows  broadband flux residuals in twenty bands for three Auriga galaxies.

\subsection{Photon packet convergence}
\label{SEDphotonsConvergence.appendix}

During the calibration of the post-processed Auriga models, we use spatially integrated broadband SEDs.
In order to verify convergence for these SEDs, we run multiple simulations with a varying number of photon packets. Fig.~\ref{Photon_and_grid_convergence.fig} shows the broadband flux residuals for three Auriga galaxies. Convergence is reached at $5 \times 10^{7}$ photon packets, which is the number of photon packets used for our calibration procedure to determine the post processing free parameters, $f_{dust}$ and $\tau_{clear}$ (see Sect.~\ref{calibration.sec}).

To determine the appropriate number of photon packets when generating high resolution broadband images, we run a number of test simulations for a limited set of Auriga galaxies. We calculate the relative error $R$ \citep{2020A&C....3100381C} on a pixel-to-pixel basis for a representative set of broadband images. 
According to \citet{2020A&C....3100381C}, the corresponding results are considered reliable for $R<0.1$. In the range $0.1 < R < 0.2$, results are questionable, and for $R > 0.2$, results are unreliable.
Fig.~{\ref{RStatistic.fig}} shows the $R$ values calculated for Auriga galaxy AU-16 in various broadband filters for a \texttt{SKIRT} simulation using $2\times10^{10}$ photon packets. In most of the bands, the SNR is sufficiently high. 
Some bands, in particular in the UV and submm spectral ranges, have $R>0.2$ in certain regions, indicative of the very low flux level of those particular regions of the image. The SNR of individual images can always be increased by spatial binning of the published images if required.

\subsection{\texttt{CIGALE} fitted fluxes: deviations from input data}
\label{CIGALE.appendix}
\begin{figure}
\centering
\includegraphics[width=\columnwidth]{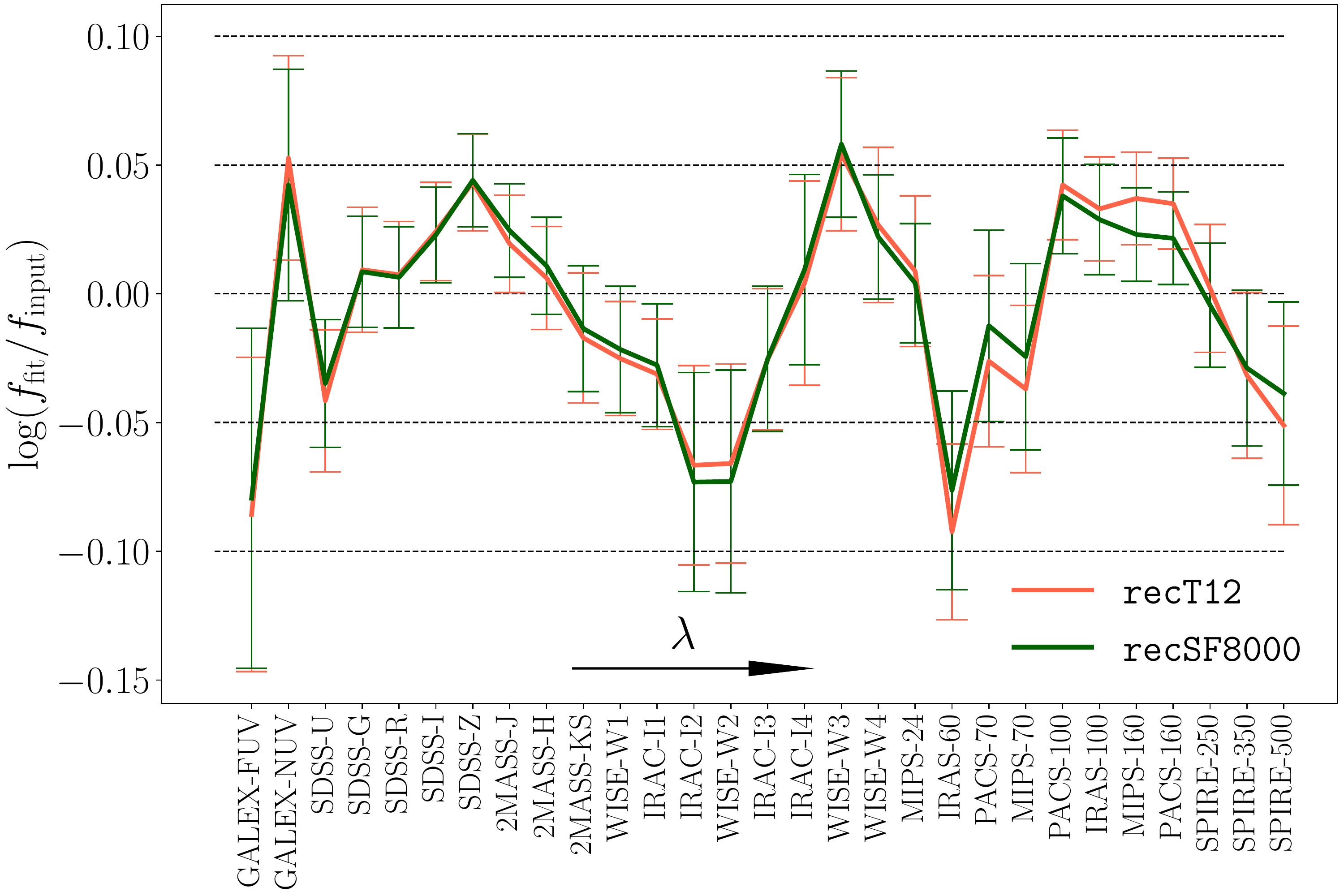}
\caption{Ratio of the fluxes obtained with \texttt{CIGALE} fitting versus the input fluxes. Error bars represent the standard deviation for each band. The mean and the standard deviation have been calculated using all thirty Auriga galaxies and all 18 orientations available per galaxy.} 
\label{CIGALEdeviations.fig}
\end{figure}

The \texttt{CIGALE} model SED provides a good match to the synthetic broadband fluxes produced by \texttt{SKIRT} for all Auriga galaxies. Fig.~\ref{CIGALEdeviations.fig} shows the mean deviation between the mock fluxes and the fitted \texttt{CIGALE} fluxes for both dust allocation recipes. Most of the fitted values deviate by less than $0.1 ~\mathrm{dex}$ from the input data. 

\subsection{\texttt{StatMorph} results with lower noise levels}
\label{StatMorph.appendix}
\begin{figure}
\centering
\includegraphics[width=\columnwidth]{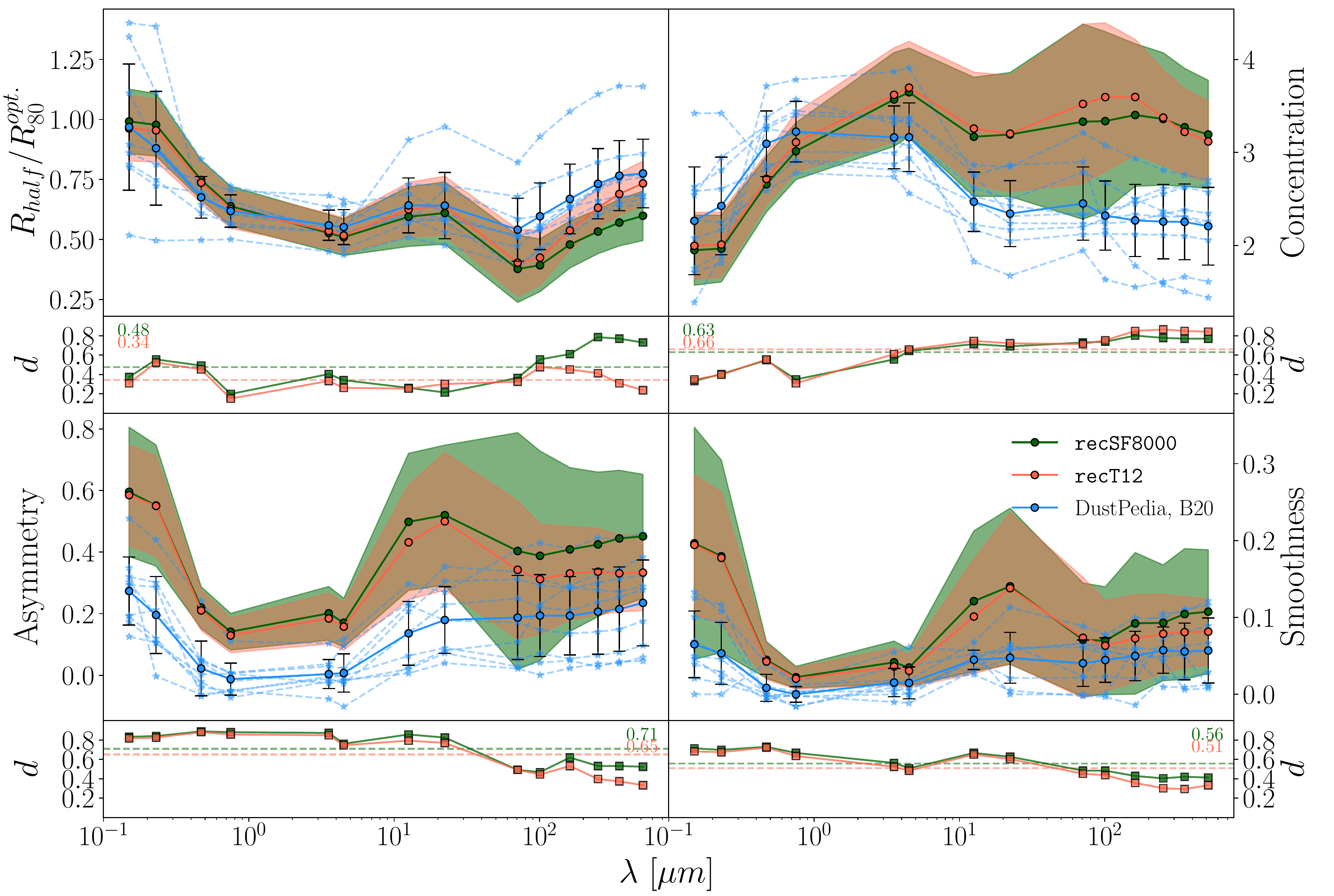}
\caption{Same as Fig.~\ref{StatMorph_CASGMvsWavelength.fig} but with a lower level of homogeneous Gaussian noise. A notable change in the asymmetry and smoothness values is seen in comparison to Fig.~\ref{StatMorph_CASGMvsWavelength.fig} .} 
\label{StatMoph_wavelength_plotLowNoise.fig}
\end{figure}

In Sect.~\ref{statMorphResults.ssec}, the noise level is normalized to the maximum flux level, which is in almost all cases located in the central pixels. 
In the results shown in Fig.~\ref{StatMoph_wavelength_plotLowNoise.fig}, we mask the central 10 kpc, and normalize the noise to the maximum level outside the central pixels.

Particularly affected by the change in noise levels is the asymmetry index and the smoothness index.
The Auriga galaxies appear more asymmetric in comparison to the observed galaxies in this case, with \texttt{recT12} showing a better agreement with the observational data in the FIR. This trend was opposite in Fig.~\ref{StatMorph_CASGMvsWavelength.fig}. 
Apart from this, a higher difference in the smoothness values is seen in the FIR bands for the two dust recipes, \texttt{recSF8000} being more clumpy, which is in line with what is seen in Fig.~\ref{recipeComaprison_dustSigma.fig}.
The indices representing the central concentration and the galaxy size remain essentially unchanged.

%%%%%%%%%%%%%%%%%%%%%%%%%%%%%%%%%%%%%%%%%%%%%%%%%%

% Don't change these lines
\bsp	% typesetting comment
\label{lastpage}
\end{document}